\documentclass[10pt]{article}
\usepackage{ragged2e}       
\usepackage{soul}           
\usepackage{xcolor}
\usepackage{graphicx}
\usepackage{hyperref}
\usepackage{amsmath}
\usepackage{subcaption}
\usepackage{amssymb}
\usepackage{booktabs}
\usepackage{fullpage}
\usepackage[section]{placeins}
\usepackage{authblk}
\usepackage{tabularx}
\usepackage{enumerate}
\usepackage{comment}
\usepackage{enumitem}
\usepackage[numbers,sort&compress,round]{natbib}%\usepackage[round]{natbib}

\newcommand{\supplementarysection}{%
  \renewcommand{\thesection}{S\arabic{section}}
}

\newcommand{\supplementaryfigure}{%
  \renewcommand{\thefigure}{S\arabic{figure}}
}

\definecolor{my_purple}{RGB}{255,0,255}
\graphicspath{{figures_Neuromech_V5/}}

\definecolor{REDCOLOR2}{RGB}{0,0,0}

\title{\textbf{\Large Neurological disorders leading to mechanical dysfunction of the esophagus: an emergent behavior of a neuromechanical dynamical system}}

\author[1]{\normalsize Guy Elisha}
\author[2,3]{\normalsize Sourav Halder}
\author[4]{\normalsize Xinyi Liu}
\author[2,3]{\normalsize Dustin A. Carlson}
\author[2,3]{\normalsize Peter J. Kahrilas}
\author[2,3]{\normalsize John E. Pandolfino}
\author[1,4]{\normalsize Neelesh A. Patankar\thanks{Corresponding author: N.~A.~Patankar (\texttt{n-patankar@northwestern.edu})}}

\affil[1]{Department of Mechanical Engineering, Northwestern University, Evanston, IL, USA \vspace{1ex}}
\affil[2]{Division of Gastroenterology and Hepatology, Feinberg School of Medicine, Northwestern University, Chicago, IL, USA \vspace{1ex}}
\affil[3]{Kenneth C. Griffin Esophageal Center, Feinberg School of Medicine, Northwestern University, Chicago, IL, USA \vspace{1ex}}
\affil[4]{Department of Engineering Sciences and Applied Mathematics, Northwestern University, Evanston, IL, USA \vspace{1ex}}

\date{}

\begin{document}

\captionsetup[figure]{labelfont={bf},name={Figure},labelsep=period}
\captionsetup[table]{labelfont={bf},name={Table},labelsep=period}
%\captionsetup[figure]{labelfont={bf},name={Figure},labelsep=space}

\maketitle 

\begin{abstract}
\noindent An understanding how neurological disorders lead to mechanical dysfunction of the esophagus requires knowledge of the neural circuit of the enteric nervous system. Historically, this has been elusive. Here, we present an empirically guided neural circuit for the esophagus. It has a chain of unidirectionally coupled relaxation oscillators, receiving excitatory signals from stretch receptors along the esophagus. The resulting neuromechanical model reveals complex patterns and behaviors that emerge from interacting components in the system. A wide variety of clinically observed normal and abnormal esophageal responses to distension are successfully predicted. Specifically, repetitive antegrade contractions (RACs) are conclusively shown to emerge from the coupled neuromechanical dynamics in response to sustained volumetric distension. Normal RACs are shown to have a robust balance between excitatory and inhibitory neuronal populations, and the mechanical input through stretch receptors. When this balance is affected, contraction patterns akin to motility disorders are observed. For example, clinically observed repetitive retrograde contractions emerge due to a hyper stretch sensitive wall. Such neuromechanical insights could be crucial to eventually develop targeted pharmacological interventions.

\end{abstract}


%%%%%%%%%%%%%%%% INTRO %%%%%%%%%%%%%%%%%
\section{Introduction}\label{Introduction}

An understanding of how neurological disorders lead to mechanical dysfunction of organs remains an open problem \cite{Trayanova2011,Du2010,buijs2016circadian,sundar2015circadian}. Specifically, gastrointestinal and esophageal motility disorders (EMDs) are neurologically driven mechanical dysfunctions affecting approximately 35 million Americans \cite{Rao2016,Drossman1993,Ouyang2007,Du2010}. These disorders have seen an alarming rise in prevalence due to the opioid epidemic. About $40\%-60\%$ of chronic opioid users experience constipation, and $33\%$ suffer from esophageal reflux \cite{Abramowitz2013,Tuteja2010Opioid}. Opioid-induced gastrointestinal motility disorder is known to be of neurological origin but with scarce understanding of how mechanical dysfunction emerges \cite{Lydon1999,Snyder2023,sobczak2014physiology}.

% Do we need the following?
These gaps in knowledge often lead to misinterpretations of disorders. Therefore, impeding the development of effective neurologically focused treatment approaches \cite{ten2008modelling,du2018progress,hardeland2012melatonin,shahriari2020emerging}. Establishing a foundational understanding of the emergent behavior of organs may play a vital role in proposing targeted solutions. For instance, the use of selected peripherally acting opioid antagonists may be recommended for clinical study \cite{thomas2008opioid,albert2016Opioid,al2011molecular}.

%The gap in knowledge of how the mechanical dysfunctions emerges due to the underlying neurological disorders often leads to misinterpretations of disorders. Therefore, impeding the development of effective neurologically focused treatment approaches \cite{ten2008modelling,du2018progress,hardeland2012melatonin,shahriari2020emerging}. Establishing a foundational understanding of the emergent behavior of organs may play a vital role in proposing targeted solutions. For instance, the use of selected peripherally acting opioid antagonists may be recommended for clinical study \cite{thomas2008opioid,albert2016Opioid,al2011molecular}.

Uncovering the pathogenesis of EMDs requires an understanding of the enteric nervous system neural circuit that controls peristalsis \cite{clave2015dysphagia,seguella2021enteric}. Historically, this has been elusive \cite{Park1999,fung2020functional,kulkarni2018advances}. 
Peristaltic motion is pervasive in the gastrointestinal tract (esophagus, stomach, intestines) \cite{Park1999,Goyal2008,spencer2020enteric,patel2023physiology}. 
Particularly in the esophagus, it has been known that there are two major types of neurons – those that stimulate contraction of esophageal muscle and another that inhibit muscle contractions \cite{Mittal2016,Sifrim2012,Woodland2013}. 
However, beyond a few overarching concepts, the peristaltic neural circuit has remained unresolved \cite{paterson2006esophageal}.

Clinically, esophageal motility has been investigated with high-resolution manometry, a technology utilizing closely spaced pressure transducers positioned along the length of the esophagus \cite{pandolfino2009high}. High-resolution manometry is an excellent method for quantifying the strength and timing of esophageal contractions \cite{Gorti2020,yadlapati2021esophageal}. 
However it cannot yield specific information about inhibition other than in tonically contracted sphincter \cite{Carlson2021Evaluating,Gorti2020,yadlapati2021esophageal}.

%However, it yields minimal information on inhibition other than in tonically contracted sphincters \cite{Carlson2021Evaluating,Gorti2020,yadlapati2021esophageal}.  

\begin{figure*}[!htb]
    \centering{{\includegraphics[trim=0 0 0 0 ,clip,width=0.95\textwidth]{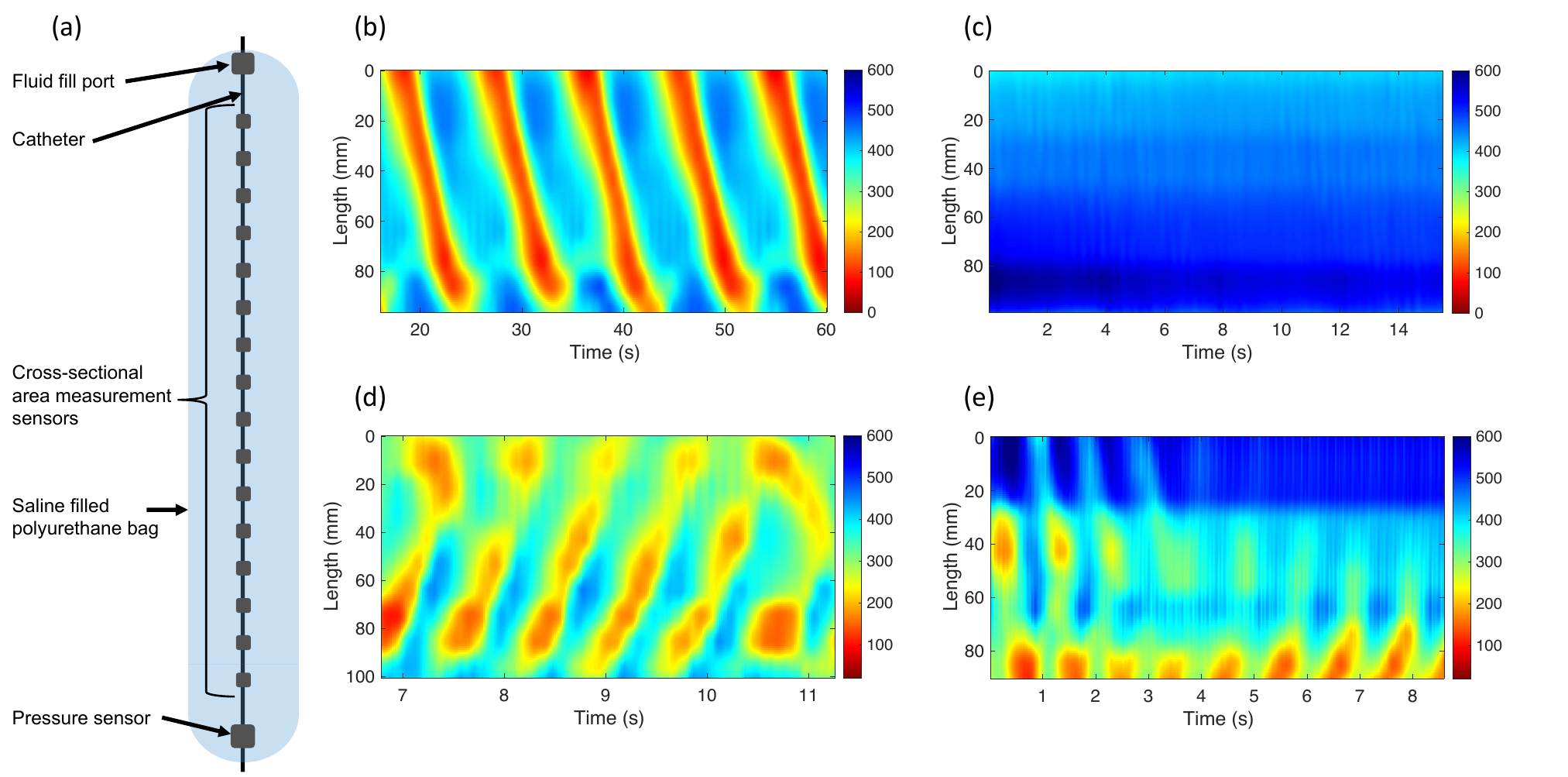}}}
    \caption{\textbf{(a)} Diagram of functional luminal imaging probe device. \textbf{(b, c, d,} and \textbf{e)} esophageal topography with color-coded cross-sectional area by axial length by time showing different distension-induced contractility patterns. \textbf{(b)} Repetitive antegrade contractions pattern from an asymptomatic control. \textbf{(c)} Absent contractile response, \textbf{(d)} repetitive retrograde contractions, and \textbf{(e)} disordered contractions characterized by sporadic or chaotic pattern, not meeting antegrade nor retrograde contractions. Figures used with permission from the Esophageal Center at Northwestern. The displayed topography exclusively represents data collected from the uppermost 60-70\% of the entire length recorded by FLIP. FLIP data includes cross-sectional area measurements at the lower esophageal sphincter, which have been intentionally omitted from these figures since the primary focus of the study is the esophagus body.}
    \label{fig:FLIP}
\end{figure*}

More recently, esophageal motility has been investigated with functional lumen imaging probe (FLIP) Panometry \cite{Donnan2020,savarino2020use}. 
The FLIP device comprises a catheter surrounded by a fluid-filled bag which is placed within the esophageal lumen, measuring cross-sectional area at different locations over time %\cite{Carlson2015}
(Fig. \ref{fig:FLIP}a). 
The normal response elicited during a FLIP study (sustained volumetric distension) is of repetitive antegrade contractions (RACs, Fig. \ref{fig:FLIP}b) \cite{Hirano2017}. 
FLIP studies exhibiting patterns other than RACs can also occur and are considered abnormal, potentially indicative of an EMD \cite{Donnan2020} (Fig. \ref{fig:FLIP}). 
Thus, FLIP has garnered attention as a potential diagnostic tool \cite{savarino2020use}.

%%%%%%%FLIP %%%%%%%%%%

It has been proposed, based on clinical data, that RACs is a form of secondary peristaltic response to non-transient esophageal distension \cite{Carlson2015Utilizing}. This response involves esophageal contractions that are provoked by local distention and that occur independently of the central nervous system \cite{Park1999,Woodland2013}. However, a complete understanding of how this pattern emerges and the involved mechanisms remains unresolved \cite{Carlson2020Repetitive}, and there is currently no known neural circuit that can explain this pattern \cite{Goyal2022}. Moreover, due to this lack of understanding, the emergence of abnormal patterns remains unclear \cite{carlson2018mechanisms,Carlson2020,Halder2022_VDL}. Clarifying the connection between neural signals and the mechanical dysfunctions observed can provide valuable insights into the nature of these disorders, crucial for developing targeted pharmacological interventions.

In this study, we propose a first-of-its-kind \textit{empirically guided} organ-scale neuromechanical model for the esophagus that predicts and explains a broad repertoire of esophageal motility patterns. Organ-scale neuromechanical models prove invaluable for unraveling the intricate patterns and behaviors that emerge from the interactions of individual components within a system, shedding light on how their failure leads to mechanical dysfunctions \cite{Mercado2022,Nishikawa2007,ijspeert2008central}.  Through this model, we aim to provide a theoretical framework capable of explaining the essential features observed in clinical FLIP studies. Further, we use this model to reveal the underlying mechanisms associated with normal and abnormal FLIP contraction patterns, hopefully providing insight into the pathogenesis of EMDs.

\section{Results}\label{Results}

\begin{figure*}[!htb]

    \centering
    \begin{subfigure}[b]{0.65\textwidth}
        \centering
        \includegraphics[trim=0 0 0 0,clip,width=\textwidth]{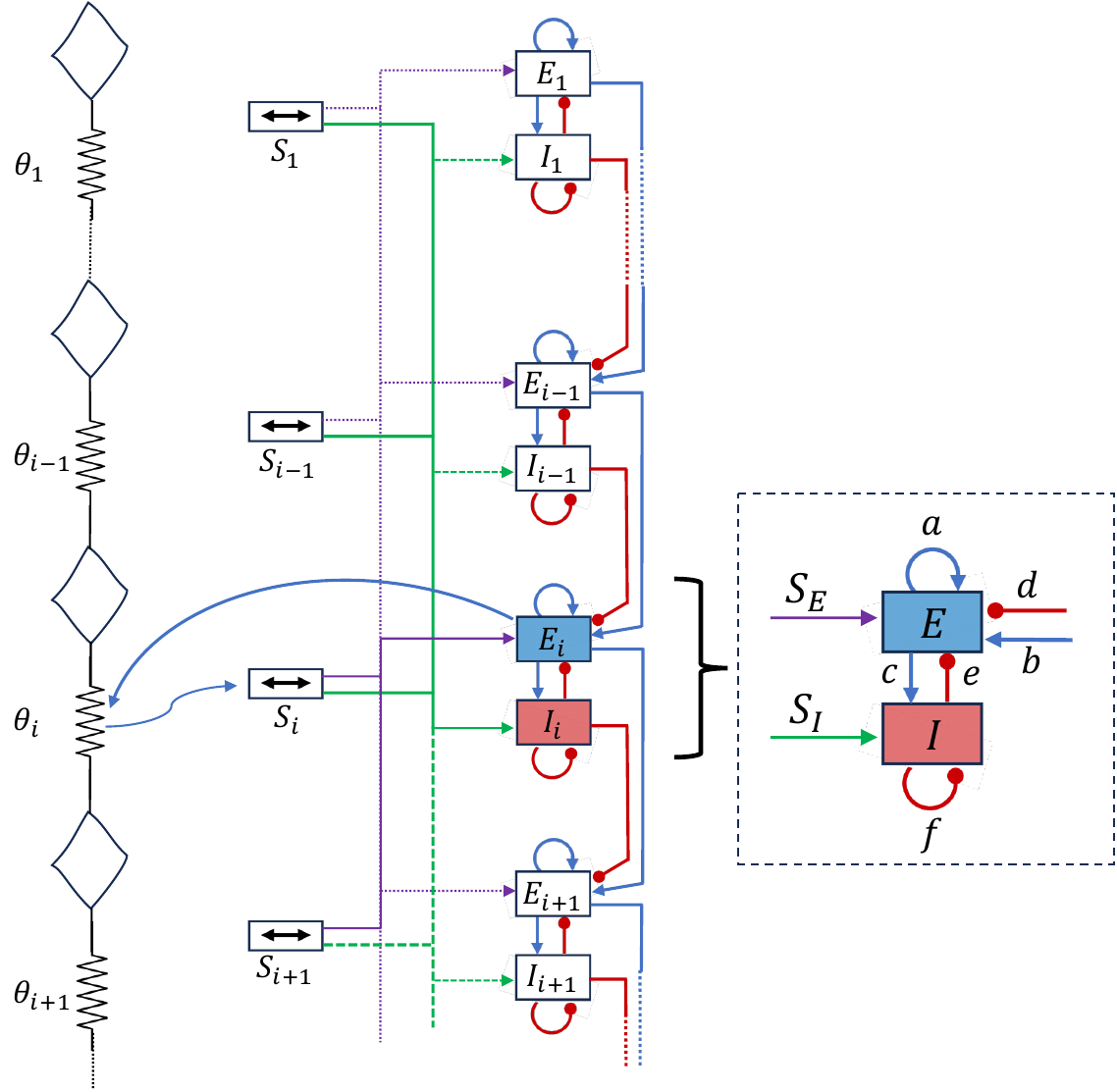}
        \caption{}
        \label{fig:circuitMain}
    \end{subfigure}
\
    \begin{subfigure}[b]{0.25\textwidth}   
        \centering 
        \includegraphics[trim=0 0 0 0,clip,width=\textwidth]{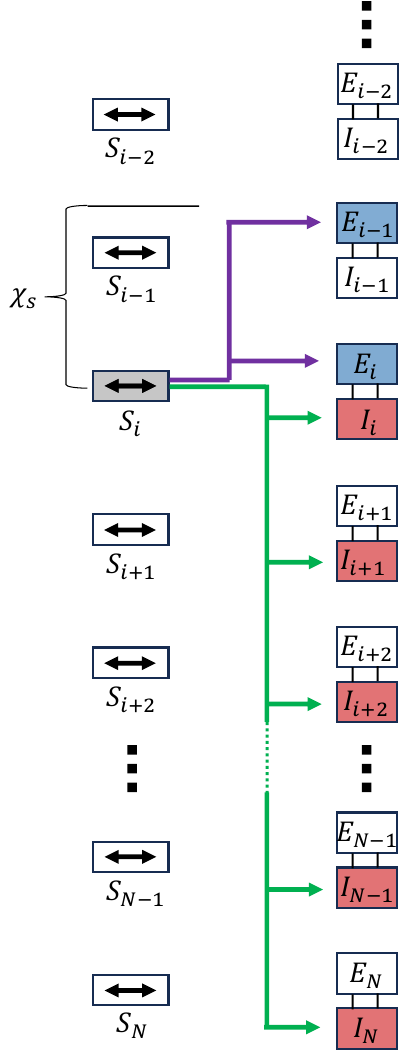}
        \caption{}
        \label{fig:circuitStretch}
    \end{subfigure}
    \caption{Schematic of the model. \textbf{(a)} An overview on the neuromechanical model consisting of the wall mechanics (chain of masses), neural circuitry ($E$ and $I$), and the coupling dynamic ($\theta$ and $S$). The neural network is composed of $N$ interconnected segments, each consisting of an excitatory ($E$) and an inhibitory ($I$) neuronal populations. Lines with circular head mark inhibitory synapses, and arrows denote excitatory synapses. The coupling mechanism is demonstrated on segment $i$, where distension activates mechanoreceptors (arrow from the muscle to the corresponding mechanoreceptors $S_i$), and local excitation actuates local body segments (arrow from $E_i$ to the corresponding muscle section). \textbf{(b)} Schematic showing the stretch induced innervation. When distension at $i$ creates sufficient strain on the walls, the corresponding mechanoreceptor ($S_i$) sends excitatory signal to distal inhibitory populations and proximal excitatory populations.}
    \label{fig:schenatic}
\end{figure*}

The results reported below are obtained based on the empirically guided neural circuit shown in Fig. \ref{fig:circuitMain} \& \ref{fig:circuitStretch}. The circuit and the corresponding neuromechanical mathematical model are described in the Methods section.

%%%%

%%%%%%%% RACs %%%%%%%%%
\subsection{Repetitive antegrade contractions} 

\begin{figure*}[!htb]
    \centering{{\includegraphics[trim=0 230 0 250 ,clip,width=0.7\textwidth]{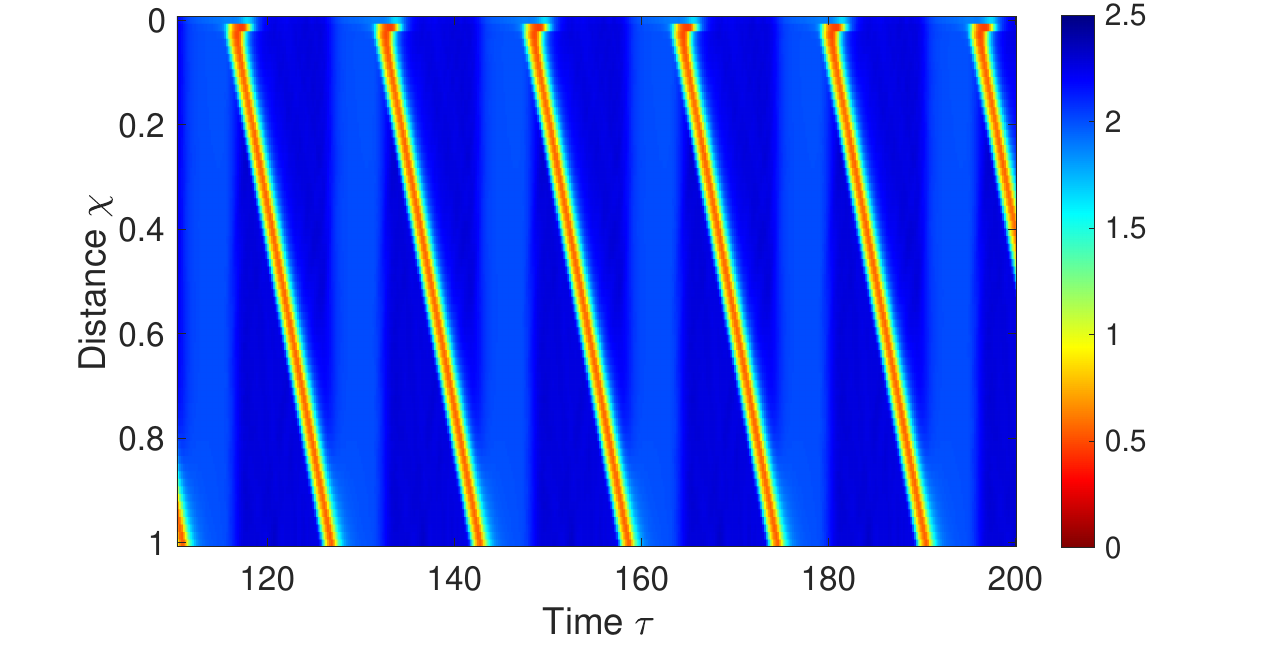}}}
    \caption{Esophageal spatio-temporal topography, depicting color-coded cross-sectional area, obtained through a neuromechanical model simulation of FLIP Manometry. The model exhibits a repetitive, antegrade contraction pattern autonomously triggered, mirroring observations from clinical FLIP measurements.}
    \label{fig:RACs}
\end{figure*}

Figure \ref{fig:RACs} presents a normal RACs pattern obtained through simulations of the mathematical model, qualitatively reproducing the key elements observed in RACs: repetitiveness, forward (antegrade) propagation, and non-overlapping contractions. The contractions emerge and are sustained autonomously and independently of the central nervous system or any externally prescribed input. We establish Fig. \ref{fig:RACs} as the baseline case. In the following sections, we use the model to explain the triggering mechanism of the pattern and the development of its essential elements, revealing the underlying dynamics of RACs.

%%%%%%%Normal RACs %%%%%%%%%%

\begin{figure*}[!htb]%[!htb]
    \centering
    \begin{subfigure}[b]{0.3\textwidth}
        \centering
        {\includegraphics[trim=30 240 60 250,clip,width=\textwidth]{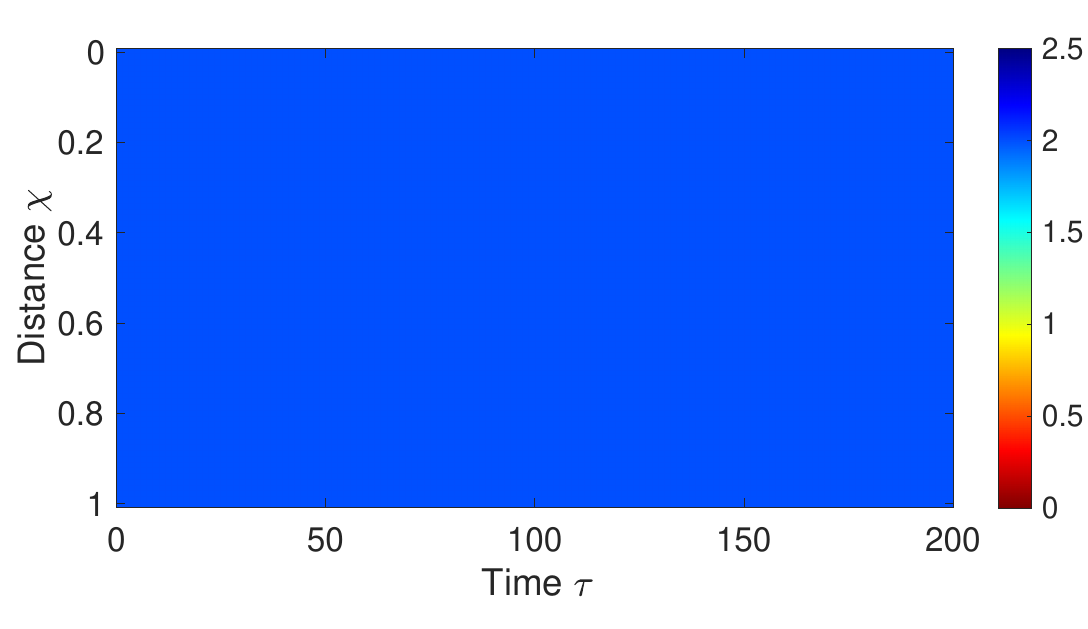}}
        \caption{}
        \label{fig:absent}
    \end{subfigure}
    \hfill
    \begin{subfigure}[b]{0.3\textwidth}  
        \centering 
        {\includegraphics[trim=30 240 60 250,clip,width=\textwidth]{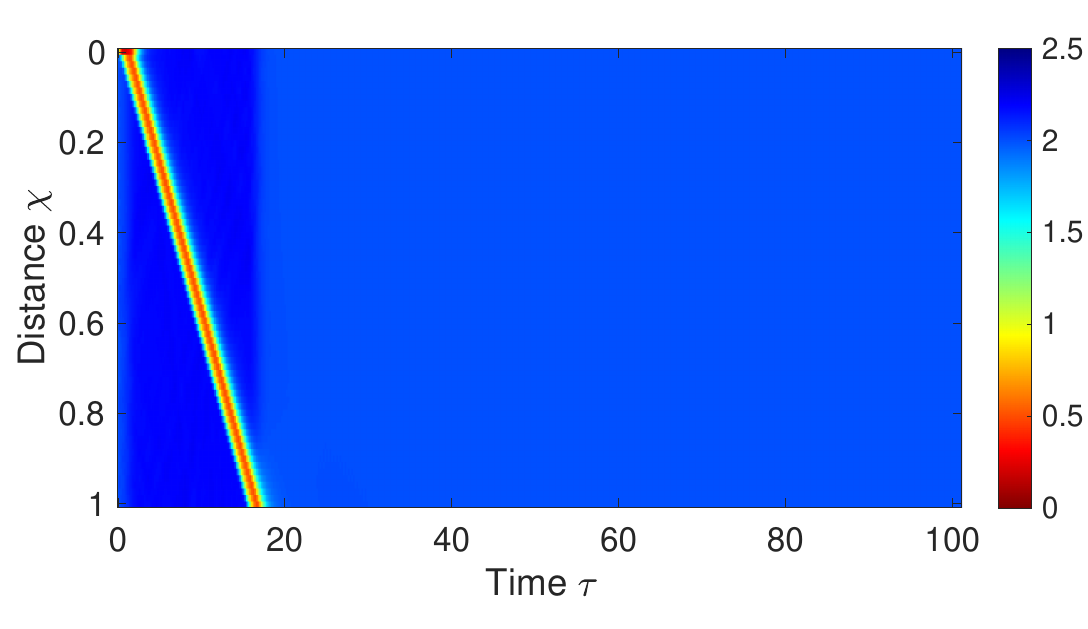}}
        \caption{}
        \label{fig:singleCont}
    \end{subfigure}
     \hfill
         \begin{subfigure}[b]{0.3\textwidth}  
        \centering 
        {\includegraphics[trim=30 240 60 250,clip,width=\textwidth]{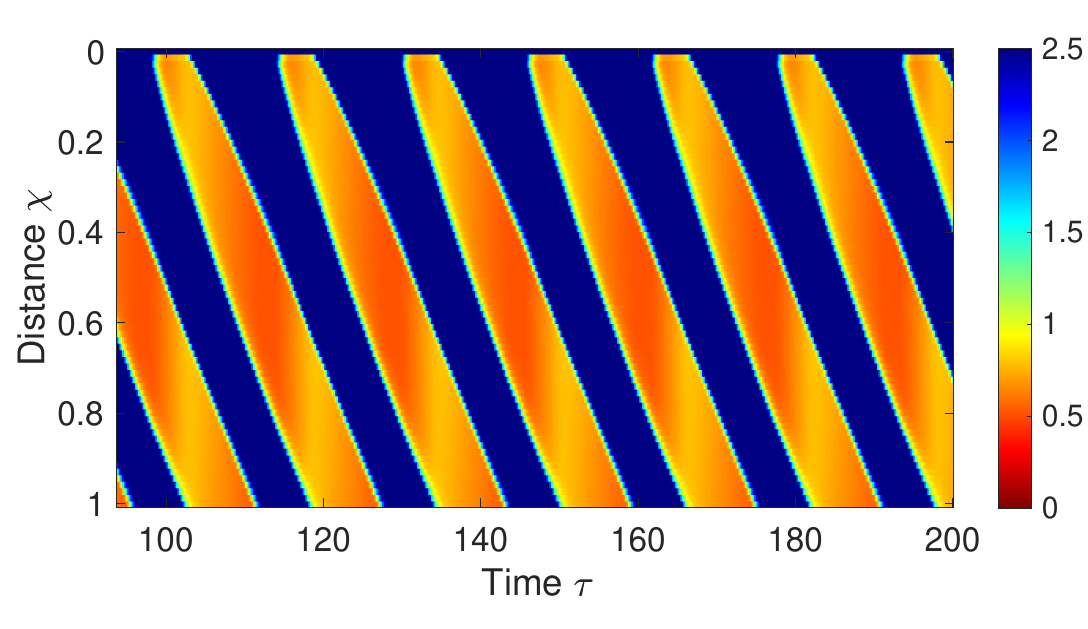}}
        \caption{}
        \label{fig:overlap1}
    \end{subfigure}
     \hfill
    \begin{subfigure}[b]{0.3\textwidth}  
        \centering 
        {\includegraphics[trim=30 240 60 250,clip,width=\textwidth]{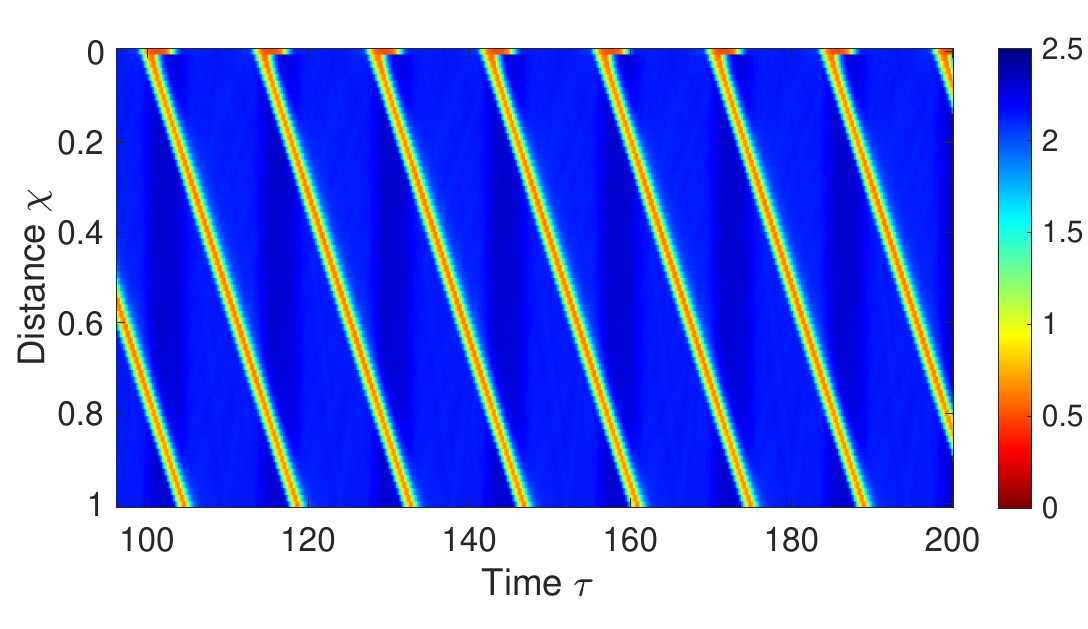}}
        \caption{}
        \label{fig:overlap2}
    \end{subfigure}
        \hfill
    \begin{subfigure}[b]{0.3\textwidth}  
        \centering 
        {\includegraphics[trim=30 240 60 250,clip,width=\textwidth]{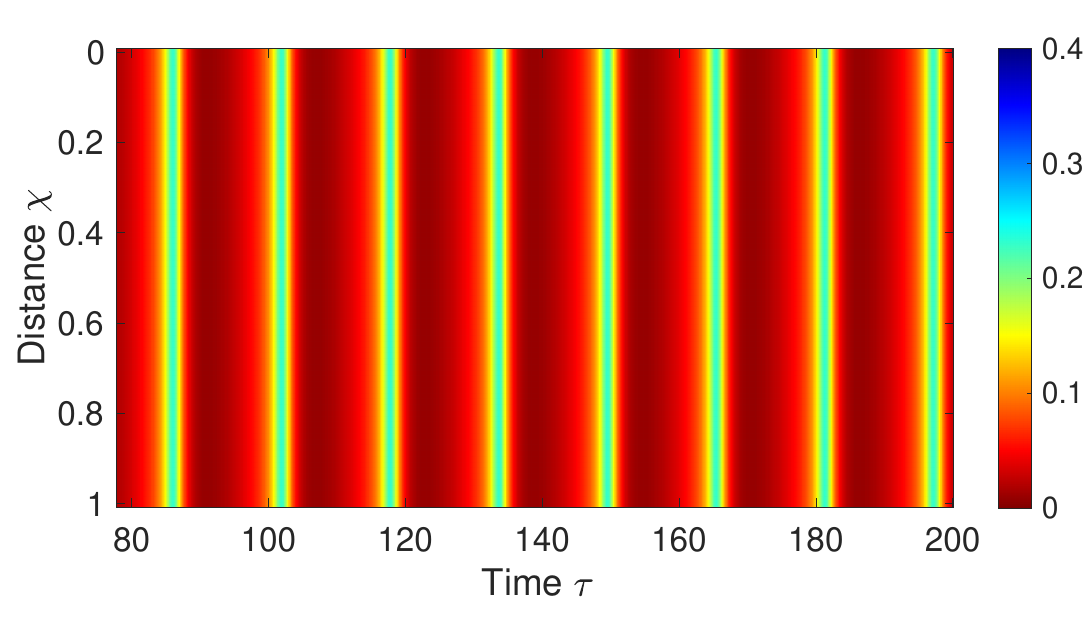}}
        \caption{}
        \label{fig:uniform}
    \end{subfigure}
     \hfill
              \begin{subfigure}[b]{0.3\textwidth}  
        \centering 
        {\includegraphics[trim=30 240 60 250,clip,width=\textwidth]{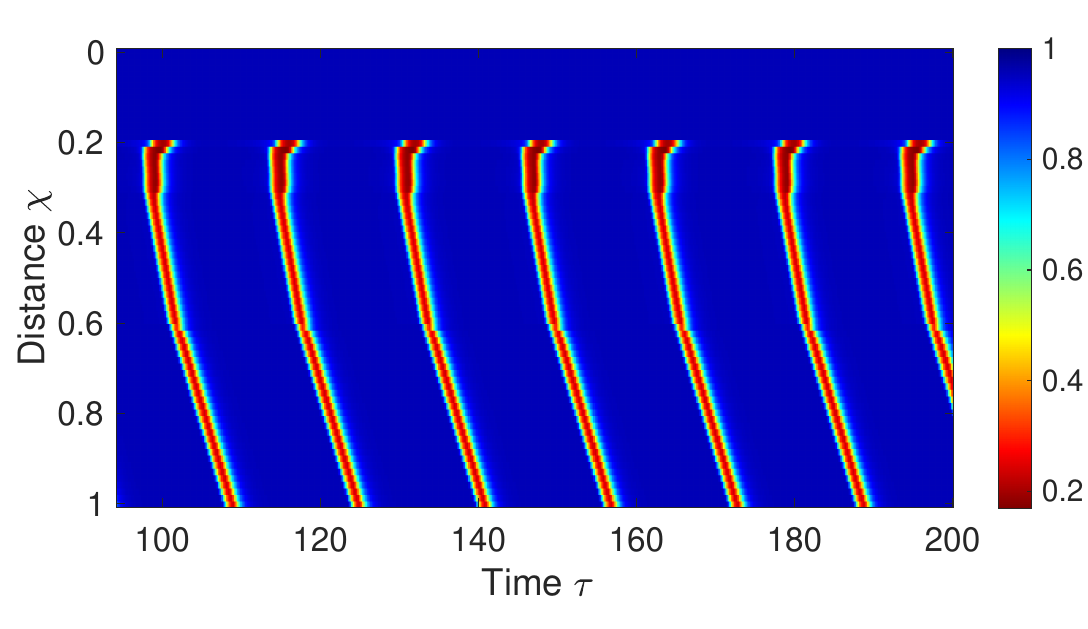}}
        \caption{}
        \label{fig:shortBagTrav}
    \end{subfigure}
        \hfill
         \begin{subfigure}[b]{0.3\textwidth}  
        \centering 
        {\includegraphics[trim=30 240 60 250,clip,width=\textwidth]{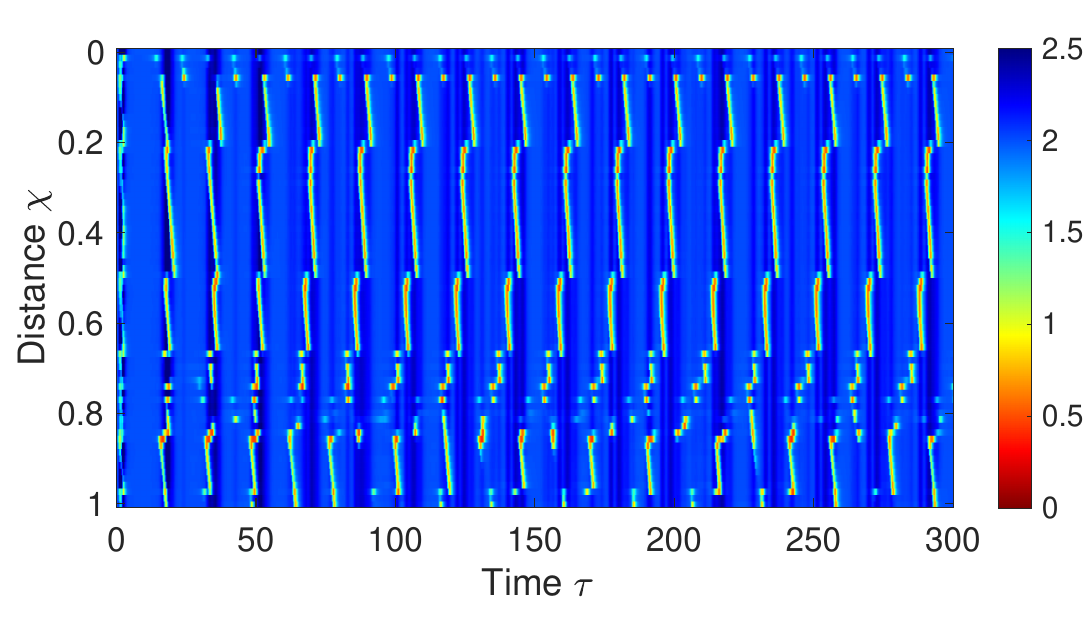}}
        \caption{}
        \label{fig:spastic1}
    \end{subfigure}
        \hfill
    \begin{subfigure}[b]{0.3\textwidth}  
        \centering 
        {\includegraphics[trim=30 240 60 250,clip,width=\textwidth]{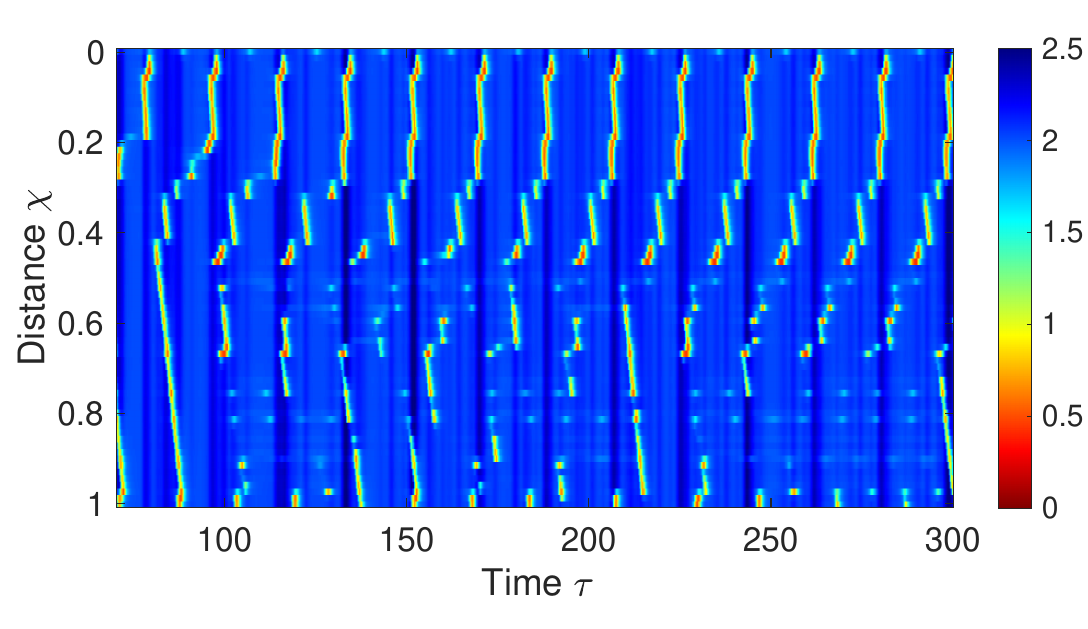}}
        \caption{}
        \label{fig:spastic2}
    \end{subfigure}
     \hfill
         \begin{subfigure}[b]{0.3\textwidth}  
        \centering 
        {\includegraphics[trim=30 240 60 250,clip,width=\textwidth]{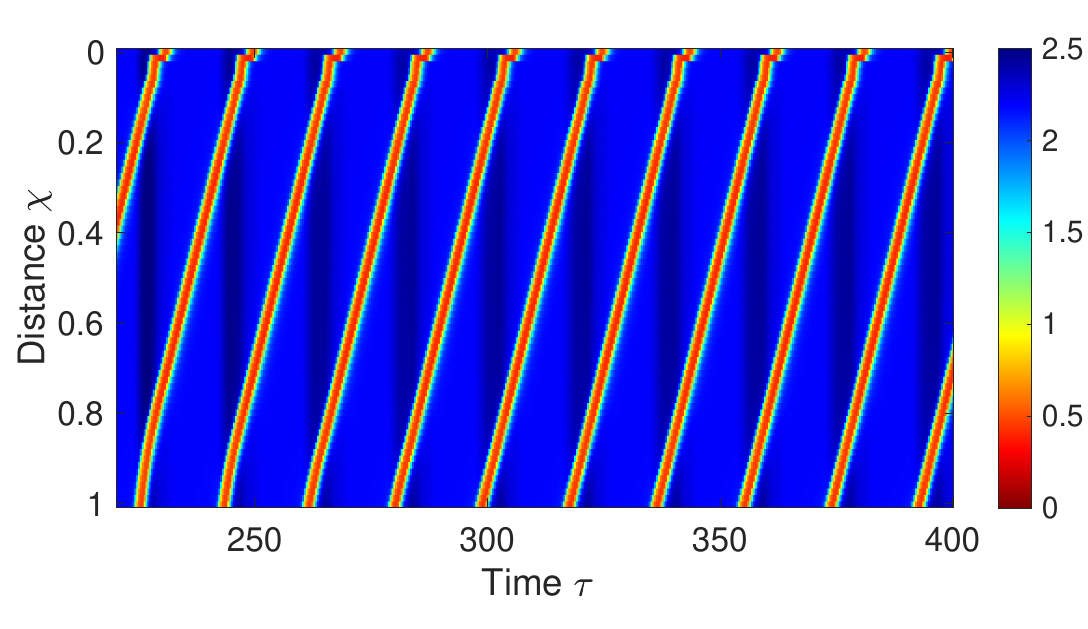}}
        \caption{}
        \label{fig:RRC}
    \end{subfigure}
    \caption{Color-coded spatio-temporal topographies depicting cross-sectional area (CSA), excitatory activity level ($E$), or muscle contraction pattern ($\theta$) obtained through a neuromechanical model mimicking esophageal distension tests. \textbf{(a)} CSA. Stretch receptors are disabled, and thus the entire esophagus remains quiescent. This is obtained through either significantly increasing sensitivity threshold ($\hat{\alpha}$) or decreasing the strength of sensory feedback to excitatory neurons ($w_E$). \textbf{(b)} CSA. Stretch receptors are disabled and an excitatory input (equivalent to the value of $w_E$) is introduced at the proximal end for a short period. One contraction emerges before the esophagus recovers and remains at rest. \textbf{(c)} CSA. Reducing inhibitory signal from anterior inhibitory population ($d$) creates overlapping contraction by increasing the activity duration of local excitatory population. \textbf{(d)} CSA. Reducing excitatory activation threshold ($\phi_E$) results in overlapping contractions though forming early excitatory spikes. \textbf{(e)} Heat map of excitatory neuronal population activity. Disabling neuronal neighboring connections (setting $b=d=0$) results in uniform excitation along the entire length.  \textbf{(f)} $\theta$. Applying sustained esophageal distension using a short bag and setting the stretch-induced innervation to inhibitory neurons locally. That is, $S_I=0$ distal to the bag. \textbf{(g)} CSA. Chaotic response created by irregularities in inhibitory neuronal pathways. \textbf{(h)} CSA. Another example of chaotic response to irregularities in inhibitory neuronal pathways (see text). \textbf{(i)} CSA. Repetitive retrograde contractions obtained by disabling excitatory innervation of inhibitory neuronal population through mechanoreceptors ($S_I$).} 
    \label{fig:contractionPatterns}
\end{figure*}

Excitatory signals from stretch receptors ($S_E$) serve as the primary initiators of RACs and play a pivotal role in sustaining the repetitive pattern. Figure \ref{fig:absent} demonstrates the consequence of disabling stretch receptors, resulting in a lack of contractility due to insufficient excitatory inputs. Figure \ref{fig:singleCont} illustrates a scenario where stretch receptors are disabled, and a brief excitatory input is introduced at the proximal end of the esophagus. In the model's context, this is equivalent to setting $S_E = S_I = 0$ everywhere except the proximal end, where $S_E\neq0$ for a short time period. As depicted, the proximal excitatory input travels down the esophageal length, leading to a single, propagating contraction. However, without stretch receptors, there is no sustained excitatory input to reinitiate additional contractions.

The emergence of the repetitive, rhythmic attribute observed in FLIP studies is mathematically elucidated through the model. Let's focus on section $i$, positioned along the length of the distended esophagus. Adequate sustained volumetric distension around $i$ triggers the activation of local stretch receptors. These receptors, in turn, innervate excitatory neurons at $i$ (via $S_{E,i}$), initiating the excitatory phase. The activated excitatory neurons then stimulate both excitatory and inhibitory cells at location $i$. Over time, the activity of inhibitory cells surpasses a threshold level, triumphing over both excitatory and inhibitory activity and inducing a refractory period where all cells cease activity. The sustained distension, resulting in constant innervation from stretch, initiates a new cycle when all cells are inactive \cite{Wilson1972}. Note that distension-induced innervation also occurs in inhibitory pathways ($S_I$), ensuring that inhibition precedes excitation.

The primary determinant governing the forward propagation of contractions is the presence of unidirectional connections ($b$ and $d$), as they are the only parameters introducing asymmetry into the system \cite{kopell1986symmetry,strogatz1993coupled}. In the absence of these connections, the system exhibits repetitive but non-propagating contractions (Fig. \ref{fig:uniform}). The constant input from mechanoreceptors to the excitatory population ($S_E$) ensures that each oscillator operates at its natural frequency, generating an individual limit cycle. Therefore, rhythmic behavior persists even without unidirectional connections (Fig. \ref{fig:uniform}).

Unidirectional coupling introduces an additional stimulus, which in an isolated case (considering a single oscillator in the chain) can be viewed as periodic perturbations. When a stable limit cycle experiences a perturbation, it introduces transient changes which quickly decay to the original oscillatory activity. However, it returns with a phase shift relative to its unperturbed cycle. This adjustment of the phase of each oscillator is termed phase resetting \cite{kopell1988coupled,smeal2010phase}. In a coupled system, the phase shift eventually stabilizes to a constant value for all subsequent perturbations, leading to a phase-locked state between coupled oscillators. The delay between each oscillator increases with distance from the first oscillator, as out-of-phase oscillations accumulate \cite{izhikevich2000phase,kopell1986symmetry}. Consequently, the signal propagates through the chain as a wave, with each oscillator oscillating with a delay relative to its proximal neighbor \cite{kopell2003chains,cohen1992modelling,schwemmer2012theory}. Since muscle contraction follows excitatory signal's pattern (Eq.(\ref{eq:tht})), muscle contraction pattern appears as a propagating wave \cite{Diamant1997,Yazaki2012,Sifrim2012}. Note that the constant phase scenario is the solution to the system. The system is not each oscillator by itself; instead, it forms a new system—a coupled chain—with its own natural frequency and a phase difference between adjacent oscillators.

Lastly, the absence of overlapping patterns observed in normal RACs results from a balance among various factors, including input to the excitatory population (such as $w_E$ and $d$), inhibitory activity levels (controlled by $c$, $f$, and $w_I$), and the excitatory activation threshold ($\phi_E$). A parametric study reveals that overlapping contractions occur when the firing of excitatory populations dominates inhibitory firing (see section \ref{sec:parametricStudy} in supplementary). This leads to excitation in the proximal esophagus before the refraction of the distal esophagus, observed in two distinct ways. The first involves extending the excitatory phase. This is exemplified, for instance, by reducing the inhibitory signal from the anterior inhibitory population ($d$) (Fig. \ref{fig:overlap1}). The second entails allowing excitatory activity to spike faster, effectively shortening the refractory period. This is demonstrated, for instance, by reducing the excitatory activation threshold ($\phi_E$), requiring less innervation to activate excitatory cells (Fig. \ref{fig:overlap2}).

%%%%%%%%%% Contraction patterns and pathologies Figures %%%%%%%%%%%%%

%%%%%%%% Comparison with clinical data %%%%%%%%%
\subsection{Comparison with clinical data} 
 
Numerous studies over the years explored esophageal response to distension, offering insights into the typical reactions of the esophagus under various conditions. However, similar to RACs, many of these responses are non-intuitive \cite{Goyal2008,Mcmahon2007,Pedersen2005,Paterson1988,gregersen2011mechanical,penagini1996effect}. In this section, we showcase the capability of the proposed neural model to reproduce these diverse distension-induced esophageal scenarios, demonstrating its versatility beyond FLIP manometry. Importantly, we leverage the results to briefly explain these non-intuitive clinical observations in a healthy esophagus. We present four distinct scenarios representing a healthy esophageal response to distension:

\begin{comment}
\begin{enumerate}[\bfseries (i)]
  \item[\textbf{Case 1.}] \textbf{Transient esophageal distension:} Results in a single contraction \cite{Paterson1988}. The short stimulus can be introduced though abrupt inflation and deflation of a distending medium or through vagal efferent nerve stimulation.
  
   \item[\textbf{Case 2.}] \textbf{Prolonged esophageal distension using short balloon followed by abrupt deflation:} During the distension, the esophagus distal to the distended section is mostly inactive. Upon abrupt deflation, one phasic contraction appears, traveling down the length of the esophagus \cite{Christensen1970,Paterson1988}. The pressure and cross-sectional area variations along the distended region is not reported in these studies.
    
  \item[\textbf{Case 3.}] \textbf{Sustained distension of a short section of the esophagus:} Results in repetitive contractions along the distended section, as observed through a pressure probe located inside the fluid-filled distending bag \cite{gregersen2011mechanical,villadsen2001oesophageal}. The contractions' frequency remains independent of the bag's length, exhibiting a consistent rate of 6 contractions per minute, akin to the FLIP scenario \cite{gregersen2011mechanical,Carlson2020Repetitive}. The pressure variations distal to the distended region is not reported in these studies.
  
   \item[\textbf{Case 4.}] \textbf{Sustained esophageal distension along its entire length:} Results in RACs \cite{Carlson2020Repetitive,Carlson2020,Mcmahon2007}.  
  
\end{enumerate}
\end{comment}

\begin{enumerate}[label=\textbf{Case \arabic*.}, align=left, leftmargin=*]
  \item \textbf{Transient esophageal distension:} Results in a single contraction \cite{Paterson1988}. The short stimulus can be introduced though abrupt inflation and deflation of a distending medium or through vagal efferent nerve stimulation.
  
  \item \textbf{Prolonged esophageal distension using a short balloon followed by abrupt deflation:} During the distension, the esophagus distal to the distended section is mostly inactive. Upon abrupt deflation, one phasic contraction appears, traveling down the length of the esophagus \cite{Christensen1970,Paterson1988}. The pressure and cross-sectional area variations along the distended region are not reported in these studies.
    
  \item \textbf{Sustained distension of a short section of the esophagus:} Results in repetitive contractions along the distended section, as observed through a pressure probe located inside the fluid-filled distending bag \cite{gregersen2011mechanical,villadsen2001oesophageal}. The contractions' frequency remains independent of the bag's length, exhibiting a consistent rate of 6 contractions per minute, akin to the FLIP scenario \cite{gregersen2011mechanical,Carlson2020Repetitive}. The pressure variations distal to the distended region are not reported in these studies.
  
  \item \textbf{Sustained esophageal distension along its entire length:} Results in RACs \cite{Carlson2020Repetitive,Carlson2020,Mcmahon2007}.  
  
\end{enumerate}

When reproducing the above scenarios, the neural circuit and its parametric values remain constant across the different simulations. The objective is to demonstrate that the properties inherent in the baseline solution, representing a typical esophageal response to FLIP (Fig. \ref{fig:RACs}), automatically encompass all clinically observed distension-induced scenarios. The simulations vary by adjusting the distension's location and duration along the esophagus.

\begin{figure*}%[!htb]
    \centering
    \begin{subfigure}[b]{0.45\textwidth}
        \centering
        {\includegraphics[trim=0 170 30 140,clip,width=\textwidth]{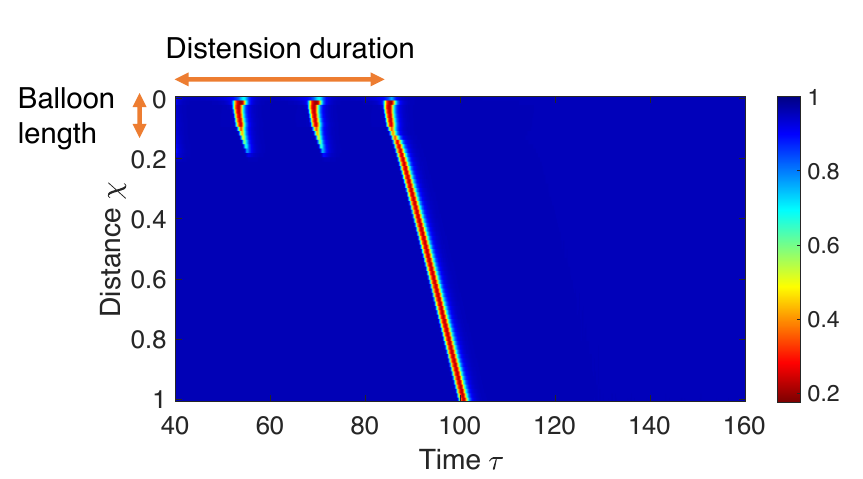}}
        \caption{}
        \label{fig:short_bag_deflate}
    \end{subfigure}
    \hfill
    \begin{subfigure}[b]{0.45\textwidth}  
        \centering 
        {\includegraphics[trim=0 170 30 140,clip,width=\textwidth]{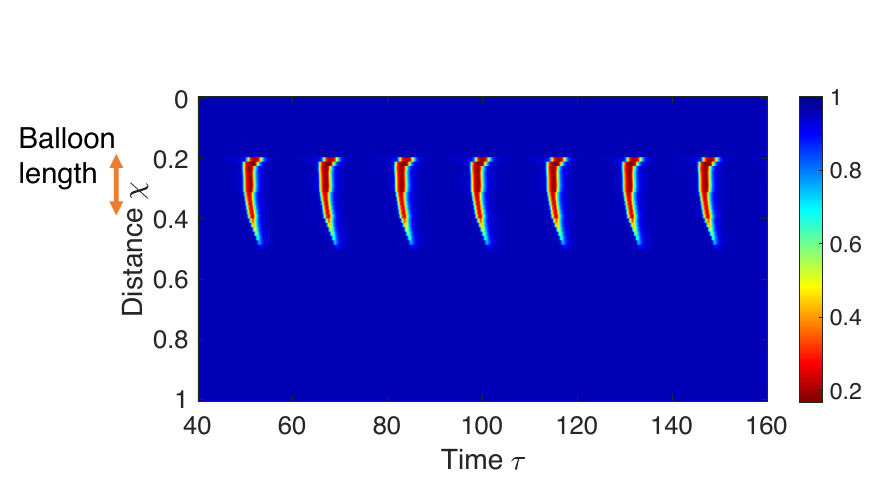}}
        \caption{}
        \label{fig:shortBag}
    \end{subfigure}
    \caption{Color-coded spatio-temporal topographies depicting muscle contraction patterns ($\theta$) obtained by the neuromechanical model, simulating two distinct esophageal distension tests. \textbf{(a)} Prolonged balloon distension followed by abrupt deflation (Case 2). \textbf{(b)} Sustained esophageal distension using a short bag (Case 3).} 
    \label{fig:distentionTests}
\end{figure*}

All four scenarios are successfully reproduced. Cases 1 and 4 are displayed in Fig. \ref{fig:singleCont} and Fig. \ref{fig:RACs}, respectively, and discussed in the preceding section. Figure \ref{fig:singleCont} corresponds to Fig. 1 in the manuscript by \citet{Paterson1988}. Cases 2 and 3 are presented in Fig. \ref{fig:distentionTests} and elaborated upon in the following discussion. Figure \ref{fig:short_bag_deflate} corresponds to Fig. 2 in the manuscript by \citet{Paterson1988}, and Fig. \ref{fig:shortBag} corresponds to Fig. 1d in the manuscript by \citet{gregersen2011mechanical}.

Note that cases 2 and 3, though similar, were reported by different research groups, conducting different studies. \citet{Paterson1988} (Case 2) focused on the esophageal response distal to the distension (air-filled balloon), providing detailed pressure readings solely distal to the balloon. \citet{gregersen2011mechanical} (Case 3) documented the pressure and cross-sectional area only inside the distending medium. From these separate studies, we infer that sustained esophageal distension using a short bag or balloon induces repetitive contractions \textit{only} along the distended section, with the rest of the esophagus mostly inactive. Upon abrupt removal of the distension, a single contraction emerges, traveling beyond the distended segment. Figure \ref{fig:distentionTests} displays the muscle contraction pattern over the entire esophageal length, capturing the behavior inside and outside the bag/balloon.

%All four scenarios are successfully reproduced. Cases 1 and 4 are displayed in Fig.\ref{fig:singleCont} and Fig. \ref{fig:RACs}, respectively, and discussed in the preceding section. Cases 2 and 3 are presented in Fig. \ref{fig:distentionTests} and elaborated upon in the following discussion. Note that these two cases, though similar, were reported by different research groups, conducting different studies. \citet{Paterson1988} (Case 2) focused on the esophageal response distal to the distension (air-filled balloon), providing detailed pressure readings solely distal to the balloon. \citet{gregersen2011mechanical} (Case 3) documented the pressure and cross-sectional area only inside the distending medium. From these separate studies, we infer that sustained esophageal distension using a short bag or balloon induces repetitive contractions \textit{only} along the distended section, with the rest of the esophagus mostly inactive. Upon abrupt removal of the distension, a single contraction emerges, traveling beyond the distended segment. Figure \ref{fig:distentionTests} displays the muscle contraction pattern over the entire esophageal length, capturing the behavior inside and outside the bag/balloon.

To explain the clinical observations in cases 2 and 3, we specifically focus on unique aspects not discussed in the preceding section, including (i) the selective appearance of contractions solely along the distended section (with a quiescent state distal to the bag), (ii) the observation of a traveling contraction upon deflation of the bag, and (iii) the explanation for the contraction frequency's independence of the bag's length.

%Note that we specifically focus on unique aspects of these cases that were not discussed in the proceeding section, including (i) the selective appearance of contractions solely along the distended section (with a quiescent state distal to the bag), (ii) the observation of a "rebound", traveling contraction upon deflation of the bag, and (iii) the explanation for the contraction frequency's independence of the bag's length.

%All three scenarios are successfully reproduced. Case 1 is displayed in Fig. \ref{fig:RACs} and discussed in the proceeding section. Cases 2 and 3 are presented in Fig. \ref{fig:distentionTests} and elaborated upon in the following discussion. Note that we specifically focus on unique aspects of these cases that were not discussed in the proceeding section, including (i) the selective appearance of contractions solely along the distended section (with a quiescent state distal to the bag), (ii) the observation of a "rebound", traveling contraction upon deflation of the bag, and (iii) the explanation for the contraction frequency's independence of the bag's length.

\begin{enumerate}[label=\textbf{(\roman*)}]%[\bfseries (i)]
  \item As previously established, muscle activation at any location along the esophagus is both initiated and sustained by local stretch receptors innervating excitatory neurons. Given that input from mechanoreceptors to the excitatory neurons is primarily local \cite{Mittal2016}, the absence of distension (proximal or distal to the bag) results in no contraction (as illustrated in Fig. \ref{fig:absent}).

However, note that excitatory neurons can also receive excitatory innervation from proximal excitatory neurons (if active) through parameter $b$. This scenario can lead to the signal propagating even without the presence of distension at a specific location, as shown in Fig. \ref{fig:singleCont}. So, what prevents these waves from propagating all the way towards the distal end of the esophagus?

The key factor is descending inhibition. When the esophagus is excited at any location, innervation is sent to inhibitory neurons distally \cite{abrahao2011swallow}, causing the esophagus to relax in anticipation of an incoming bolus \cite{paterson2006esophageal}. Thus, introducing enough inhibition which stops the wave from propagating beyond the distended region. In our model, this is represented by $S_{I,i}\neq0$, where $i$ is a location distal to the distended section. If we set the stretch-induced innervation to inhibitory neurons only around the distended region, as in Fig. \ref{fig:shortBagTrav}, contractions distal to the distended region are present.

  \item Note that the inhibitory activity levels distal to distension are low, allowing a quick recovery once distension is eliminated. Conversely, excitatory activity levels along the distended section are high, taking longer to recover. Therefore, upon abrupt emptying of the bag or balloon, the excitatory signal is free to travel down through neighboring communication (parameter $b$), unimpeded by inhibitory activity (as depicted in Fig. \ref{fig:singleCont}). This propagating excitatory signal translates into muscle contraction, resulting in the observed phasic contraction.
  
  \item Examining the rate of propagating contractions is equivalent to studying the frequency of muscle contraction at a single location, dependent on the excitation-refractory cycle of neuronal populations at that specific location \cite{Carlson2020,gregersen2011mechanical}. The time elapsed from the initiation of the excitatory phase to the end of the refractory period is dictated by the neural architecture and the strength of excitatory input from stretch ($w_E$) \cite{Wilson1972}. Since these parameters are independent of the bag’s length, the contraction frequency is not expected to change with the bag’s length.
\end{enumerate}

%%%%%%%%%% Comparing opening Speed Figure - loop and work %%%%%%%%%%%%%

%%%%%%%% Esophageal motility disorders%%%%%%%%%
\subsection{Esophageal motility disorders} 

In the following section, we showcase how alterations in specific parameters disrupt the RACs pattern. These disruptions result in solutions reminiscent of established esophageal motility disorders, as defined by the Chicago Classification \cite{Carlson2020,yadlapati2021esophageal}. This demonstration not only exposes the underlying mechanisms of these disorders but also emphasizes that they are emergent behaviors. Through this exploration, we establish the model's consistency with pathologies, providing valuable insights into the dynamic interplay of components influencing esophageal motility disorders. The results are obtained though parametric and robustness studies available in section \ref{sec:parametricStudy} in supplementary.

%%%%%%% Absent Contractile %%%%%%%%%%%%%
\textbf{Absent contractile response.}

As discussed earlier, an absent contractile response emerges when the excitatory population’s activity ($E$), responsible for controlling muscle contraction, is either reduced or absent (Fig. \ref{fig:absent}). This response may be attributed to dysfunctions in the normal operation of stretch receptors ($S_E$ and $S_I$), particularly when there are alterations in the sensitivity threshold of the mechanoreceptors ($\hat{\alpha}$). Changes in esophageal stiffness and dilation are suspected to impact esophageal sensitivity to distension, a phenomenon associated with motility disorders characterized by an absent contractile response, such as achalasia I \cite{Gregersen2018Pathophysiology}.

The results in Fig. \ref{fig:singleCont} might provide insights into cases where patients exhibit normal primary peristalsis (as measured by high-resolution manometry) but abnormal findings in FLIP studies \cite{Carlson2021Evaluating,Carlson2015}. Such discrepancies could be explained by abnormal responses to distension, leading to a lack of an excitatory signal and, consequently, an inability to initiate secondary peristalsis.

%%%%%%%%%%% Spastic %%%%%%%%%%%
\textbf{Disordered non-occluding contractions.}

The introduction of local variations to the parameters defining the neuronal circuitry results in the observation of a sporadic or chaotic contraction pattern, as illustrated in Fig. \ref{fig:spastic1} and Fig. \ref{fig:spastic2}. Given the inherent variability of biological systems, small irregularities are immanent. Implementing such small irregularities in the model does not significantly alter the contraction pattern, showcasing its robustness (see supplementary Fig. \ref{fig:sens_varia}). However, significant local variations in the system's parameters (such as $a$, $c$ and $e$) may disrupt the desired excitatory-inhibitory balance required for proper RACs, resulting in pronounced irregularities in the contraction pattern (Fig. \ref{fig:spastic1}). 

It is crucial to emphasize the significance of the system's robustness. If the system is not initially robust, it becomes more sensitive to small irregularities. Such cases have parametric values allowing the solution to exhibit regular RACs. Nevertheless, the excitatory-inhibitory balance is not as stable. Thus, these scenarios are more prone to sporadic or chaotic contraction patterns when introducing small irregularities, which would not otherwise trigger such responses (Fig. \ref{fig:spastic2})

%%%% Sustained %%%%%%
\textbf{Sustained panesophageal contractions.}

\begin{figure*}[!htb]
    \centering{{\includegraphics[trim=0 170 0 210 ,clip,width=0.7\textwidth]{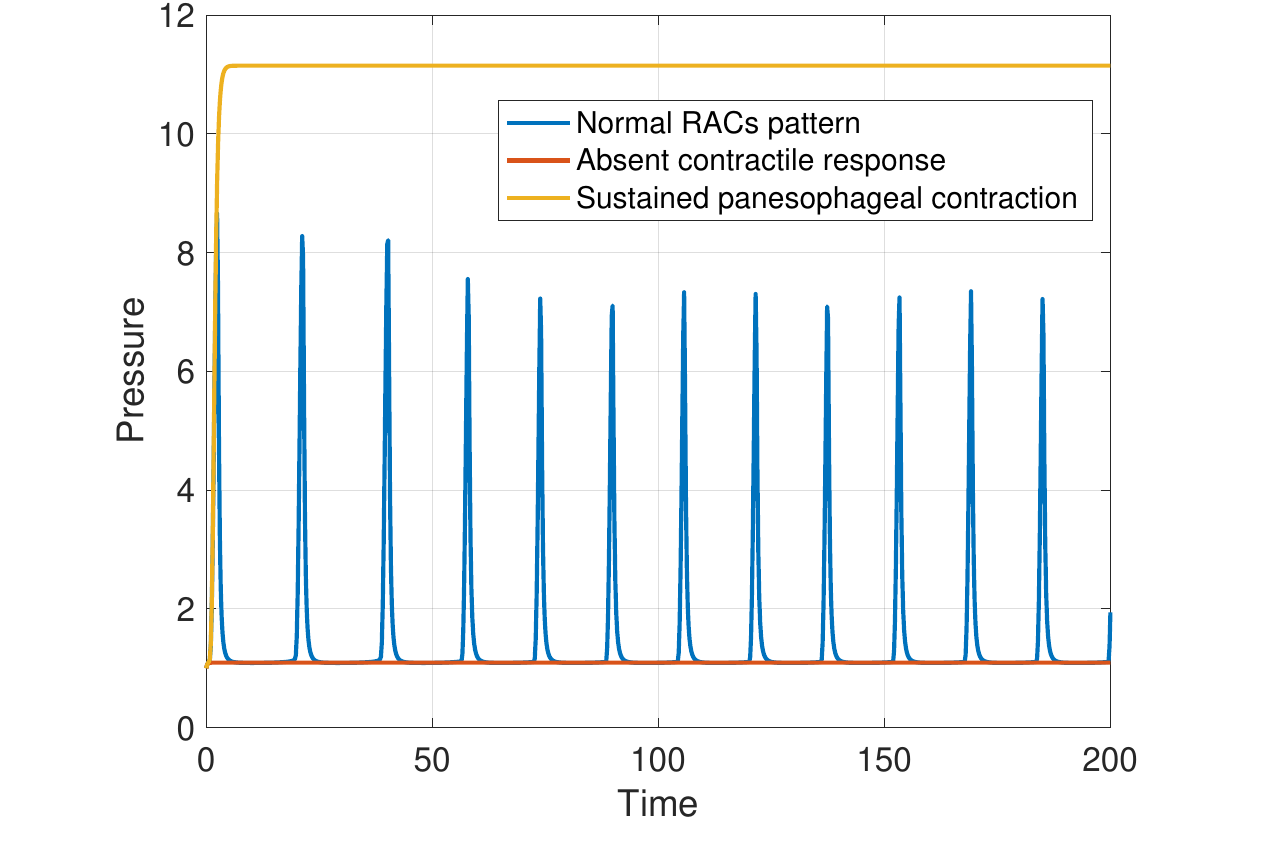}}}
    \caption{Internal bag pressure over time obtained in three distinct scenarios. Pressure is recorded at a single location along the bag length. In the case of normal RACs, the bag pressure fluctuates, reaching its peak during maximum contraction. Conversely, in absent contractile response, where no contraction occurs, the pressure remains consistently low throughout volumetric distension. In sustained panesophageal contraction, the entire esophagus contracts uniformly, increasing bag pressure that persists at a high level throughout the volumetric distension.}
    \label{fig:pressurePlot}
\end{figure*}

Sustained, non-propagating contractions leading to an increase in FLIP pressure are observed when the inhibitory signal to the excitatory population is weakened. Thus, preventing inhibitory activity from overcoming excitatory activity, keeping excitatory cells active. This scenario is reproduced by reducing the inhibitory signal to adjacent excitatory neurons ($e$) or by decreasing the excitation of inhibitory neurons ($c$ or $w_I$). Note that since the FLIP’s fluid is incompressible, the color-coded topography of cross-sectional area may resemble absent contractile response (Fig. \ref{fig:absent}). However, the pressure profile differs, as the uniform muscle contraction increases the pressure in the bag, as illustrated in Fig. \ref{fig:pressurePlot}.

%%%%%%% Pressure over time Figure %%%%%%%%%%

%%%%%%%%%%% RRC %%%%%%%%%%%
\textbf{Repetitive retrograde contractions.}

Repetitive retrograde contractions emerge with decreased innervation of inhibitory neurons via mechanoreceptors ($S_I$) or increased innervation of excitatory neurons via mechanoreceptors ($S_E$) (Fig. \ref{fig:RRC}). The retrograde pattern results from a phase shift in the excitatory signal, opposing the antegrade pattern. Importantly, this phenomenon is not due to a retrograde traveling signal, as the innervation pathways remain unidirectional. Further details on reversing propagation direction are discussed extensively in our recent work \cite{elisha2024direct}.

\section{Discussion}\label{Discussion}

This work closes the gap of clarifying the connection between the aberrant neural circuitry and the emergent mechanical dysfunction. The organ-scale neuromechanical model proves useful in giving clarity into the specific neural pathways, connections, and interactions between different types of neurons in the esophagus. The main objectives were to unveil the underlying mechanism of RACs by providing a neural circuitry that reproduces this pattern and to elucidate the mechanisms leading to neurologically driven EMDs. Our results suggest that abnormal contraction patterns are likely to originate from neuronal imbalance. This is aligned with prior speculations concerning the manifestation of esophageal motility disorders \cite{Carlson2020,Gregersen2018Pathophysiology,carlson2018mechanisms}.

Additionally, we addressed open questions about the emergence of RACs. By examining scenarios in which stretch receptors are disabled, we demonstrated that RACs are locally triggered, supporting the notion that they are a form of secondary peristalsis. Thus, showcasing how neuromechanical models can bring clarity to clinical observations.

However, it is essential to acknowledge certain limitations. The parametric values in our model, while effective in reproducing observed behaviors, need calibration. Extracting these values experimentally is a challenging task \cite{yeoh2017modelling}, especially considering the current lack of a fully explained and quantified esophageal neuronal pattern. As such, we relied on qualitative clinical observations reported in the literature to guide our model parameters. Hence, future research should focus on experimentally validating these parameters.

Advancing towards experimental validation can be achieved through additional pharmacological studies, electrophysiological recordings, and by studying the neural circuity at the molecular level though immunohistochemical, optogenetics, neuronal tracing, and calcium imaging \cite{Park1999,fung2020functional,kulkarni2018advances,yang2023role,rajendran2019identification,jean2001brain,harrington2007immunohistochemical,koh2022propulsive}. A recent study on the peristaltic reflex of the colon provides a promising template, revealing the physiological and neural elements involved using multiple experimental methods \cite{koh2022propulsive}. 

Another limitation concerns the absence of the lower esophageal sphincter (LES) in our model, particularly regarding its potential influence on the emergence of repetitive retrograde contractions. Repetitive retrograde contractions pattern has previously been theorized to arise from either impaired inhibitory innervation or esophageal outflow obstruction \cite{carlson2018mechanisms}. A recent clinical study supports the former hypothesis, attributing repetitive retrograde contractions to neural imbalance leading to excess excitation \cite{carlson2018mechanisms}. Our investigation aligns with this perspective. However, the current work does not exclude the second possibility. The effect of esophageal outflow obstruction should be point of future investigation, with a proper LES model.

Nevertheless, despite these challenges, our work opens a new area of investigation in esophagology that is based on mechanophysiology (mechanics-based organ function). One of the leading diagnostic protocols for EMDs is the Chicago Classification scheme \cite{yadlapati2021esophageal}. This work is a milestone in our effort to develop the first mechanics-guided disease classification and diagnostic protocol for EMDs \cite{AcharyaEsoWork2020,halder2023assessing}.

There is also a potential for broad impact beyond the esophagus. Mechanical dysfunctions caused by neurological disorders expand beyond the gastrointestinal tract and are prominent in other organs where an emergent behavior occurs such as the heart, lungs, bladder, and uterus \cite{Jie2010,Trayanova2011,Jelinvcic2022breathing,andersson2004urinary,Berridge2008smooth}. For example, acute respiratory distress syndrome and chronic obstructive pulmonary disease are associated with disturbance in the lungs’ rhythmic function \cite{sundar2015circadian,jung2021early}. In the heart, asynchronous in electrical activation can cause abnormalities in perfusion and pump function, leading to arrhythmias or even heart failure \cite{Leclercq2002,Usyk2003,Niederer2011}. 

Organ-scale neuromechanical models can mark a significant step forward in bridging the gap between clinical observations and mechanistic understanding, thereby aiding the development of effective neurologically focused treatment approaches. This work provides a template to interrogate neurologically driven mechanophysiological pathologies of other organs.

\section{Methods}\label{Methods}

\subsection{Overview of mathematical model}\label{Overview}

In this section, we provide a brief overview of esophageal neuromechanics and describe how the different components are represented in our empirically guided model. The mathematical details are discussed in the succeeding section. 

The esophagus is a multilayered tubular organ that connects between the mouth and the stomach, transporting swallowed material through peristaltic contraction \cite{Yazaki2012}. This coordinated movement results from sequential contraction of the circular muscle, that is neurally controlled \cite{paterson2006esophageal,Orvar1993,Neuhuber2016}. Peristalsis can be triggered by swallowing, initiated by the central nervous system (primary peristalsis) \cite{Diamant1997,Goyal2008}, or locally, provoked by esophageal distention and is independent of the central nervous system (secondary peristalsis) \cite{Park1999,Woodland2013}. Since esophageal response to FLIP is known to be involuntary and assumed to be triggered locally, the model is such that the innervation is lead by local feedback \cite{Goyal2022,Carlson2015Utilizing,Carlson2022heterogeneity}.

The esophageal wall and FLIP mechanics are modeled as one, close ended, flexible tube filled with incompressible, viscous fluid \cite{Acharya2021Pumping}. The esophageal neural circuitry is constructed as a chain of unidirectionally coupled relaxation oscillators, often used to model central pattern generators \cite{ijspeert2008central}. The neural circuitry receives excitatory signals from stretch receptors along the esophageal body (Fig. \ref{fig:circuitMain}).  Since esophageal contraction wave is coordinated by a sequential excitation \cite{Diamant1997,Yazaki2012,Sifrim2012}, the coupling between the neurological model and the body mechanics is done by initiating contraction when the amount of local excitation rises above a threshold. The innervation pathways for each neuronal population (excitatory and inhibitory) are discussed next.

\textbf{Excitatory input from stretch.} 
When the FLIP bag inside the esophagus is inflated, it causes the esophagus to distend, activating the proprioceptive channel that is mediated by stretch receptors, which respond to changes in strain \cite{Lang2019,Sengupta2000}. These mechanoreceptors are parts of the esophageal motor neurons and are distributed along the length of the esophagus. Thus, in our neurological model, mechanoreceptors are activated when they sense that local strain increases above a threshold and send excitatory signals to both excitatory and inhibitory neuronal populations. 

Since the esophagus contracts proximal to a bolus and relaxes distal to a bolus \cite{Mittal2016}, activated mechanoreceptors are set to send excitatory signals to proximal excitatory populations and distal inhibitory populations \cite{paterson2006esophageal}. 
Additionally, it has been reported that during esophageal distension, the portion of the esophageal body that is located distal to the distended balloon typically remains inactive or quiescent
%and only contracts once the balloon is deflated 
\cite{Christensen1970,Paterson1988}. The esophagus distal to the excited portion in inhibited to accommodate for the incoming bolus \cite{Mittal2016,abrahao2011swallow,paterson2006esophageal}. Therefore, the input from stretch receptors to inhibitory neuronal population affects all inhibitory neurons distal to the sensed distension. On the other hand, the input from mechanoreceptors to the excitatory neuronal population is more local, such that sensed distension only inputs excitatory signal to proximal excitatory neurons that are within 10$\%$ of esophageal length.

\textbf{Input from adjacent neuronal populations.} It is believed that depolarization of one muscle cell in the esophagus will result in electronic increase of neighboring muscle cells in an aboral direction \cite{Sifrim2012,Omari2022}. Hence, excitatory populations in the model also receive signals from adjacent proximal populations. Nearest-neighbor connections of two types are introduced \cite{Pehlevan2016,Gjorgjieva2013,Hughes2007}. First, unidirectional, aboral excitatory connections between excitatory populations of neighboring segments leading to a propagation of activity along the chain of segments (parameter $b$ in Fig. \ref{fig:circuitMain}). Second, inhibitory connections from the inhibitory population in one segment to the neighboring (anterior) excitatory populations (parameter $d$ in Fig. \ref{fig:circuitMain}).

\textbf{Intrasegmental synaptic inputs.} Lastly, each segment’s excitatory and inhibitory populations exchange local signals, imitating the excitatory and inhibitory synapses per cell in the excitatory or inhibitory populations (parameters $a$, $c$, $e$, and $f$ in Fig. \ref{fig:circuitMain}) \cite{Wilson1972}. These are trivial elements of the system when modeling the electrical activity using phenomenological approach \cite{Nash2004}.

Our study is built upon and extends the models recently developed for larval locomotion which used the Wilson-Cowan model \cite{Pehlevan2016,Gjorgjieva2013,Wilson1972}. It is the first to connect crawling in Drosophila models to the contraction pattern of the esophagus. However, the biggest and most unique advance is in how the initial excitation occurs and the way in which the rhythmic motion emerges. In the locomotion models, the forward waves were induced by applying an external input to distal excitatory population. In the current model, no external neuronal signaling is needed to initiate the contraction nor to sustain the rhythmic contractions.

\subsection{Body mechanics and fluid equations}\label{BodyMech}
The flow inside the FLIP device placed in the esophageal lumen is modeled as a one-dimensional, fluid-filled, flexible tube that is closed on both ends \cite{Acharya2021Pumping,Elisha2022Regime,Elisha2022Sphincter}. The conservation of mass and momentum equations are

\begin{equation} \label{eq:continuity}
    \frac{\partial A}{\partial t}+\frac{\partial\left(Au\right)}{\partial x} = 0,
\end{equation}

\noindent and

\begin{equation} \label{eq:momentum}
    \frac{\partial u}{\partial t} + u\frac{\partial u}{\partial x} = 
    -\frac{1}{\rho}\frac{\partial P}{\partial x}-\frac{8\pi\mu u}{\rho A},
\end{equation}

\noindent respectively \cite{Ottesen2003}. A parabolic flow is assumed everywhere. In the equations above, $A(x,t)$, $u(x,t)$, $P(x,t)$,  $\rho$ and $\mu$ are the tube cross-sectional area, fluid velocity (averaged at each cross-sectional area), pressure inside the tube, fluid density, and fluid viscosity, respectively. We introduce a linear constitutive relation

\begin{equation}
{{P}}={K_{\scriptscriptstyle e}}\left(\frac{A}{A_{\scriptscriptstyle o}\theta}-1\right) + P_o.
\label{eqn:tube_law_theta}
\end{equation}

\noindent to complete the system \cite{Whittaker2010,Kwiatek2011}. Here, ${P}_o$, $K_e$, and $A_o$ are external pressure, tube stiffness, and undeformed reference area (cross-sectional area of the tube when $P=P_o$), respectively. Lastly, the neurally controlled contraction of the esophageal lumen is set through dynamically varying the rest cross-sectional area of the tube by a factor $\theta(x,t)$ \cite{Ottesen2003,Manopoulos2006,Bringley2008}. $\theta(x,t)$ captures the muscular dynamic, induced by the neural activity discussed next.

%% section name up to 60 characters
\subsection{Neural circuitry: excitation, inhibition, and muscle contraction pattern}\label{NeuralCircuity}
Relaxation oscillators are often used to model cardiac and gastro-intestinal electrical activity \cite{Nash2004,Pullan2004,Chambers2008,Du2010}. To represent the electrical activity of the entire organ, we introduce a system of locally coupled Wilson–Cowan oscillators, distributed uniformly along the length of the esophagus.  Each unit consists of two neuronal populations, excitatory ($E$) and inhibitory ($I$) (Fig. \ref{fig:circuitMain}). The values of $E_i(t)$ and $I_i(t)$ at each oscillator depict the activity levels of the excitatory and inhibitory neuron populations. The differential equations for the time-dependent variation of averaged excitatory and inhibitory neuronal activities at node $i$ introduced by \cite{Wilson1972} with the unidirectional coupling introduced by \cite{Gjorgjieva2013} is

\begin{equation} \label{eq:Exc}
    \tau_{E}\dot{E}_i=-E_i+(1-E_i)\sigma_{E}[aE_i+bE_{i-1}-eI_i-dI_{i-1}+S_{E,i}],
\end{equation}
and
\begin{equation} \label{eq:Inh}
\tau_{I}\dot{I}_i=-I_i+(1-I_i)\sigma_{I}[cE_i-fI_i+S_{I,i}].
\end{equation}

\noindent The intrasegmental connectivity parameters $a$, $e$, $c$, and $f$ represent the average number of excitatory and inhibitory synapses per cell in the excitatory or inhibitory population. The connectivity coefficient $b$ denotes the unidirectional excitatory connections between the excitatory populations of neighboring segments (excitatory signal from nearest posterior excitatory population). The connectivity coefficient $d$ signifies the unidirectional inhibitory connections from a segment's inhibitory population to the nearest anterior segment's excitatory population. For the relaxation oscillators located at the proximal end, $b=d=0$. The time constants $\tau_{E}$ and $\tau_{I}$ dictate the decay of the excitatory and inhibitory activities, and determine the activities' timescale. The sigmoid function characterizes the switching threshold defined as 

\begin{equation} \label{eq:sigmoid}
\sigma_{E/I}[x]=\frac{1}{1+\text{exp}[-\lambda_{E/I}(x-\phi_{E/I})]}-\frac{1}{1+\text{exp}(\lambda_{E/I}\phi_{E/I})}.
\end{equation}

\noindent where $\lambda$ is activation speed (slope of sigmoid), and $\phi$ is activation threshold (location of sigmoid's maximum slope). Lastly, $S_{E,i}$ and $S_{I,i}$ symbolize the local excitatory inputs to each population at oscillator $i$, capturing the mechano-sensory feedback coming from the stretch receptors \cite{Pehlevan2016,Gjorgjieva2013}, discussed next. 
 
In our model, the mechano-receptors are activated when they sense that local strain increases above a threshold ($\hat{\alpha}$). As discussed in a prior section, when activated, a sensory stretch receptor sends excitatory signals to proximal excitatory populations and excitatory signals to distal inhibitory populations. The input to excitatory neurons from stretch receptors is more local, meaning that stretch excites excitatory neurons within the proximal vicinity of the activated receptor. On the other hand, the input from stretch receptors to inhibitory neurons is applied to all inhibitory neurons distal to the activated receptor. Thus, $S_{E}$ and $S_{I}$ input for a segment located in $x_i$ are defined as

\begin{equation} \label{eq:S_E_S_I}
    S_{E,i}=w_{E}\sigma_{S}\bigg[\int_{x_i}^{L} h(y)\beta_E(x_i-y) \,dy\bigg]\qquad\text{and}\qquad S_{I,i}=w_{I}\sigma_{S}\bigg[\int_{0}^{x_i} h(y)\beta_I(x_i-y) \,dy\bigg],
\end{equation}

\noindent where $\sigma_{S}[x]=\text{tanh}[g_Sx]$, $\beta_E(x)=0.5+0.5 \text{tanh}(g_E(x+x_s))$, $\beta_I=1$, $h(x)=max[A(x,t)/A_o\theta(x,t)-\hat{\alpha},0]$, $w_{E}$ and $w_{I}$ are the strength of the net sensory feedback coming from the stretch receptors, $L$ is esophageal length, $g_S$ and $g_E$ are gains, and $x_s$ is horizontal shift. The schematic of the stretch induced innervation is presented in Fig. \ref{fig:circuitStretch}.

%$\hat{\alpha}$ is segmental strain threshold,

Using the neural coupling above, we can solve for the activation function $\theta$ such that
\begin{equation} \label{eq:tht}
    \tau_{\theta}\dot{\theta}_i=1-\theta_i-\sigma_{\theta}[E_i-\hat{E}],
\end{equation}
where $\tau_{\theta}$ is a time constant, $\hat{E}$ a threshold for muscle activation, and $\sigma_{\theta}[x]=0.5[1-\theta_o+(1-\theta_o)\text{tanh}(g_\theta x)]$, where $g_\theta$ is the gain, and $\theta_o$ is the maximum contraction strength. 

\subsection{Non-dimensional form}\label{NonDim}

The dynamic system of equations described above is non-dimensionalized using
\begin{equation}
A=\alpha A_{o}, \qquad t=\tau\tau_E, \qquad u=UL/\tau_E, \qquad P=pK_e, \qquad \text{and} \qquad x=\chi L,
\end{equation}
where $\alpha$, $\tau$, $U$, $p$, and $\chi$ are the non-dimensional variables for cross-sectional area, time, velocity, pressure, and position, respectively \cite{Acharya2021Pumping,Elisha2022Regime}. Thus, dimensional equations ((\ref{eq:continuity}), (\ref{eq:momentum}), (\ref{eqn:tube_law_theta}), (\ref{eq:Exc}), (\ref{eq:Inh}), and (\ref{eq:tht})) become

\begin{equation} \label{eq:continuity_nondim}
    \frac{\partial\alpha}{\partial\tau}+\frac{\partial\left(\alpha U\right)}{\partial\chi} = 0, 
\end{equation}

\begin{equation} \label{eq:momentum_nondim}
    \frac{\partial U}{\partial\tau} + U\frac{\partial U}{\partial\chi} + 
    \psi\frac{\partial p}{\partial\chi}
    + \beta\frac{U}{\alpha} = 0 , 
\end{equation}

\begin{equation} \label{eq:tube_law_nondim}
    \textcolor{REDCOLOR2}{p=\left( \frac{\alpha}{\theta}-1 \right)},
\end{equation}

\begin{equation} \label{eq:ExcNonD}
    \dot{E}_i=-E_i+(1-E_i)\sigma_{E}[aE_i+bE_{i-1}-eI_i-dI_{i-1}+S_{E,i}],
\end{equation}

\begin{equation} \label{eq:InhNonD}
\hat{\tau}_{I}\dot{I}_i=-I_i+(1-I_i)\sigma_{I}[cE_i-fI_i+S_{I,i}]\qquad \text{and}
\end{equation}

\begin{equation} \label{eq:thtNonD}
    \hat{\tau}_{\theta}\dot{\theta}_i=1-\theta_i-\sigma_{\theta}[E_i-\hat{E}],
\end{equation}

\noindent respectively, where $\psi=K_e\tau_E^2/\rho {L}^2$, $\beta = 8\pi\mu L\tau_E/\rho A_o$, $\hat{\tau}_{\theta}={\tau}_{\theta}/{\tau}_{E}$, and $\hat{\tau}_{I}={\tau}_{I}/{\tau}_{E}$.

\subsection{Initial and boundary conditions}\label{IBC_Conditions}
The fluid in the tube is initially at rest and the cross-sectional area is uniform, hence

\begin{equation} \label{eq:velareaIC}
    U\left(\chi,\tau=0\right)=0  \qquad\qquad  \alpha\left(\chi,\tau=0\right)=S_{\text{IC}}\theta(\chi,\tau=0),
\end{equation}
where $S_{\text{IC}}=\text{Volume}/A_o$  \cite{Elisha2023EGJLoop,Elisha2022Sphincter}. Additionally, 

\begin{equation} \label{eq:E_I_tht_IC}
    E_{\tau=0}=0  \qquad\qquad  I_{\tau=0}=0 \qquad\qquad  \theta_{\tau=0}=1.
\end{equation}
\noindent since the neuronal system is at rest. Lastly, recall that the tube is closed on both ends, thus, 
\begin{equation} \label{eq:velBC}
    U\left(\chi=0,\tau\right)=0\qquad\text{and}\qquad U\left(\chi=1,\tau\right)=0.
\end{equation}

To obtain a boundary condition for $\alpha$, we plug equation (\ref{eq:tube_law_nondim}) and the velocity boundary condition in equation (\ref{eq:velBC}) into equation (\ref{eq:momentum_nondim}), which yields
\begin{equation} \label{eq:areaBC}
    \left.\frac{\partial}{\partial \chi} \left(\frac{\alpha}{\theta}\right)\right|_{\chi=0,\tau}=0\qquad\mathrm{and}\qquad
    \left.\frac{\partial}{\partial \chi} \left(\frac{\alpha}{\theta}\right)\right|_{\chi=1,\tau}=0.
\end{equation}

\subsection{Numerical implementation }\label{Numerical_Impl}
The system of equations is solved using MATLAB $\tt{ode15s}$ function for the time derivatives and central difference discretization for the spatial derivatives. The discrete expressions for the external source terms in equation (\ref{eq:S_E_S_I}) are
\begin{align}
\label{eq:S_E_S_I_desc}
\begin{split}
 S_{E,i} &=w_{E}\sigma_{S}\bigg[\sum_{k=i}^{N} max(\alpha_k/\theta_k-\hat{\alpha},0)\beta_{E}(\chi_i-\chi_k) \,\Delta \chi\bigg], \qquad\text{and}\qquad
\\
 S_{I,i} &=w_{I}\sigma_{S}\bigg[\sum_{k=1}^{i} max(\alpha_k/\theta_k-\hat{\alpha},0) \,\Delta \chi\bigg],
\end{split}
\end{align}
where $N$ denotes the number of nodes.

Table \ref{table:param} lists the model's parameters and their values. Values for the parameters defining esophageal body mechanics and fluid properties are approximated based on clinical data and previous computational studies 
\cite{Gregersen2008,Kou2015ajpgi,Halder_2021,Elisha2023EsophBodyLoop}. 
The parameters involved in the neural equations are chosen such that the system has a single unstable fixed point and a stable limit cycle in response to the constant external stimulus $S_E$ and $S_I$. Depending on the values of the parameters in the system, one can obtain multiple equilibria with different stability properties. For simplicity, we choose the values introduced by \cite{Wilson1972} with small variations. %to decrease the limit cycle frequency and accommodate for the chain system.
For example, theoretical work concerning Wilson-Cowan oscillators often set $S_I~=0$ for simplicity \cite{Wilson1972}. By setting $S_I\neq0$, we ensure that both excitatory and inhibitory populations "turn-on" when distension is applied, independent of one other. A theoretical overview of relaxation oscillators, limit cycle solution, and the constrains introduced by modeling esophageal response is provided in Section \ref{sec:limitCycleTheory} of the supplementary.

Note that obtaining a limit cycle solution is not the only constrain in this case. The parameters (specially $b$, $d$, $w_E$, and $w_I$) are adjusted such that the model captures the qualitative patterns observed in clinical esophageal distension studies. For instance, high value for $b$ makes the propagation more distinct and the system more robust \cite{Pehlevan2016,Gjorgjieva2013}. However, it also breaks the constrain in which the rhythmic contractions only take place along the distended section. For high $b$, excitation from proximal cells dominate, allowing the wave to propagate even without local excitation from mechanoreceptors.

\begin{table}
                \caption{List of parameters and their values}
            \footnotesize
                \begin{tabular}[t]{|p{2.2cm}|p{2.2cm}|}
                    \hline
                    $\psi$&3000\\ \hline
                    $\beta$&100 \\ \hline
                    $\theta_o$&0.05\\ \hline
                    $S_{\text{IC}}$&2 \\ \hline  
                    $\hat{\alpha}$&1.5 \\ \hline   
                    $x_s$&0.1\\ \hline
                    $\hat{\tau}_{\theta}$&0.2 \\ \hline
                    $\hat{\tau}_{I}$&4\\ \hline   
                \end{tabular}
                \hfill
              \begin{tabular}[t]{|p{2.2cm}|p{2.2cm}|}
                    \hline
                    $a$&16\\ \hline
                    $b$&20 \\ \hline
                    $c$&12\\ \hline
                    $d$&40 \\ \hline   
                    $e$&15\\ \hline
                    $f$&3 \\ \hline
                     $w_E$&1.6 \\ \hline
                     $w_I$&1.35\\ \hline     
                \end{tabular}
                \hfill
              \begin{tabular}[t]{|p{2.2cm}|p{2.2cm}|}
                    \hline
                    $\phi_E$&4\\ \hline
                    $\phi_I$&3.7 \\ \hline
                    $\lambda_E$&1.3\\ \hline
                    $\lambda_I$&2 \\ \hline 
                    $g_S$&1000 \\ \hline   
                    $g_E$&1000 \\ \hline   
                    $g_{\theta}$&5\\ \hline
                    $\hat{E}$&0.3\\ \hline
                \end{tabular}
                \label{table:param}
            \end{table}

\subsection*{Acknowledgments}
This work was funded by the by the National Institutes of Health (NIDDK grants DK079902 \& DK117824 and NIDDK grants DK079902), and National Science Foundation (OAC grants 1450374 \& 1931372)
% Thank Dr. Lior Tiroshi and Hannah

\subsection*{Author contributions}
GE wrote the main manuscript text, prepared all the figures, and wrote the code; NP, and GE developed the mathematical formulation; DC contributed to data acquisition; GE, SH, XL, DC, PK, JP, and NP interpreted results of calculations; PK, JP, and NP contributed to obtaining funding, critical revision of the manuscript and final approval.

\subsection*{Competing interests} 
PK and JP hold shared intellectual property rights and ownership surrounding FLIP panometry systems, methods, and apparatus with Medtronic Inc. DC: Medtronic (Speaking, Consulting) PK: Ironwood (Consulting), Reckitt (Consulting), Johnson \& Johnson (Consulting)

% supplementary section command
\supplementarysection
\supplementaryfigure

\section{Supplementary} 

 %%%%%%%%% Sensitivity testing and robustness analysis %%%%%%%
\subsection{Sensitivity testing and robustness analysis}\label{sec:parametricStudy} 

In the pursuit of a comprehensive understanding of the intricate dynamics within the coupled model, we perform a parametric study. This study involves systematically varying individual parameters by increments of $\pm 5$, $\pm 10$, $\pm 15$, $\pm 20\%$, allowing us to dissect the relationship between these parameters and the model's response. For some parameters, greater increments are considered. The parametric study or sensitivity test has a dual purpose. Firstly, to evaluate the robustness of the model to varying parameters. Secondly, to explore how different parameters affect the solution to the extent that the emerging response resembles an esophageal motility disorder. The impact of varying individual parameters on the simulation results is assessed both qualitatively and quantitatively, providing a comprehensive view of the model's behavior under various conditions.

The quantitative evaluation focuses on the following criteria:
\begin{itemize}
    \item Maximum values of $E$ and $I$ – amplitude of $E$ and $I$ oscillations.
    \item Period of $E$ and $I$ oscillations.
    \item Contraction duration (length/wave speed) – how long it takes the wave to propagate along the entire length.
    \item Intersegmental phase lag – time delay in the propagation of the neural signal between adjacent segments.
    \item Segment activity duration – how long an oscillator is active.
\end{itemize}

The qualitative assessment is concerned with examining the color-coded topography of the solution, and comparing it with the one of the baseline case (Fig. \ref {fig:RACs}). Are there repetitive, forward propagating contractions? If so, do they decay with time or remain intact as long as distension is applied?

\begin{figure*}[!htb]

    \centering
    \begin{subfigure}[b]{0.3\textwidth}
        \centering
        \includegraphics[trim=30 250 60 260,clip,width=\textwidth]{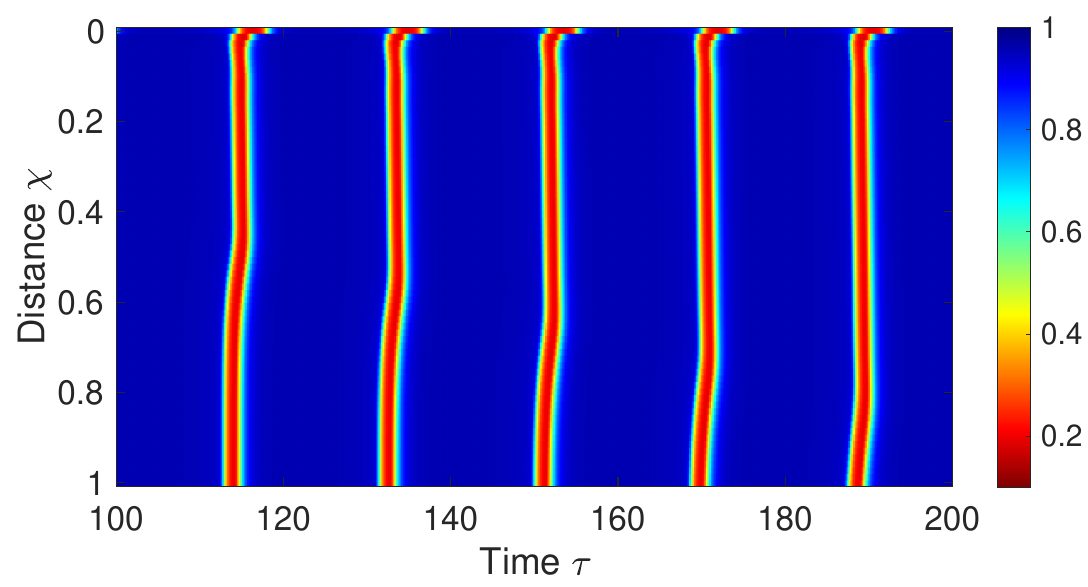}
        \caption{}
        \label{fig:eDecrease}
    \end{subfigure}
\
    \begin{subfigure}[b]{0.3\textwidth}   
        \centering 
        \includegraphics[trim=30 250 60 260,clip,width=\textwidth]{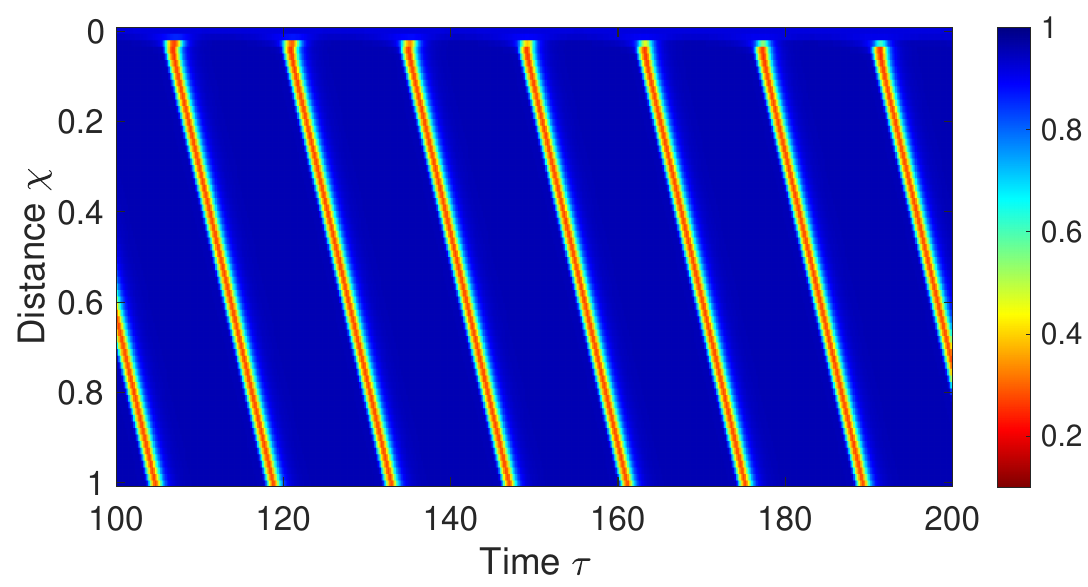}
        \caption{}
        \label{fig:eIncrease}
    \end{subfigure}
    \
    \begin{subfigure}[b]{0.3\textwidth}   
        \centering 
        \includegraphics[trim=30 250 60 260,clip,width=\textwidth]{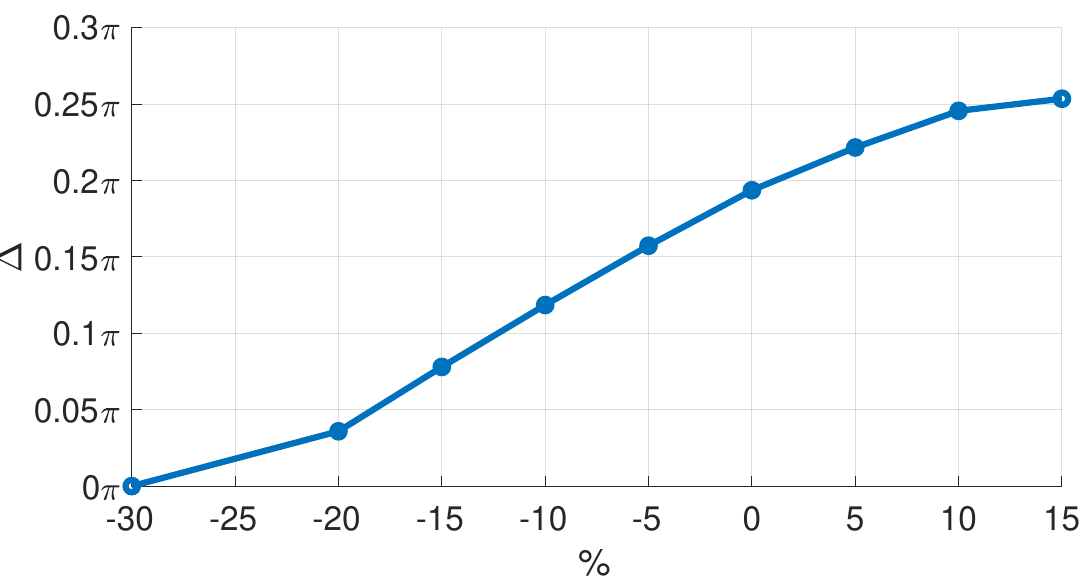}
        \caption{}
        \label{fig:IPL_e}
    \end{subfigure}
    \caption{Solution sensitivity to varying the value of $e$. \textbf{(a)} Color-coded topography of muscle contraction pattern ($\theta$) with the parameter $e$ modified to $0.7e$, representing a 30\% decrease from the original value of $e$.  \textbf{(b)} Color-coded topography of muscle contraction pattern ($\theta$) with the parameter $e$ modified to $1.2e$, representing a 20\% increase from the original value of $e$. \textbf{(c)} Plot of the intersegmental phase lag as a function of percentage deviation from the baseline value of parameter $e$ where 0\% corresponds to the baseline $e$ value ($e=15$).}
    \label{fig:sens_e}
\end{figure*}

\begin{figure*}[!htb]

    \centering
    \begin{subfigure}[b]{0.3\textwidth}
        \centering
        \includegraphics[trim=30 250 60 260,clip,width=\textwidth]{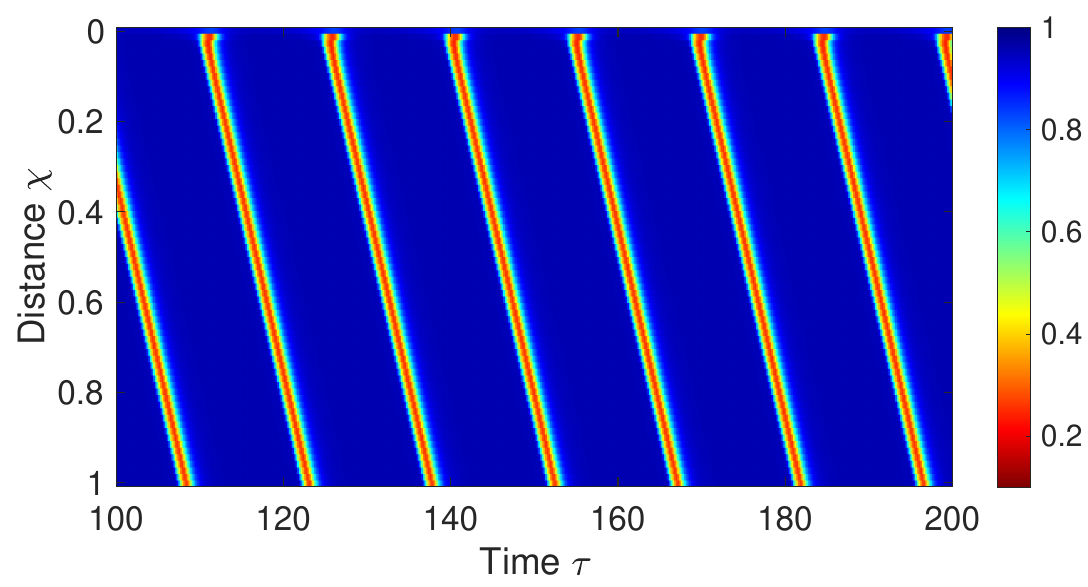}
        \caption{}
        \label{fig:fDecrease}
    \end{subfigure}
\
    \begin{subfigure}[b]{0.3\textwidth}   
        \centering 
        \includegraphics[trim=30 250 60 260,clip,width=\textwidth]{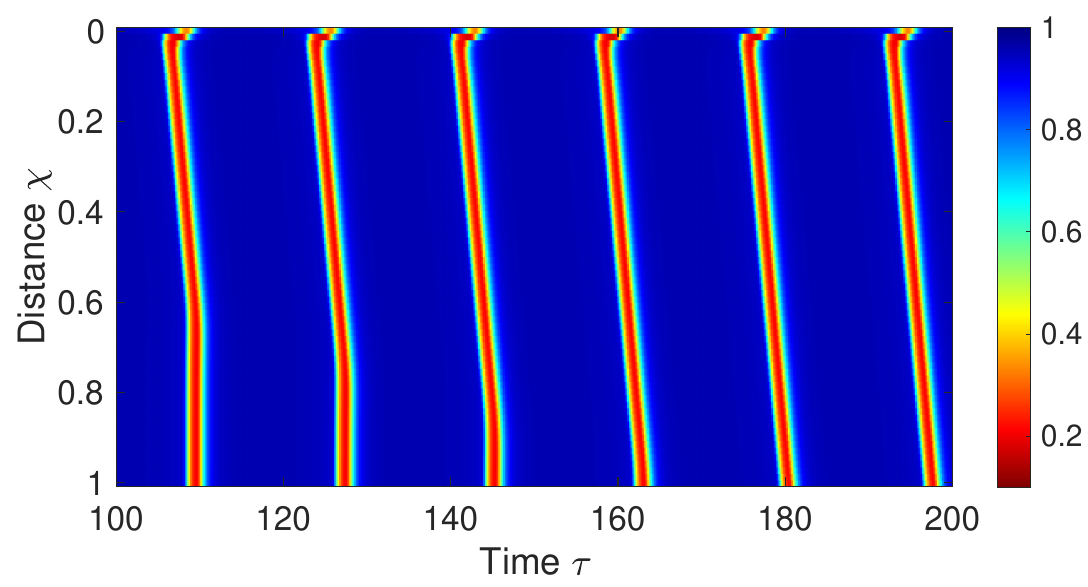}
        \caption{}
        \label{fig:fIncrease}
    \end{subfigure}
    \
    \begin{subfigure}[b]{0.3\textwidth}   
        \centering 
        \includegraphics[trim=30 250 60 260,clip,width=\textwidth]{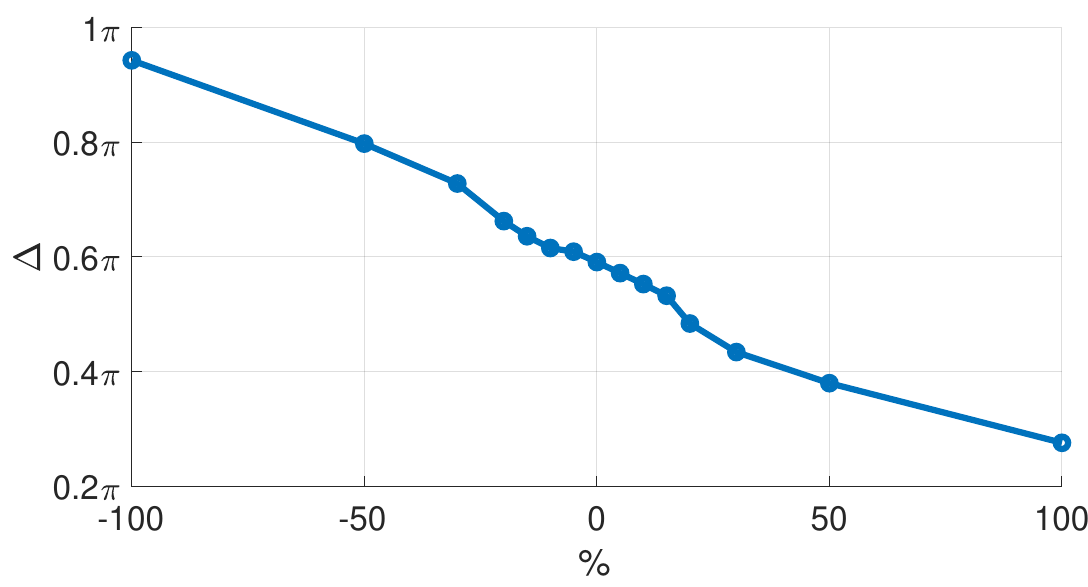}
        \caption{}
        \label{fig:IPL_f}
    \end{subfigure}
    \caption{Solution sensitivity to varying the value of $f$. \textbf{(a)} Color-coded topography of muscle contraction pattern ($\theta$) with the parameter $f$ modified to $0.5f$, representing a 50\% decrease from the original value of $f$.  \textbf{(b)} Color-coded topography of muscle contraction pattern ($\theta$) with the parameter $f$ modified to $2f$, representing a 100\% increase from the original value of $f$. \textbf{(c)} Plot of the intersegmental phase lag as a function of percentage deviation from the baseline value of parameter $f$ where 0\% corresponds to the baseline $f$ value ($f=3$).}
    \label{fig:sens_f}
\end{figure*}

\begin{figure*}[!htb]

    \centering
    \begin{subfigure}[b]{0.3\textwidth}
        \centering
        \includegraphics[trim=30 250 60 260,clip,width=\textwidth]{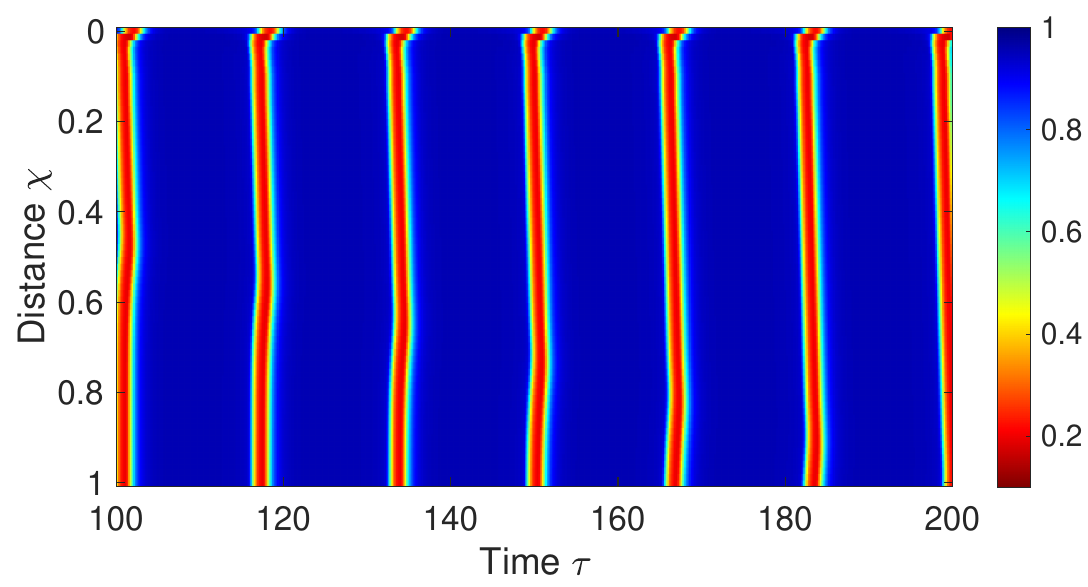}
        \caption{}
        \label{fig:w_I_Decrease}
    \end{subfigure}
\
    \begin{subfigure}[b]{0.3\textwidth}   
        \centering 
        \includegraphics[trim=30 250 60 260,clip,width=\textwidth]{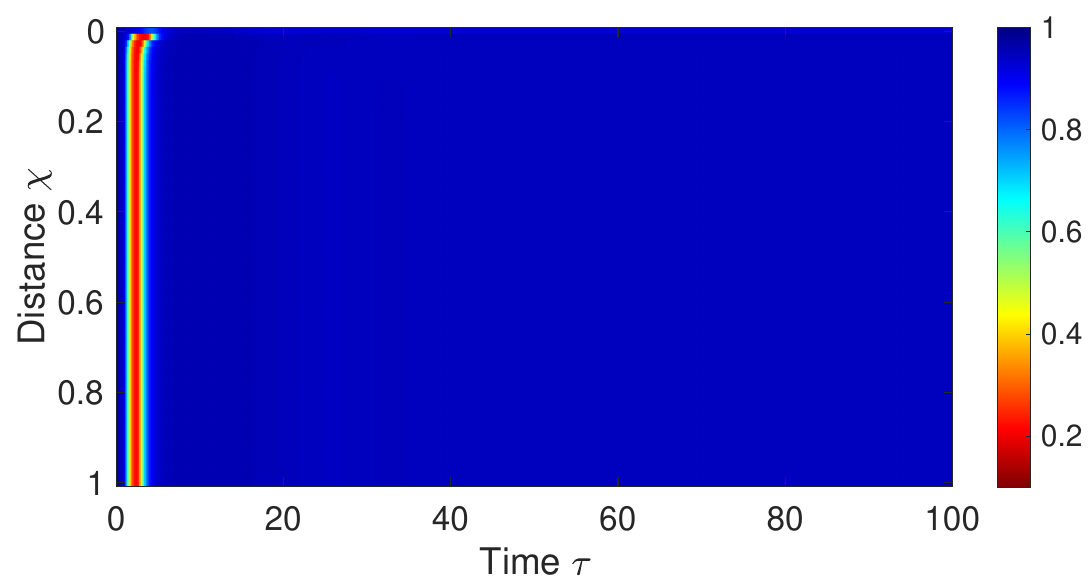}
        \caption{}
        \label{fig:w_I_Increase}
    \end{subfigure}
    \
    \begin{subfigure}[b]{0.3\textwidth}   
        \centering 
        \includegraphics[trim=30 250 60 260,clip,width=\textwidth]{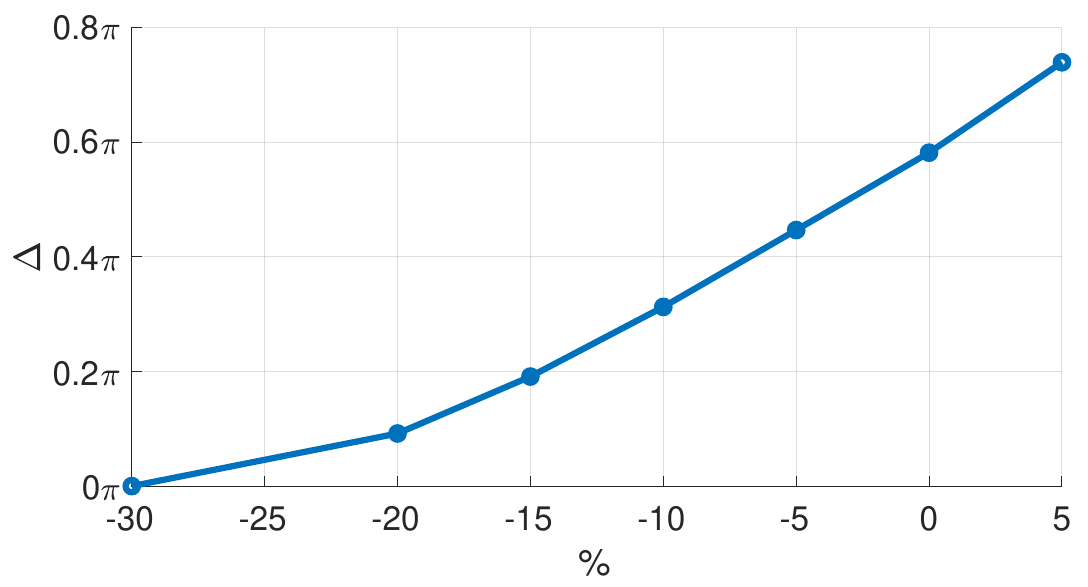}
        \caption{}
        \label{fig:IPL_w_I}
    \end{subfigure}
    \caption{Solution sensitivity to varying the value of $w_I$. \textbf{(a)} Color-coded topography of muscle contraction pattern ($\theta$) with the parameter $w_I$ modified to $0.5w_I$, representing a 50\% decrease from the original value of $w_I$.  \textbf{(b)} Color-coded topography of muscle contraction pattern ($\theta$) with the parameter $w_I$ modified to $1.2w_I$, representing a 20\% increase from the original value of $w_I$. \textbf{(c)} Plot of the contraction propagating speed as a function of percentage deviation from the baseline value of parameter $w_I$ where 0\% corresponds to the baseline $w_I$ value ($w_I=1.35$).}
    \label{fig:sens_w_I}
\end{figure*}

\begin{figure*}[!htb]

    \centering
    \begin{subfigure}[b]{0.4\textwidth}
        \centering
        \includegraphics[trim=30 250 60 260,clip,width=\textwidth]{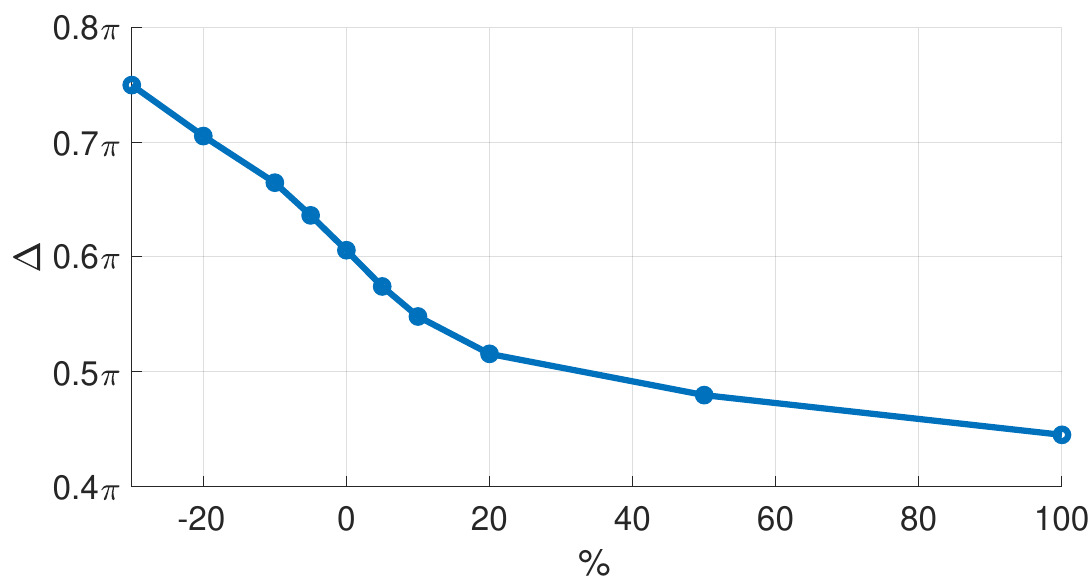}
        \caption{}
        \label{fig:b_IPL}
    \end{subfigure}
\
    \begin{subfigure}[b]{0.4\textwidth}   
        \centering 
        \includegraphics[trim=30 250 60 260,clip,width=\textwidth]{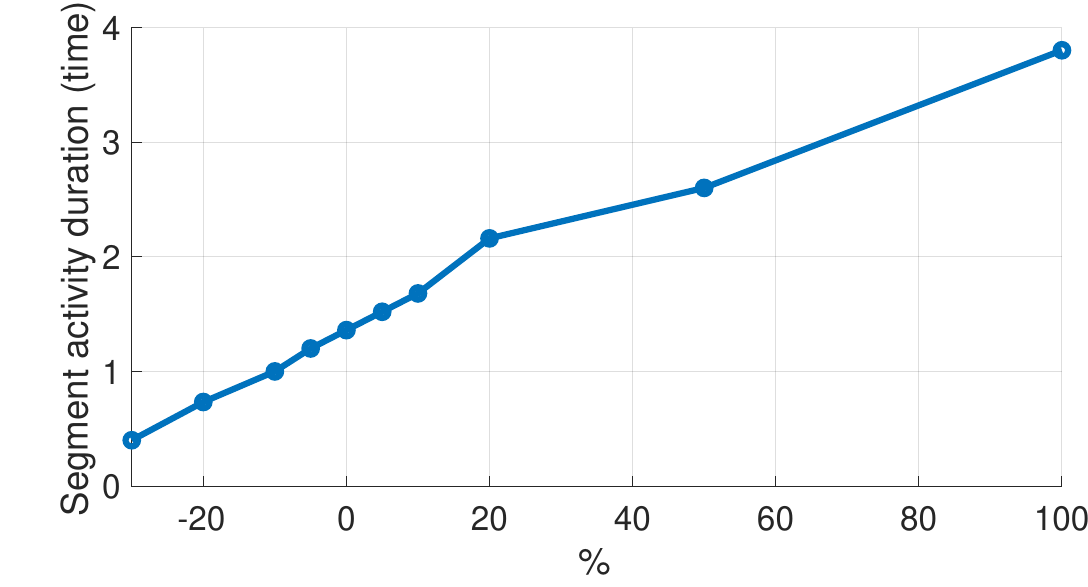}
        \caption{}
        \label{fig:b_actLevel}
    \end{subfigure}
    \
    \begin{subfigure}[b]{0.4\textwidth}   
        \centering 
        \includegraphics[trim=30 250 60 260,clip,width=\textwidth]{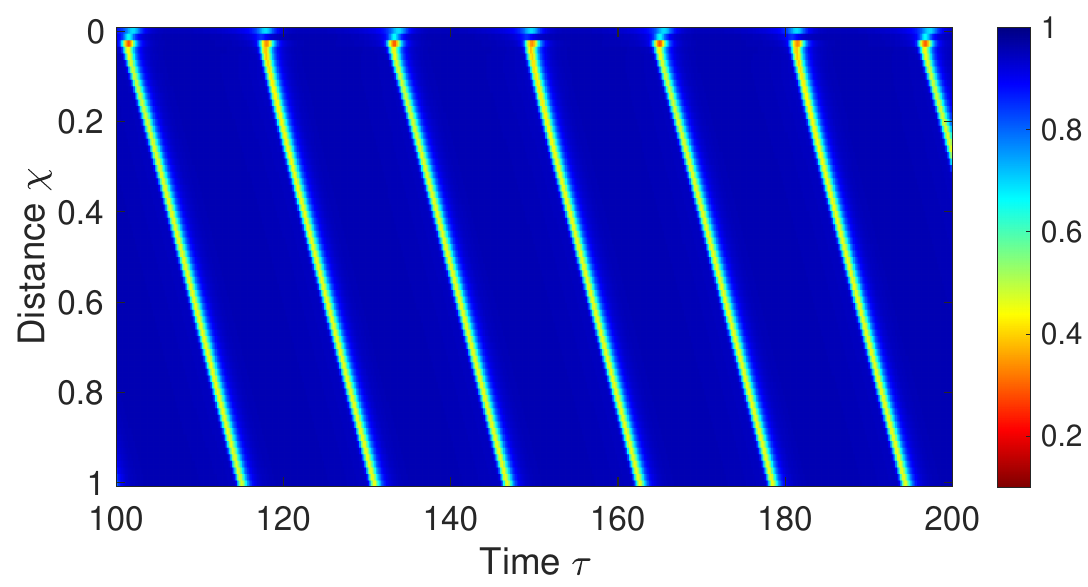}
        \caption{}
        \label{fig:bDecrease}
    \end{subfigure}
    \
    \begin{subfigure}[b]{0.4\textwidth}   
        \centering 
        \includegraphics[trim=30 250 60 260,clip,width=\textwidth]{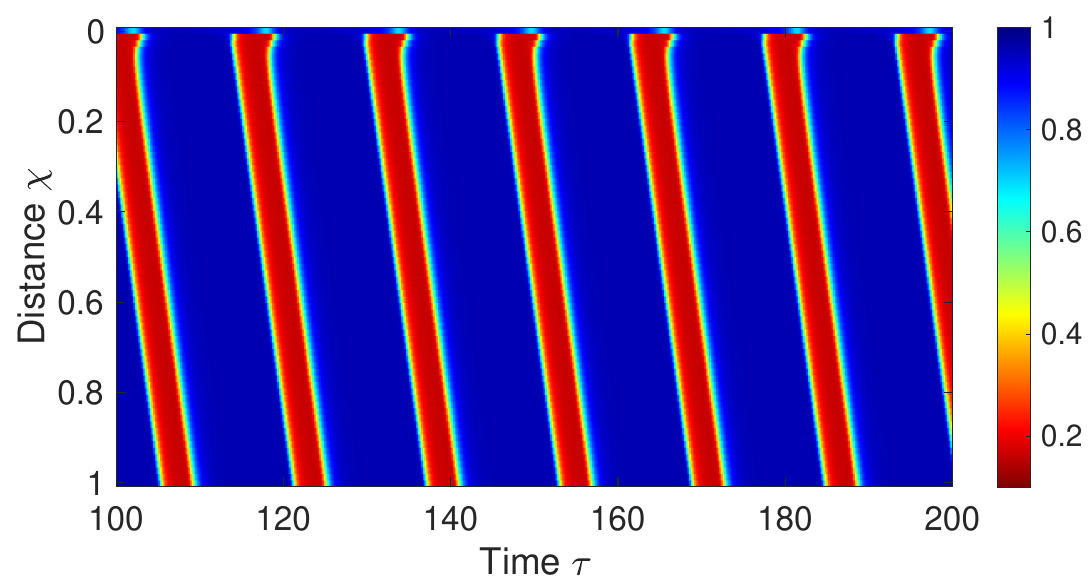}
        \caption{}
        \label{fig:bIncrease}
    \end{subfigure}
    \caption{Solution sensitivity to varying the value of $b$. \textbf{(a)} Plot of the intersegmental phase lag as a function of percentage deviation from the baseline value of parameter $b$ where 0\% corresponds to the baseline $b$ value ($b=20$). \textbf{(b)} Plot of the segment activity duration (how long an oscillator is active) as a function of percentage deviation from the baseline value of parameter $b$. Due to the relation between muscle contraction pattern ($\theta$) and the excitatory activity level ($E$) displayed in Eq. (\ref {eq:tht}), segment activity duration is directly related to contraction strength. \textbf{(c)} Color-coded topography of muscle contraction pattern ($\theta$) with the parameter $b$ modified to $0.7b$, representing a 30\% decrease from the original value of $b$.  \textbf{(d)} Color-coded topography of muscle contraction pattern ($\theta$) with the parameter $b$ modified to $2b$, representing a 100\% increase from the original value of $b$.}
    \label{fig:sens_b}
\end{figure*}

\begin{figure*}[!htb]
    \begin{subfigure}[b]{0.45\textwidth}   
        \centering 
        \includegraphics[trim=30 250 60 260,clip,width=\textwidth]{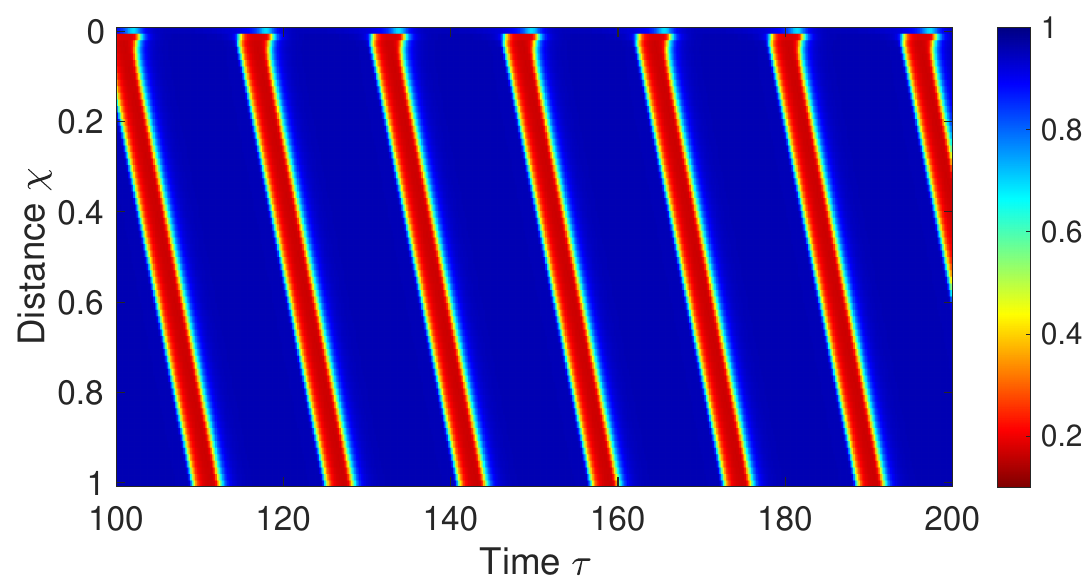}
        \caption{}
        \label{fig:dDecrease}
    \end{subfigure}
    \
    \begin{subfigure}[b]{0.45\textwidth}   
        \centering 
        \includegraphics[trim=30 250 60 260,clip,width=\textwidth]{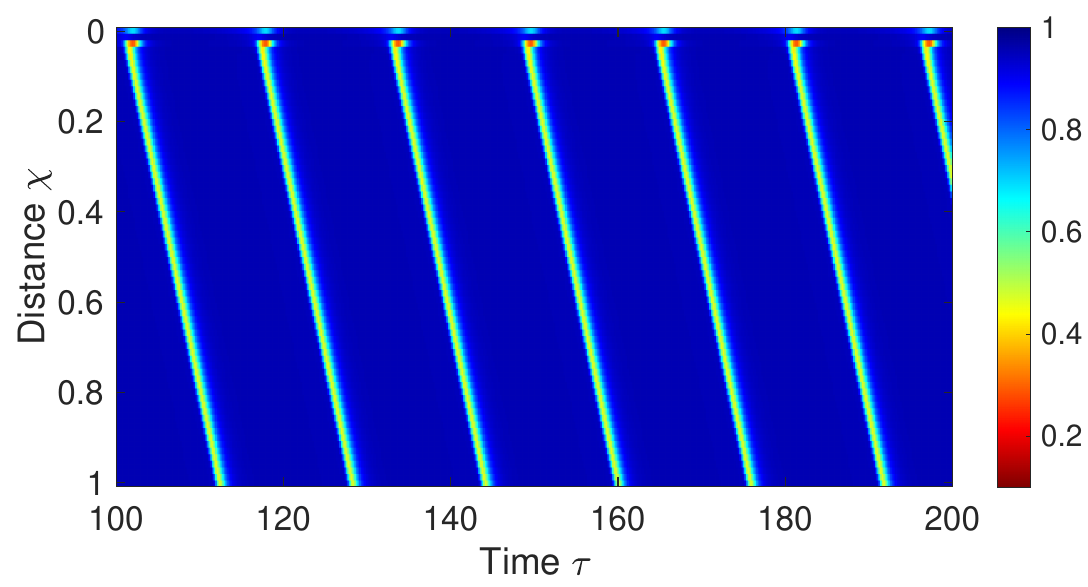}
        \caption{}
        \label{fig:dIncrease}
    \end{subfigure}
    \caption{Solution sensitivity to varying the value of $d$. \textbf{(a)} Color-coded topography of muscle contraction pattern ($\theta$) with the parameter $d$ modified to $0.5d$, representing a 50\% decrease from the original value of $d$.  \textbf{(b)} Color-coded topography of muscle contraction pattern ($\theta$) with the parameter $d$ modified to $1.5d$, representing a 50\% increase from the original value of $d$.}
    \label{fig:sens_d}
\end{figure*}

The results of the sensitivity testing reveals that the model is in fact robust to varying individual parameters. Additionally, it shows that the model is more sensitive to some parameters over others. Lastly, the results uncover interesting patterns which provides insights into the role each parameter plays in creating the desired RACs pattern, with the constrains discussed in the main manuscript. Samples for the qualitative and quantitative analysis are presented graphically or though topographic maps in Figs \ref{fig:sens_e},\ref{fig:sens_f},\ref{fig:sens_w_I},\ref{fig:sens_b}, and \ref{fig:sens_d} are discussed next. Table – provides a summary of the parametric investigation.

Figure \ref{fig:sens_e} displays the simulations solutions to different values of $e$. Smaller $e$ value implies less inhibition of excitatory neurons, increasing overall excitatory activity. This is translated into uniform or fast propagating contractions, and increase intersegmental phase lag (Fig. \ref{fig:eDecrease}). We explain this observation though the following concepts. When excitation is increased, excitatory neurons fire more frequently and with greater intensity, leading to a stronger signal being transmitted along the neural pathways. Stronger signal can overcome the differences in physical properties between segments, and propagate more quickly along the neural pathways, thus, reduce intersegmental phase lag. For the exact same reason, we see an opposite pattern for increasing $e$ value, and thus reducing excitatory activity, which results in greater phase lag (Fig. \ref{fig:eIncrease}).

%%%%% sensitivity in parameter e %%%%%%%%%%%%

Figure \ref{fig:sens_f} presents the results obtained by the simulations with different $f$ values. Since $f$ is responsible for the inhibition of inhibitory neuronal population, the pattern is opposite to the one obtained by varying $e$. As shown in Fig. \ref{fig:sens_f}, increasing $f$ results in fast contractions whereas decreasing $f$ results in slower propagation, more slanted contractions. Increasing $f$ means less excitation of inhibitory neurons, thus less inhibition on excitatory population, and therefore more uniform firing.

%%%%% sensitivity in parameter f %%%%%%%%%%%%

Figure \ref{fig:sens_w_I} displays the results obtained using various values of the parameter $w_I$. These results demonstrate that reducing this parameter, indicating less overall inhibition, leads to a significant decrease in intersegmental phase lag, in line with our expectations. However, when examining the system's response to increased values of $w_I$, it becomes evident that the system converges to an absent contractile response. Initially, there is a spike, known as the excitable regime, which quickly decays to the system's resting state. This outcome is due to the fact that the $w_I$ value chosen for the baseline case ($w_I=1.35$) is close to the upper limit for this parameter for which a limit cycle solution exists. Increasing $w_I$ shifts the $I$ nullcline (represented by the black curve in Fig. \ref{fig:phaseDiag}) to the left on the phase diagram.

%%%%% sensitivity in parameter w_I %%%%%%%%%%%%

Other interesting parameters are $b$ and $d$, which were discussed in great details in \cite{Gjorgjieva2013} and \cite{Pehlevan2016}. Figures \ref{fig:sens_b}and 5 present the results obtained by the simulations with different $b$ and $d$ values, respectively. Both \cite{Gjorgjieva2013} and \cite{Pehlevan2016} concluded that increasing intersegmental coupling has a stabilizing effect on propagation, where larger values of $b$ and $d$ increase the model’s robustness. However, our scenario has additional constrains that were not considered in previous work.
While making the solution more stable though increasing $b$ and $d$, one minimizes the effect of the other parameters on solution. Specifically, the interplay between $b$ and $w_I$ is essential to capture the unique features of the esophagus (discussed in the main manuscript and shown in Fig \ref{fig:distentionTests}). If $b$ is too large, it can cause in the signal to propagate even without the present of distension (a solution which looks like Fig. \ref{fig:shortBagTrav} rather than Fig. \ref{fig:shortBag}).

As shown in Fig. \ref{fig:b_IPL}, increasing $b$ values is translated into reduction of intersegmental phase lag. However, varying $b$ has an additional, much more obvious effect on the solution, shown in Fig. \ref{fig:b_actLevel}. Increasing $b$ implies stronger and longer contraction as segment activity duration increases. On the other hand, when $b$ is reduced, there is not enough excitatory signal passing, leading very weak contractions. These patterns can be observed qualitatively in Fig. \ref{fig:bDecrease} and Fig. \ref{fig:bIncrease}.

%%%%% sensitivity in parameter b %%%%%%%%%%%%

An opposite pattern is observed when increasing or decreasing $d$. Increasing inhibitory signal to excitatory population from nearest-neighbor ($d$) implies weaker and shorter contraction duration as segment activity duration decreases. This is presented qualitatively in Fig. \ref{fig:sens_d}.

%%%%% sensitivity in parameterd %%%%%%%%%%%%

Lastly, note that the system’s response to the change of strength of input parameter to excitatory population, $w_E$, is as the center of a different study, which we encourage the readers to visit \cite{elisha2024direct}.

\begin{figure*}[!htb]

    \centering
    \begin{subfigure}[b]{0.4\textwidth}
        \centering
        \includegraphics[trim=30 250 60 260,clip,width=\textwidth]{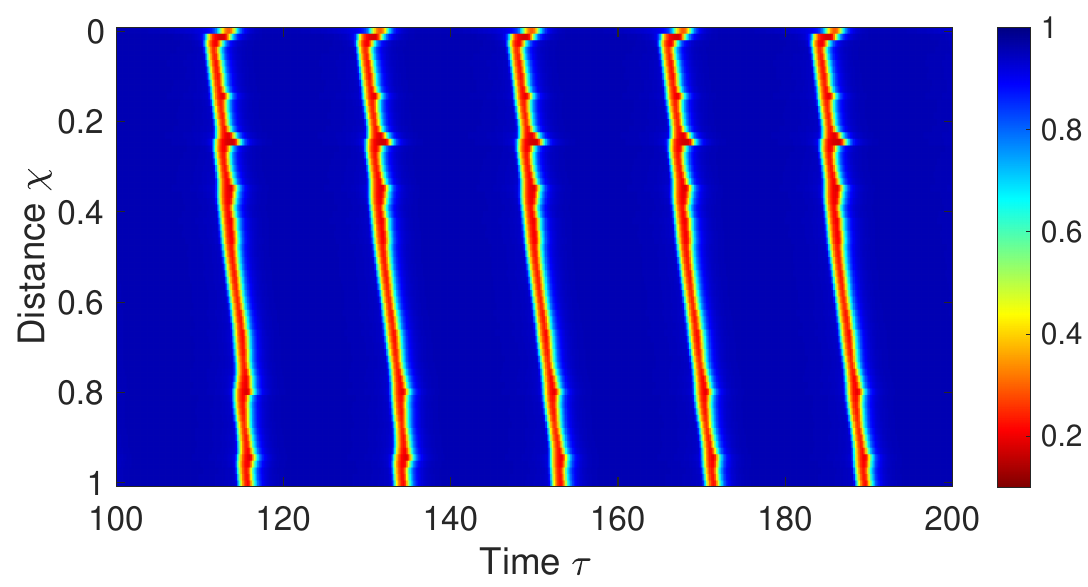}
        \caption{}
        \label{fig:c_var}
    \end{subfigure}
\
    \begin{subfigure}[b]{0.4\textwidth}   
        \centering 
        \includegraphics[trim=30 250 60 260,clip,width=\textwidth]{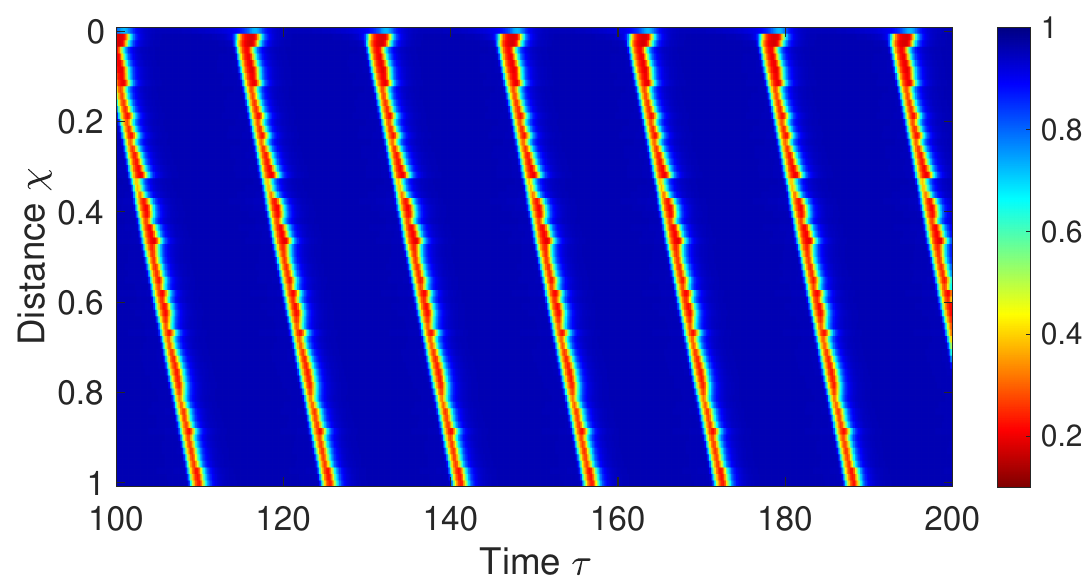}
        \caption{}
        \label{fig:e_var}
    \end{subfigure}
    \
    \begin{subfigure}[b]{0.4\textwidth}   
        \centering 
        \includegraphics[trim=30 250 60 260,clip,width=\textwidth]{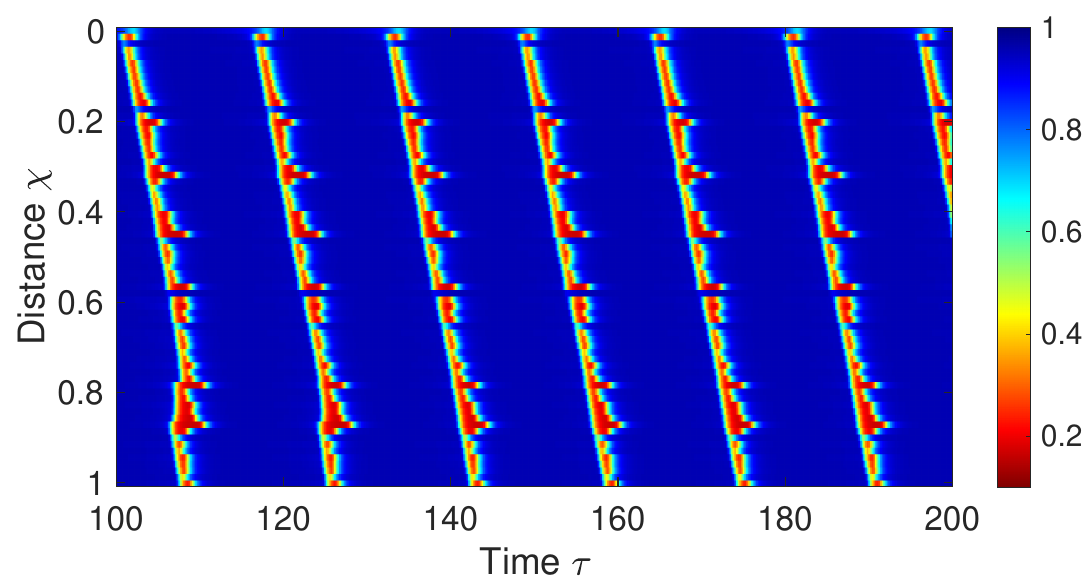}
        \caption{}
        \label{fig:d_var}
    \end{subfigure}
    \
    \begin{subfigure}[b]{0.4\textwidth}   
        \centering 
        \includegraphics[trim=30 250 60 260,clip,width=\textwidth]{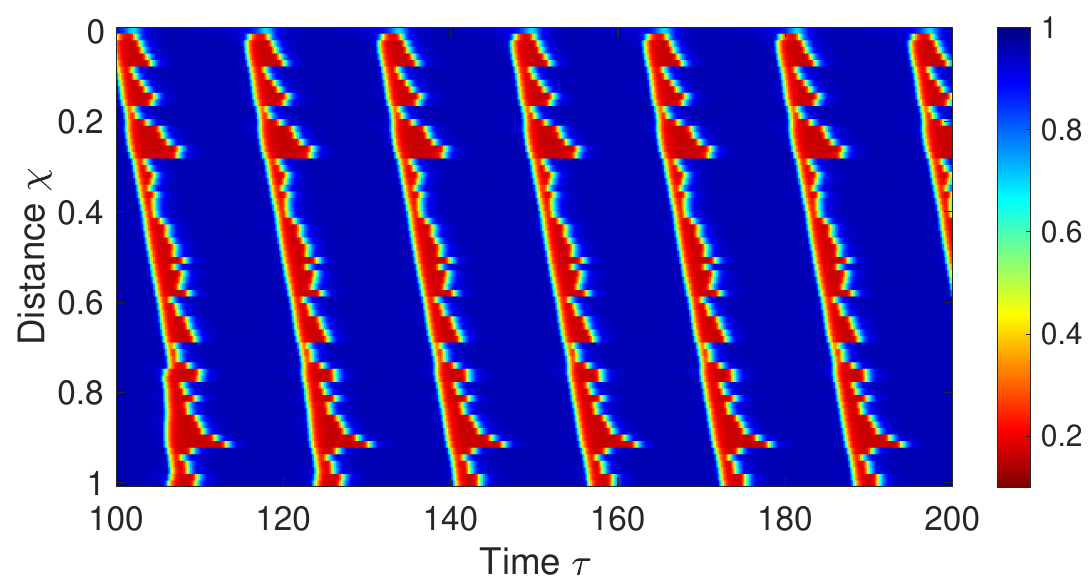}
        \caption{}
        \label{fig:b_var}
    \end{subfigure}
    \caption{Color-coded muscle contraction pattern ($\theta$) topography obtained by the neuromechanical model with randomly sample values from a Gaussian distribution with specific mean and variance. \textbf{(a)} Irregularities introduced to $c$ with mean=12 and variance = 3. \textbf{(b)} Irregularities introduced to $e$ with mean=15 and variance = 10. \textbf{(b)} Irregularities introduced to $d$ with mean=40 and variance = 20. \textbf{(d)} Irregularities introduced to $b$ with mean=20 and variance = 10.}
    \label{fig:sens_varia}
\end{figure*}

As indicated in the main text, biological systems are naturally prone to variation, making it essential to test the model's ability to withstand minor disruptions without altering its contraction pattern. To achieve this, we systematically assess the model's robustness by introducing small irregularities into its parameters along the length. For each parameter, we randomly sample values from a Gaussian distribution with specific mean and variance, and examine whether the model can maintain its contraction pattern when subjected to these parameter variations. Several examples are presented in Fig. \ref{fig:sens_varia}. The figure indicates that the model is robust for the parametric values chosen as the baseline solution.

Note that variability may not come from randomness but due to structural changes in the esophagus.

%%%%% Varry parameter Gaussian %%%%%%%%%%%%

\subsection{Background and theory - relaxation oscillators and limit cycle solution}\label{sec:limitCycleTheory} 
A relaxation oscillator is a mathematical model of electronic circuit that can exhibit a variety of behaviors that depend on the intrinsic and input parameters defining it \cite{wang1999relaxation}. Relaxation oscillators have been used for years to model the electrical activity of the heart due to their ability to display rhythmic, repetitive patterns known as limit cycles \cite{Trayanova2011, cherubini2008electromechanical, Du2010}. The shape, frequency, and amplitude of the oscillations depend on the properties of the oscillator \cite{wang1999relaxation}.

Unlike the heart, the esophagus does not exhibit a consistent rhythmic pattern in its natural state. In the absence of any electrical input or stimulation, the esophagus remains at rest \cite{Goyal2008}. When subjected to short-term stimuli, the esophagus briefly becomes active, undergoing a single cycle of activity before gradually returning to its resting state \cite{Paterson1988}. Lastly, under sustained stimuli (through sustained volumetric distension), a rhythmic pattern emerges in its behavior \cite{Carlson2020Repetitive}. By definition, relaxation oscillators can exhibit these distinct behaviors based on the presence and nature of external inputs, as discussed next.

\begin{figure*}[!htb]

    \centering
    \begin{subfigure}[b]{0.3\textwidth}
        \centering
        \includegraphics[trim=60 150 65 210,clip,width=\textwidth]{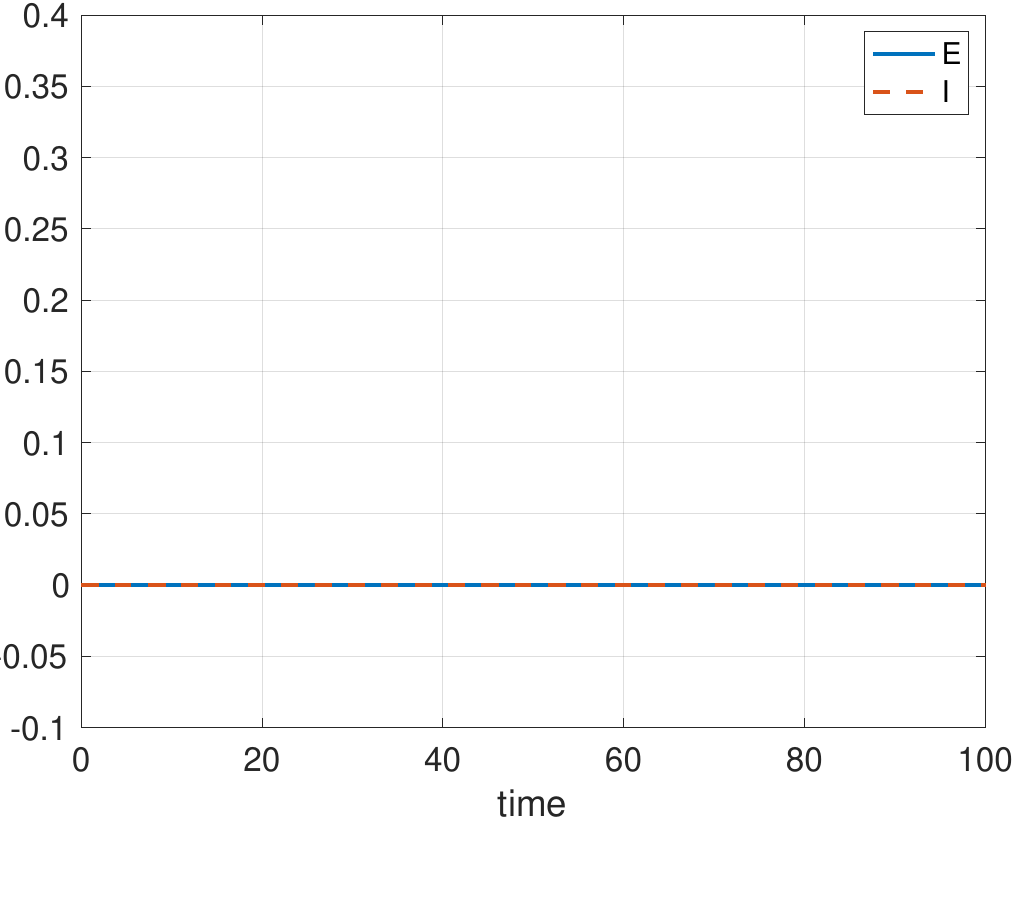}
        \caption{}
        \label{fig:sup_E_I_noInput}
    \end{subfigure}
\
    \begin{subfigure}[b]{0.3\textwidth}   
        \centering 
        \includegraphics[trim=60 150 65 210,clip,width=\textwidth]{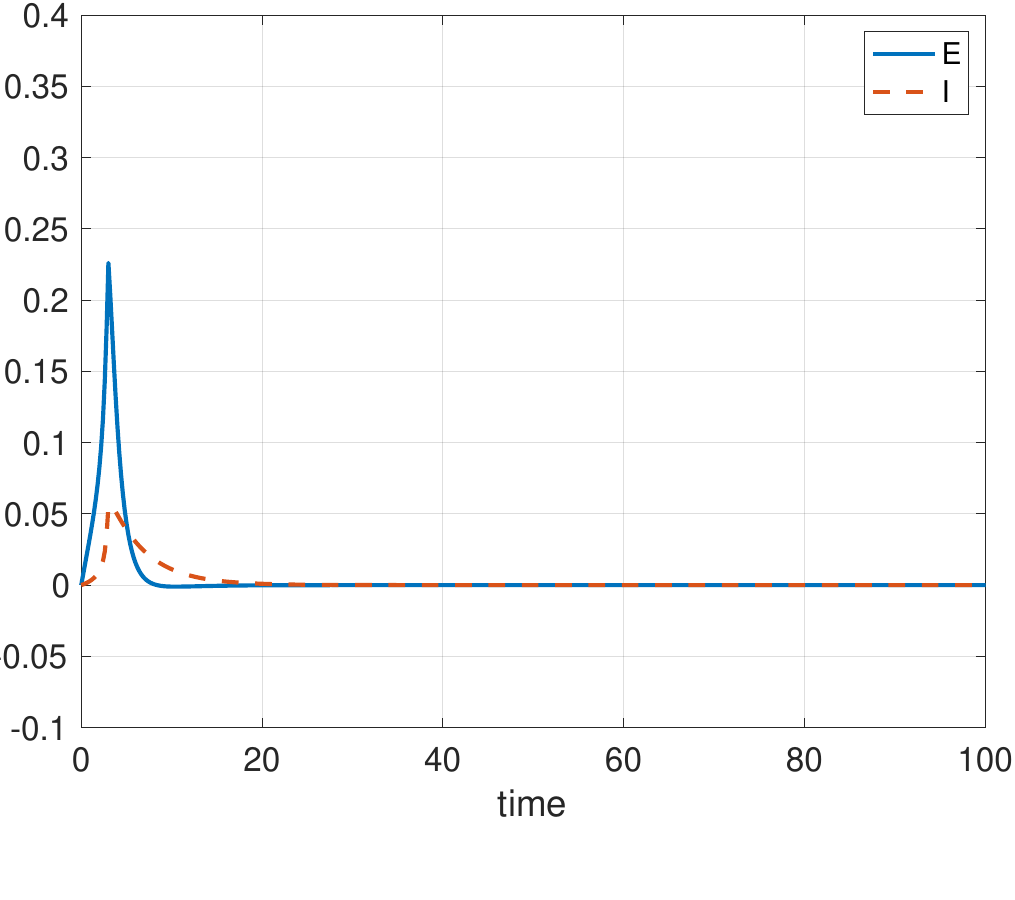}
        \caption{}
        \label{fig:sup_E_I_shortInput}
    \end{subfigure}
    \
    \begin{subfigure}[b]{0.3\textwidth}   
        \centering 
        \includegraphics[trim=60 150 65 210,clip,width=\textwidth]{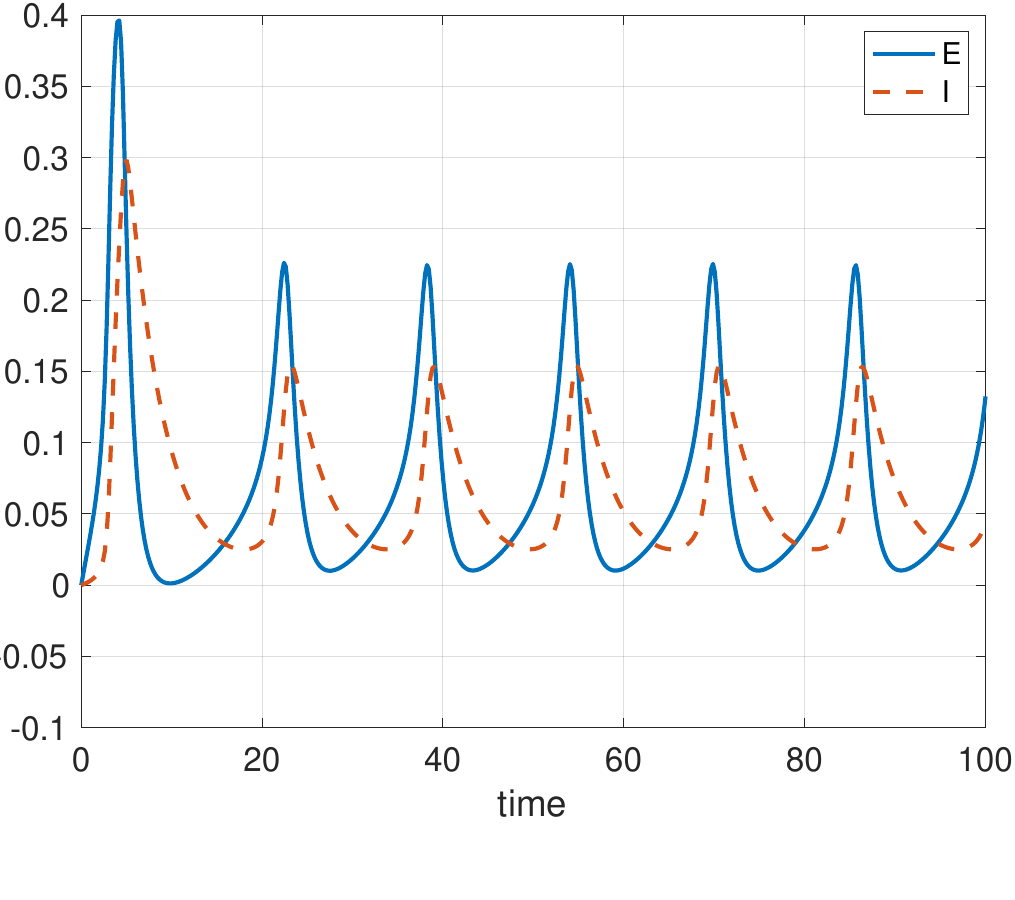}
        \caption{}
        \label{fig:sup_E_I_constInput}
    \end{subfigure}
    \caption{Solution of a relaxation oscillator with different inputs. Plots of the state variables ($E$ and $I$) over time. \textbf{(a)} No input is introduced ($S_E=0$), so the system remains at rest, where $E$ and $I$ do not change over time, remaining at zero. \textbf{(b)} System’s response to short-term stimuli creating excitable regime, in which $E$ and $I$ increase in response to the transient stimuli before decaying to rest values. \textbf{(c)} System’s response to sustained stimuli, creating a limit cycle solution, in which $E$ and $I$ fluctuate at constant pattern over time.}
    \label{fig:solType}
\end{figure*}

In the absence of any prescribed input, a relaxation oscillator remains in a resting state. During this time, the system’s output remains constant (Fig. \ref{fig:sup_E_I_noInput}). When a transient input is introduced, a relaxation oscillator responds by producing a transient and distinguishable output. Following the excitation, the system gradually decays to its rest state (Fig. \ref{fig:sup_E_I_shortInput}). Lastly, if the input is sustained and exceeds a certain threshold, the relaxation oscillator can transition into a self-sustained, rhythmic oscillatory behavior known as a limit cycle solution (Fig. \ref{fig:sup_E_I_constInput}). 

\begin{figure*}[!htb]
    \centering{{\includegraphics[trim=0 180 0 200 ,clip,width=0.7\textwidth]{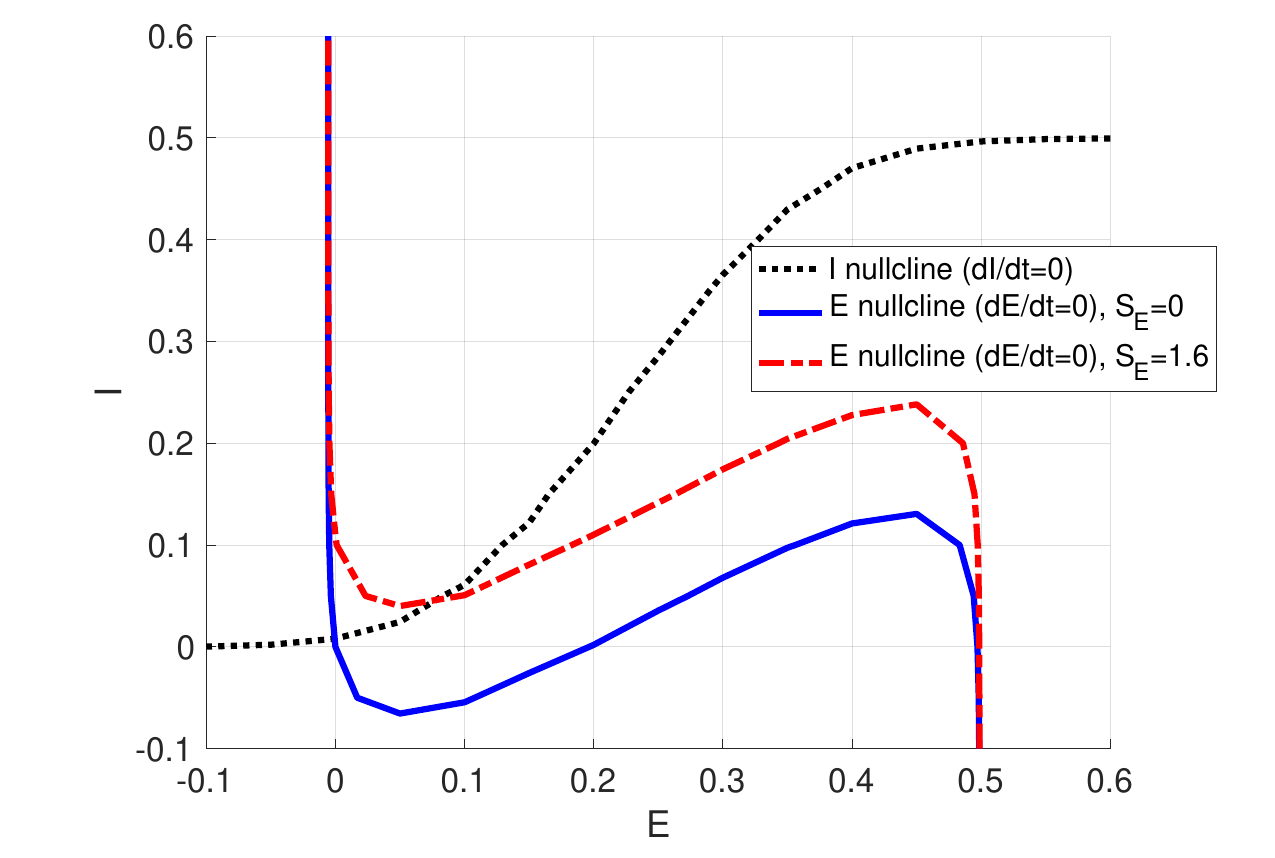}}}
    \caption{Phase diagram of a single Wilson-Cowan oscillator with the input parameters in table \ref{table:param}, including the $I$ (black) and two $E$ nullclines (blue $S_E=0$ and red $S_E=1.6$). For $S_E=1.6$, a limit cycle solution emerges (Fig. \ref{fig:sup_E_I_constInput}).}%, and the temporal response is plotted in purple.}
    \label{fig:phaseDiag}
\end{figure*}

Since this work is mostly concerned with esophageal rhythmic response, we wish to further explain limit cycle oscillations. To understand the emergence of limit cycle oscillations, we employ phase analysis, displayed in Fig. \ref{fig:phaseDiag}. By doing so, we can clearly identify the threshold value the input needs to surpass to obtain transition from rest state (at $S_E=0$) to limit cycle oscillations. The axis are the state variables, and the curves are the nullclines ($dE/dt=0$ and $dI/dt=0$). As $S_E$ increases, the $E$ nullcline shifts up, such that the intersection of the two curves is at the middle branch of the $E$ nullcline. As a results, a limit cycle solution emerges.  

%%%%%%%Phase diagram %%%%%%%%%%

 For additional information on dynamic systems, refer to \cite{wang1999relaxation, smeal2010phase, grasman2012asymptotic}.
 
 A chain of limit cycle oscillators is a well-studied system \cite{daniel1994relaxation, jones2006local}. The oscillators in such a system are coupled, often identical, and each oscillator exhibits a limit cycle solution, independent of the coupling \cite{cohen1992modelling,collins1993coupled,izhikevich2000phase}. Such systems often exhibit “frequency pulling”, where all oscillators converge into an entrained frequency. However, they are not synchronized, as there is phase shift between adjacent oscillators \cite{daniel1994relaxation, wang1995emergent}. At steady state, the oscillators phase lock, meaning that they oscillate at constant phase from one another, creating a special case of synchrony \cite{kopell1986symmetry,campbell1996synchronization}. This emerging pattern, known to be the fundamental solution of coupled system in equilibrium state, is the key element in creating the propagating attribute of the system \cite{strogatz1993coupled,kopell2003chains}.

\bibliographystyle{biophysj}
\bibliography{neuromech_bib_V5} 

\begin{thebibliography}{118}
\providecommand{\url}[1]{\texttt{#1}}
\providecommand{\urlprefix}{ }

\bibitem[Trayanova et~al.(2011)Trayanova, Constantino, and
  Gurev]{Trayanova2011}
Trayanova, N.~A., J.~Constantino, and V.~Gurev, 2011.
\newblock Electromechanical models of the ventricles.
\newblock \emph{American Journal of Physiology-Heart and Circulatory
  Physiology} 301:H279--H286.

\bibitem[Du et~al.(2010)Du, O'Grady, Davidson, Cheng, and Pullan]{Du2010}
Du, P., G.~O'Grady, J.~B. Davidson, L.~K. Cheng, and A.~J. Pullan, 2010.
\newblock Multiscale modeling of gastrointestinal electrophysiology and
  experimental validation.
\newblock \emph{Critical Reviews™ in Biomedical Engineering} 38.

\bibitem[Buijs et~al.(2016)Buijs, Le{\'o}n-Mercado, Guzm{\'a}n-Ruiz,
  Guerrero-Vargas, Romo-Nava, and Buijs]{buijs2016circadian}
Buijs, F.~N., L.~Le{\'o}n-Mercado, M.~Guzm{\'a}n-Ruiz, N.~N. Guerrero-Vargas,
  F.~Romo-Nava, and R.~M. Buijs, 2016.
\newblock The circadian system: a regulatory feedback network of periphery and
  brain.
\newblock \emph{Physiology} 31:170--181.

\bibitem[Sundar et~al.(2015)Sundar, Yao, Sellix, and
  Rahman]{sundar2015circadian}
Sundar, I.~K., H.~Yao, M.~T. Sellix, and I.~Rahman, 2015.
\newblock Circadian clock--coupled lung cellular and molecular functions in
  chronic airway diseases.
\newblock \emph{American journal of respiratory cell and molecular biology}
  53:285--290.

\bibitem[Rao and Gershon(2016)]{Rao2016}
Rao, M., and M.~D. Gershon, 2016.
\newblock The bowel and beyond: the enteric nervous system in neurological
  disorders.
\newblock \emph{Nature reviews Gastroenterology \& hepatology} 13:517--528.

\bibitem[Drossman et~al.(1993)Drossman, Li, Andruzzi, Temple, Talley,
  Grant~Thompson, Whitehead, Janssens, Funch-Jensen, Corazziari,
  et~al.]{Drossman1993}
Drossman, D.~A., Z.~Li, E.~Andruzzi, R.~D. Temple, N.~J. Talley,
  W.~Grant~Thompson, W.~E. Whitehead, J.~Janssens, P.~Funch-Jensen,
  E.~Corazziari, et~al., 1993.
\newblock US householder survey of functional gastrointestinal disorders:
  prevalence, sociodemography, and health impact.
\newblock \emph{Digestive diseases and sciences} 38:1569--1580.

\bibitem[Ouyang and Locke~III(2007)]{Ouyang2007}
Ouyang, A., and G.~R. Locke~III, 2007.
\newblock Overview of neurogastroenterology-gastrointestinal motility and
  functional GI disorders: classification, prevalence, and epidemiology.
\newblock \emph{Gastroenterology clinics of North America} 36:485--498.

\bibitem[Abramowitz et~al.(2013)Abramowitz, B{\'e}ziaud, Labreze, Giardina,
  Causs{\'e}, Chuberre, Allaert, and Perrot]{Abramowitz2013}
Abramowitz, L., N.~B{\'e}ziaud, L.~Labreze, V.~Giardina, C.~Causs{\'e},
  B.~Chuberre, F.~Allaert, and S.~Perrot, 2013.
\newblock Prevalence and impact of constipation and bowel dysfunction induced
  by strong opioids: a cross-sectional survey of 520 patients with cancer pain:
  DYONISOS study.
\newblock \emph{Journal of medical economics} 16:1423--1433.

\bibitem[Tuteja et~al.(2010)Tuteja, Biskupiak, Stoddard, and
  Lipman]{Tuteja2010Opioid}
Tuteja, A., J.~Biskupiak, G.~Stoddard, and A.~Lipman, 2010.
\newblock Opioid-induced bowel disorders and narcotic bowel syndrome in
  patients with chronic non-cancer pain.
\newblock \emph{Neurogastroenterology \& Motility} 22:424--e96.

\bibitem[Lydon et~al.(1999)Lydon, Cooke, Duggan, and Shorten]{Lydon1999}
Lydon, A.~M., T.~Cooke, F.~Duggan, and G.~D. Shorten, 1999.
\newblock Delayed postoperative gastric emptying following intrathecal morphine
  and intrathecal bupivacaine.
\newblock \emph{Canadian Journal of Anesthesia} 46:544--549.

\bibitem[Snyder and Vela(2023)]{Snyder2023}
Snyder, D.~L., and M.~F. Vela, 2023.
\newblock Impact of opioids on esophageal motility.
\newblock \emph{Neurogastroenterology \& Motility} e14587.

\bibitem[Sobczak et~al.(2014)Sobczak, Sa{\l}aga, Storr, and
  Fichna]{sobczak2014physiology}
Sobczak, M., M.~Sa{\l}aga, M.~A. Storr, and J.~Fichna, 2014.
\newblock Physiology, signaling, and pharmacology of opioid receptors and their
  ligands in the gastrointestinal tract: current concepts and future
  perspectives.
\newblock \emph{Journal of gastroenterology} 49:24--45.

\bibitem[Ten~Tusscher and Panfilov(2008)]{ten2008modelling}
Ten~Tusscher, K., and A.~V. Panfilov, 2008.
\newblock Modelling of the ventricular conduction system.
\newblock \emph{Progress in biophysics and molecular biology} 96:152--170.

\bibitem[Du et~al.(2018)Du, Calder, Angeli, Sathar, Paskaranandavadivel,
  O'Grady, and Cheng]{du2018progress}
Du, P., S.~Calder, T.~R. Angeli, S.~Sathar, N.~Paskaranandavadivel, G.~O'Grady,
  and L.~K. Cheng, 2018.
\newblock Progress in mathematical modeling of gastrointestinal slow wave
  abnormalities.
\newblock \emph{Frontiers in physiology} 8:1136.

\bibitem[Hardeland et~al.(2012)Hardeland, Madrid, Tan, and
  Reiter]{hardeland2012melatonin}
Hardeland, R., J.~A. Madrid, D.-X. Tan, and R.~J. Reiter, 2012.
\newblock Melatonin, the circadian multioscillator system and health: the need
  for detailed analyses of peripheral melatonin signaling.
\newblock \emph{Journal of pineal research} 52:139--166.

\bibitem[Shahriari et~al.(2020)Shahriari, Rosenfeld, and
  Anikeeva]{shahriari2020emerging}
Shahriari, D., D.~Rosenfeld, and P.~Anikeeva, 2020.
\newblock Emerging frontier of peripheral nerve and organ interfaces.
\newblock \emph{Neuron} 108:270--285.

\bibitem[Thomas(2008)]{thomas2008opioid}
Thomas, J., 2008.
\newblock Opioid-induced bowel dysfunction.
\newblock \emph{Journal of pain and symptom management} 35:103--113.

\bibitem[Albert-Vartanian et~al.(2016)Albert-Vartanian, Boyd, Hall, Morgado,
  Nguyen, Nguyen, Patel, Russo, Shao, and Raffa]{albert2016Opioid}
Albert-Vartanian, A., M.~Boyd, A.~Hall, S.~Morgado, E.~Nguyen, V.~Nguyen,
  S.~Patel, L.~Russo, A.~Shao, and R.~Raffa, 2016.
\newblock Will peripherally restricted kappa-opioid receptor agonists (pKORA s)
  relieve pain with less opioid adverse effects and abuse potential?
\newblock \emph{Journal of Clinical Pharmacy and Therapeutics} 41:371--382.

\bibitem[Al-Hasani and Bruchas(2011)]{al2011molecular}
Al-Hasani, R., and M.~R. Bruchas, 2011.
\newblock Molecular mechanisms of opioid receptor-dependent signaling and
  behavior.
\newblock \emph{The Journal of the American Society of Anesthesiologists}
  115:1363--1381.

\bibitem[Clav{\'e} and Shaker(2015)]{clave2015dysphagia}
Clav{\'e}, P., and R.~Shaker, 2015.
\newblock Dysphagia: current reality and scope of the problem.
\newblock \emph{Nature Reviews Gastroenterology \& Hepatology} 12:259--270.

\bibitem[Seguella and Gulbransen(2021)]{seguella2021enteric}
Seguella, L., and B.~D. Gulbransen, 2021.
\newblock Enteric glial biology, intercellular signalling and roles in
  gastrointestinal disease.
\newblock \emph{Nature Reviews Gastroenterology \& Hepatology} 18:571--587.

\bibitem[Park and Conklin(1999)]{Park1999}
Park, H., and J.~L. Conklin, 1999.
\newblock Neuromuscular control of esophageal peristalsis.
\newblock \emph{Current gastroenterology reports} 1:186--197.

\bibitem[Fung and Vanden~Berghe(2020)]{fung2020functional}
Fung, C., and P.~Vanden~Berghe, 2020.
\newblock Functional circuits and signal processing in the enteric nervous
  system.
\newblock \emph{Cellular and Molecular Life Sciences} 77:4505--4522.

\bibitem[Kulkarni et~al.(2018)Kulkarni, Ganz, Bayrer, Becker, Bogunovic, and
  Rao]{kulkarni2018advances}
Kulkarni, S., J.~Ganz, J.~Bayrer, L.~Becker, M.~Bogunovic, and M.~Rao, 2018.
\newblock Advances in enteric neurobiology: the “brain” in the gut in
  health and disease.
\newblock \emph{Journal of Neuroscience} 38:9346--9354.

\bibitem[Goyal and Chaudhury(2008)]{Goyal2008}
Goyal, R.~K., and A.~Chaudhury, 2008.
\newblock Physiology of normal esophageal motility.
\newblock \emph{Journal of clinical gastroenterology} 42:610.

\bibitem[Spencer and Hu(2020)]{spencer2020enteric}
Spencer, N.~J., and H.~Hu, 2020.
\newblock Enteric nervous system: sensory transduction, neural circuits and
  gastrointestinal motility.
\newblock \emph{Nature reviews Gastroenterology \& hepatology} 17:338--351.

\bibitem[Patel and Thavamani(2023)]{patel2023physiology}
Patel, K.~S., and A.~Thavamani, 2023.
\newblock Physiology, peristalsis.
\newblock \emph{In} StatPearls [Internet], StatPearls Publishing.

\bibitem[Mittal(2016)]{Mittal2016}
Mittal, R.~K., 2016.
\newblock Regulation and dysregulation of esophageal peristalsis by the
  integrated function of circular and longitudinal muscle layers in health and
  disease.
\newblock \emph{American Journal of Physiology-Gastrointestinal and Liver
  Physiology} 311:G431--G443.

\bibitem[Sifrim and Jafari(2012)]{Sifrim2012}
Sifrim, D., and J.~Jafari, 2012.
\newblock Deglutitive inhibition, latency between swallow and esophageal
  contractions and primary esophageal motor disorders.
\newblock \emph{Journal of neurogastroenterology and motility} 18:6.

\bibitem[Woodland et~al.(2013)Woodland, Sifrim, Krarup, Brock, Fr{\o}kj{\ae}r,
  Lottrup, Drewes, Swanstrom, and Farmer]{Woodland2013}
Woodland, P., D.~Sifrim, A.~L. Krarup, C.~Brock, J.~B. Fr{\o}kj{\ae}r,
  C.~Lottrup, A.~M. Drewes, L.~L. Swanstrom, and A.~D. Farmer, 2013.
\newblock The neurophysiology of the esophagus.
\newblock \emph{Annals of the New York Academy of Sciences} 1300:53--70.

\bibitem[Paterson(2006)]{paterson2006esophageal}
Paterson, W.~G., 2006.
\newblock Esophageal peristalsis.
\newblock \emph{GI Motility online} .

\bibitem[Pandolfino et~al.(2009)Pandolfino, Fox, Bredenoord, and
  Kahrilas]{pandolfino2009high}
Pandolfino, J., M.~Fox, A.~Bredenoord, and P.~Kahrilas, 2009.
\newblock High-resolution manometry in clinical practice: utilizing pressure
  topography to classify oesophageal motility abnormalities.
\newblock \emph{Neurogastroenterology \& Motility} 21:796--806.

\bibitem[Gorti et~al.(2020)Gorti, Samo, Shahnavaz, and Qayed]{Gorti2020}
Gorti, H., S.~Samo, N.~Shahnavaz, and E.~Qayed, 2020.
\newblock Distal esophageal spasm: Update on diagnosis and management in the
  era of high-resolution manometry.
\newblock \emph{World Journal of Clinical Cases} 8:1026.

\bibitem[Yadlapati et~al.(2021)Yadlapati, Kahrilas, Fox, Bredenoord,
  Prakash~Gyawali, Roman, Babaei, Mittal, Rommel, Savarino,
  et~al.]{yadlapati2021esophageal}
Yadlapati, R., P.~J. Kahrilas, M.~R. Fox, A.~J. Bredenoord, C.~Prakash~Gyawali,
  S.~Roman, A.~Babaei, R.~K. Mittal, N.~Rommel, E.~Savarino, et~al., 2021.
\newblock Esophageal motility disorders on high-resolution manometry: Chicago
  classification version 4.0{\copyright}.
\newblock \emph{Neurogastroenterology \& Motility} 33:e14058.

\bibitem[Carlson et~al.(2021)Carlson, Baumann, Donnan, Krause, Kou, and
  Pandolfino]{Carlson2021Evaluating}
Carlson, D.~A., A.~J. Baumann, E.~N. Donnan, A.~Krause, W.~Kou, and J.~E.
  Pandolfino, 2021.
\newblock Evaluating esophageal motility beyond primary peristalsis: assessing
  esophagogastric junction opening mechanics and secondary peristalsis in
  patients with normal manometry.
\newblock \emph{Neurogastroenterology \& Motility} 33:e14116.

\bibitem[Donnan and Pandolfino(2020)]{Donnan2020}
Donnan, E.~N., and J.~E. Pandolfino, 2020.
\newblock EndoFLIP in the esophagus: assessing sphincter function, wall
  stiffness, and motility to guide treatment.
\newblock \emph{Gastroenterology Clinics} 49:427--435.

\bibitem[Savarino et~al.(2020)Savarino, Di~Pietro, Bredenoord, Carlson, Clarke,
  Khan, Vela, Yadlapati, Pohl, Pandolfino, et~al.]{savarino2020use}
Savarino, E., M.~Di~Pietro, A.~J. Bredenoord, D.~A. Carlson, J.~O. Clarke,
  A.~Khan, M.~F. Vela, R.~Yadlapati, D.~Pohl, J.~E. Pandolfino, et~al., 2020.
\newblock Use of the functional lumen imaging probe in clinical esophagology.
\newblock \emph{The American journal of gastroenterology} 115:1786.

\bibitem[Hirano et~al.(2017)Hirano, Pandolfino, and Boeckxstaens]{Hirano2017}
Hirano, I., J.~E. Pandolfino, and G.~E. Boeckxstaens, 2017.
\newblock Functional lumen imaging probe for the management of esophageal
  disorders: expert review from the clinical practice updates committee of the
  AGA Institute.
\newblock \emph{Clinical Gastroenterology and Hepatology} 15:325--334.

\bibitem[Carlson et~al.(2015{\natexlab{a}})Carlson, Lin, Rogers, Lin, Kahrilas,
  and Pandolfino]{Carlson2015Utilizing}
Carlson, D., Z.~Lin, M.~Rogers, C.~Lin, P.~Kahrilas, and J.~Pandolfino, 2015.
\newblock Utilizing functional lumen imaging probe topography to evaluate
  esophageal contractility during volumetric distention: a pilot study.
\newblock \emph{Neurogastroenterology \& Motility} 27:981--989.

\bibitem[Carlson et~al.(2020{\natexlab{a}})Carlson, Kou, Masihi, Acharya,
  Baumann, Donnan, Kahrilas, and Pandolfino]{Carlson2020Repetitive}
Carlson, D.~A., W.~Kou, M.~Masihi, S.~Acharya, A.~J. Baumann, E.~N. Donnan,
  P.~J. Kahrilas, and J.~E. Pandolfino, 2020.
\newblock Repetitive antegrade contraction: a novel response to sustained
  esophageal distension is modulated by cholinergic influence.
\newblock \emph{American Journal of Physiology-Gastrointestinal and Liver
  Physiology} 319:G696--G702.

\bibitem[Goyal(2022)]{Goyal2022}
Goyal, R.~K., 2022.
\newblock EndoFLIP Topography: Motor Patterns in an Obstructed Esophagus.
\newblock \emph{Gastroenterology} 163:552--555.

\bibitem[Carlson et~al.(2018)Carlson, Kahrilas, Ritter, Lin, and
  Pandolfino]{carlson2018mechanisms}
Carlson, D.~A., P.~J. Kahrilas, K.~Ritter, Z.~Lin, and J.~E. Pandolfino, 2018.
\newblock Mechanisms of repetitive retrograde contractions in response to
  sustained esophageal distension: a study evaluating patients with
  postfundoplication dysphagia.
\newblock \emph{American Journal of Physiology-Gastrointestinal and Liver
  Physiology} 314:G334--G340.

\bibitem[Carlson et~al.(2020{\natexlab{b}})Carlson, Kou, and
  Pandolfino]{Carlson2020}
Carlson, D.~A., W.~Kou, and J.~E. Pandolfino, 2020.
\newblock The rhythm and rate of distension-induced esophageal contractility: a
  physiomarker of esophageal function.
\newblock \emph{Neurogastroenterology \& Motility} 32:e13794.

\bibitem[Halder et~al.(2022)Halder, Yamasaki, Acharya, Kou, Elisha, Carlson,
  Kahrilas, Pandolfino, and Patankar]{Halder2022_VDL}
Halder, S., J.~Yamasaki, S.~Acharya, W.~Kou, G.~Elisha, D.~A. Carlson, P.~J.
  Kahrilas, J.~E. Pandolfino, and N.~A. Patankar, 2022.
\newblock Virtual disease landscape using mechanics-informed machine learning:
  Application to esophageal disorders.
\newblock \emph{Artificial Intelligence in Medicine} 102435.

\bibitem[Mercado-Perez and Beyder(2022)]{Mercado2022}
Mercado-Perez, A., and A.~Beyder, 2022.
\newblock Gut feelings: mechanosensing in the gastrointestinal tract.
\newblock \emph{Nature Reviews Gastroenterology \& Hepatology} 19:283--296.

\bibitem[Nishikawa et~al.(2007)Nishikawa, Biewener, Aerts, Ahn, Chiel, Daley,
  Daniel, Full, Hale, Hedrick, et~al.]{Nishikawa2007}
Nishikawa, K., A.~A. Biewener, P.~Aerts, A.~N. Ahn, H.~J. Chiel, M.~A. Daley,
  T.~L. Daniel, R.~J. Full, M.~E. Hale, T.~L. Hedrick, et~al., 2007.
\newblock Neuromechanics: an integrative approach for understanding motor
  control.
\newblock \emph{Integrative and comparative biology} 47:16--54.

\bibitem[Ijspeert(2008)]{ijspeert2008central}
Ijspeert, A.~J., 2008.
\newblock Central pattern generators for locomotion control in animals and
  robots: a review.
\newblock \emph{Neural networks} 21:642--653.

\bibitem[Wilson and Cowan(1972)]{Wilson1972}
Wilson, H.~R., and J.~D. Cowan, 1972.
\newblock Excitatory and inhibitory interactions in localized populations of
  model neurons.
\newblock \emph{Biophysical journal} 12:1--24.
\newblock \urlprefix\url{https://doi.org/10.1016/S0006-3495(72)86068-5}.

\bibitem[Kopell and Ermentrout(1986)]{kopell1986symmetry}
Kopell, N., and G.~B. Ermentrout, 1986.
\newblock Symmetry and phaselocking in chains of weakly coupled oscillators.
\newblock \emph{Communications on Pure and Applied Mathematics} 39:623--660.

\bibitem[Strogatz and Stewart(1993)]{strogatz1993coupled}
Strogatz, S.~H., and I.~Stewart, 1993.
\newblock Coupled oscillators and biological synchronization.
\newblock \emph{Scientific american} 269:102--109.

\bibitem[Kopell and Ermentrout(1988)]{kopell1988coupled}
Kopell, N., and G.~B. Ermentrout, 1988.
\newblock Coupled oscillators and the design of central pattern generators.
\newblock \emph{Mathematical biosciences} 90:87--109.

\bibitem[Smeal et~al.(2010)Smeal, Ermentrout, and White]{smeal2010phase}
Smeal, R.~M., G.~B. Ermentrout, and J.~A. White, 2010.
\newblock Phase-response curves and synchronized neural networks.
\newblock \emph{Philosophical Transactions of the Royal Society B: Biological
  Sciences} 365:2407--2422.

\bibitem[Izhikevich(2000)]{izhikevich2000phase}
Izhikevich, E.~M., 2000.
\newblock Phase equations for relaxation oscillators.
\newblock \emph{SIAM Journal on Applied Mathematics} 60:1789--1804.

\bibitem[Kopell and Ermentrout(2003)]{kopell2003chains}
Kopell, N., and G.~B. Ermentrout, 2003.
\newblock Chains of oscillators in motor and sensory systems.

\bibitem[Cohen et~al.(1992)Cohen, Ermentrout, Kiemel, Kopell, Sigvardt, and
  Williams]{cohen1992modelling}
Cohen, A.~H., G.~B. Ermentrout, T.~Kiemel, N.~Kopell, K.~A. Sigvardt, and T.~L.
  Williams, 1992.
\newblock Modelling of intersegmental coordination in the lamprey central
  pattern generator for locomotion.
\newblock \emph{Trends in neurosciences} 15:434--438.

\bibitem[Schwemmer and Lewis(2012)]{schwemmer2012theory}
Schwemmer, M.~A., and T.~J. Lewis, 2012.
\newblock The theory of weakly coupled oscillators.
\newblock \emph{Phase response curves in neuroscience: theory, experiment, and
  analysis} 3--31.

\bibitem[Diamant(1997)]{Diamant1997}
Diamant, N.~E., 1997.
\newblock Neuromuscular mechanisms of primary peristalsis.
\newblock \emph{The American journal of medicine} 103:40S--43S.

\bibitem[Yazaki and Sifrim(2012)]{Yazaki2012}
Yazaki, E., and D.~Sifrim, 2012.
\newblock Anatomy and physiology of the esophageal body.
\newblock \emph{Diseases of the Esophagus} 25:292--298.

\bibitem[McMahon et~al.(2007)McMahon, Fr{\o}kj{\ae}r, Kunwald, Liao,
  Funch-Jensen, Drewes, and Gregersen]{Mcmahon2007}
McMahon, B.~P., J.~B. Fr{\o}kj{\ae}r, P.~Kunwald, D.~Liao, P.~Funch-Jensen,
  A.~M. Drewes, and H.~Gregersen, 2007.
\newblock The functional lumen imaging probe (FLIP) for evaluation of the
  esophagogastric junction.
\newblock \emph{American Journal of Physiology-Gastrointestinal and Liver
  Physiology} 292:G377--G384.

\bibitem[Pedersen et~al.(2005)Pedersen, Drewes, and Gregersen]{Pedersen2005}
Pedersen, J., A.~M. Drewes, and H.~Gregersen, 2005.
\newblock New analysis for the study of the muscle function in the human
  oesophagus.
\newblock \emph{Neurogastroenterology \& Motility} 17:767--772.

\bibitem[Paterson et~al.(1988)Paterson, Rattan, and Goyal]{Paterson1988}
Paterson, W.~G., S.~Rattan, and R.~Goyal, 1988.
\newblock Esophageal responses to transient and sustained esophageal
  distension.
\newblock \emph{American Journal of Physiology-Gastrointestinal and Liver
  Physiology} 255:G587--G595.

\bibitem[Gregersen et~al.(2011)Gregersen, Villadsen, and
  Liao]{gregersen2011mechanical}
Gregersen, H., G.~E. Villadsen, and D.~Liao, 2011.
\newblock Mechanical characteristics of distension-evoked peristaltic
  contractions in the esophagus of systemic sclerosis patients.
\newblock \emph{Digestive diseases and sciences} 56:3559--3568.

\bibitem[Penagini et~al.(1996)Penagini, Picone, and
  Bianchi]{penagini1996effect}
Penagini, R., A.~Picone, and P.~A. Bianchi, 1996.
\newblock Effect of morphine and naloxone on motor response of the human
  esophagus to swallowing and distension.
\newblock \emph{American Journal of Physiology-Gastrointestinal and Liver
  Physiology} 271:G675--G680.

\bibitem[Christensen(1970)]{Christensen1970}
Christensen, J., 1970.
\newblock Patterns and origin of some esophageal responses to stretch and
  electrical stimulation.
\newblock \emph{Gastroenterology} 59:909--916.

\bibitem[Villadsen et~al.(2001)Villadsen, Storkholm, Zachariae, Hendel,
  Bendtsen, and Gregersen]{villadsen2001oesophageal}
Villadsen, G., J.~Storkholm, H.~Zachariae, L.~Hendel, F.~Bendtsen, and
  H.~Gregersen, 2001.
\newblock Oesophageal pressure--cross-sectional area distributions and
  secondary peristalsis in relation to subclassification of systemic sclerosis.
\newblock \emph{Neurogastroenterology \& Motility} 13:199--210.

\bibitem[Abrahao~Jr et~al.(2011)Abrahao~Jr, Bhargava, Babaei, Ho, and
  Mittal]{abrahao2011swallow}
Abrahao~Jr, L., V.~Bhargava, A.~Babaei, A.~Ho, and R.~Mittal, 2011.
\newblock Swallow induces a peristaltic wave of distension that marches in
  front of the peristaltic wave of contraction.
\newblock \emph{Neurogastroenterology \& Motility} 23:201--e110.

\bibitem[Gregersen and Lo(2018)]{Gregersen2018Pathophysiology}
Gregersen, H., and K.~M. Lo, 2018.
\newblock Pathophysiology and treatment of achalasia in a muscle mechanical
  perspective.
\newblock \emph{Annals of the New York Academy of Sciences} 1434:173--184.

\bibitem[Carlson et~al.(2015{\natexlab{b}})Carlson, Lin, Kahrilas, Sternbach,
  Donnan, Friesen, Listernick, Mogni, and Pandolfino]{Carlson2015}
Carlson, D.~A., Z.~Lin, P.~J. Kahrilas, J.~Sternbach, E.~N. Donnan, L.~Friesen,
  Z.~Listernick, B.~Mogni, and J.~E. Pandolfino, 2015.
\newblock The Functional Lumen Imaging Probe Detects Esophageal Contractility
  Not Observed With Manometry in Patients With~Achalasia.
\newblock \emph{Gastroenterology} 149:1742--1751.
\newblock \urlprefix\url{https://doi.org/10.1053/j.gastro.2015.08.005}.

\bibitem[Elisha et~al.(2024)Elisha, Gast, Halder, Solla, Kahrilas, Pandolfino,
  and Patankar]{elisha2024direct}
Elisha, G., R.~Gast, S.~Halder, S.~A. Solla, P.~J. Kahrilas, J.~E. Pandolfino,
  and N.~A. Patankar, 2024.
\newblock Direct and retrograde signal propagation in unidirectionally coupled
  Wilson-Cowan oscillators.
\newblock \emph{arXiv preprint arXiv:2402.18100} .

\bibitem[Yeoh et~al.(2017)Yeoh, Corrias, and Buist]{yeoh2017modelling}
Yeoh, J.~W., A.~Corrias, and M.~L. Buist, 2017.
\newblock Modelling human colonic smooth muscle cell electrophysiology.
\newblock \emph{Cellular and molecular bioengineering} 10:186--197.

\bibitem[Yang et~al.(2023)Yang, Almanzar, and Chiu]{yang2023role}
Yang, D., N.~Almanzar, and I.~M. Chiu, 2023.
\newblock The role of cellular and molecular neuroimmune crosstalk in gut
  immunity.
\newblock \emph{Cellular \& Molecular Immunology} 1--11.

\bibitem[Rajendran et~al.(2019)Rajendran, Challis, Fowlkes, Hanna, Tompkins,
  Jordan, Hiyari, Gabris-Weber, Greenbaum, Chan,
  et~al.]{rajendran2019identification}
Rajendran, P.~S., R.~C. Challis, C.~C. Fowlkes, P.~Hanna, J.~D. Tompkins, M.~C.
  Jordan, S.~Hiyari, B.~A. Gabris-Weber, A.~Greenbaum, K.~Y. Chan, et~al.,
  2019.
\newblock Identification of peripheral neural circuits that regulate heart rate
  using optogenetic and viral vector strategies.
\newblock \emph{Nature communications} 10:1944.

\bibitem[Jean(2001)]{jean2001brain}
Jean, A., 2001.
\newblock Brain stem control of swallowing: neuronal network and cellular
  mechanisms.
\newblock \emph{Physiological reviews} 81:929--969.

\bibitem[Harrington et~al.(2007)Harrington, Hutson, and
  Southwell]{harrington2007immunohistochemical}
Harrington, A., J.~Hutson, and B.~Southwell, 2007.
\newblock Immunohistochemical localisation of cholinergic muscarinic receptor
  subtype 1 (M1r) in the guinea pig and human enteric nervous system.
\newblock \emph{Journal of chemical neuroanatomy} 33:193--201.

\bibitem[Koh et~al.(2022)Koh, Drumm, Lu, Kim, Ryoo, Kim, Lee, Rhee, Wang,
  Gould, et~al.]{koh2022propulsive}
Koh, S.~D., B.~T. Drumm, H.~Lu, H.~J. Kim, S.-B. Ryoo, H.-U. Kim, J.~Y. Lee,
  P.-L. Rhee, Q.~Wang, T.~W. Gould, et~al., 2022.
\newblock Propulsive colonic contractions are mediated by inhibition-driven
  poststimulus responses that originate in interstitial cells of Cajal.
\newblock \emph{Proceedings of the National Academy of Sciences}
  119:e2123020119.

\bibitem[Acharya et~al.(2021{\natexlab{a}})Acharya, Halder, Carlson, Kou,
  Kahrilas, Pandolfino, and Patankar]{AcharyaEsoWork2020}
Acharya, S., S.~Halder, D.~A. Carlson, W.~Kou, P.~J. Kahrilas, J.~E.
  Pandolfino, and N.~A. Patankar, 2021.
\newblock Assessment of esophageal body peristaltic work using functional lumen
  imaging probe panometry.
\newblock \emph{American Journal of Physiology-Gastrointestinal and Liver
  Physiology} 320:G217--G226.

\bibitem[Halder et~al.(2023)Halder, Pandolfino, Kahrilas, Koop, Schauer,
  Araujo, Elisha, Kou, Patankar, and Carlson]{halder2023assessing}
Halder, S., J.~E. Pandolfino, P.~J. Kahrilas, A.~Koop, J.~Schauer, I.~K.
  Araujo, G.~Elisha, W.~Kou, N.~A. Patankar, and D.~A. Carlson, 2023.
\newblock Assessing mechanical function of peristalsis with functional lumen
  imaging probe panometry: Contraction power and displaced volume.
\newblock \emph{Neurogastroenterology \& Motility} 35:e14692.

\bibitem[Jie et~al.(2010)Jie, Gurev, and Trayanova]{Jie2010}
Jie, X., V.~Gurev, and N.~Trayanova, 2010.
\newblock Mechanisms of mechanically induced spontaneous arrhythmias in acute
  regional ischemia.
\newblock \emph{Circulation research} 106:185--192.

\bibitem[Jelin{\v{c}}i{\'c} et~al.(2022)Jelin{\v{c}}i{\'c}, Van~Diest, Torta,
  and von Leupoldt]{Jelinvcic2022breathing}
Jelin{\v{c}}i{\'c}, V., I.~Van~Diest, D.~M. Torta, and A.~von Leupoldt, 2022.
\newblock The breathing brain: The potential of neural oscillations for the
  understanding of respiratory perception in health and disease.
\newblock \emph{Psychophysiology} 59:e13844.

\bibitem[Andersson and Arner(2004)]{andersson2004urinary}
Andersson, K.-E., and A.~Arner, 2004.
\newblock Urinary bladder contraction and relaxation: physiology and
  pathophysiology.
\newblock \emph{Physiological reviews} 84:935--986.

\bibitem[Berridge(2008)]{Berridge2008smooth}
Berridge, M.~J., 2008.
\newblock Smooth muscle cell calcium activation mechanisms.
\newblock \emph{The Journal of physiology} 586:5047--5061.

\bibitem[Jung and Vij(2021)]{jung2021early}
Jung, T., and N.~Vij, 2021.
\newblock Early diagnosis and real-time monitoring of regional lung function
  changes to prevent chronic obstructive pulmonary disease progression to
  severe emphysema.
\newblock \emph{Journal of Clinical Medicine} 10:5811.

\bibitem[Leclercq et~al.(2002)Leclercq, Faris, Tunin, Johnson, Kato, Evans,
  Spinelli, Halperin, McVeigh, and Kass]{Leclercq2002}
Leclercq, C., O.~Faris, R.~Tunin, J.~Johnson, R.~Kato, F.~Evans, J.~Spinelli,
  H.~Halperin, E.~McVeigh, and D.~A. Kass, 2002.
\newblock Systolic improvement and mechanical resynchronization does not
  require electrical synchrony in the dilated failing heart with left
  bundle-branch block.
\newblock \emph{Circulation} 106:1760--1763.

\bibitem[Usyk and Mcculloch(2003)]{Usyk2003}
Usyk, T.~P., and A.~D. Mcculloch, 2003.
\newblock Relationship between regional shortening and asynchronous electrical
  activation in a three-dimensional model of ventricular electromechanics.
\newblock \emph{Journal of cardiovascular electrophysiology} 14:S196--S202.

\bibitem[Niederer et~al.(2011)Niederer, Plank, Chinchapatnam, Ginks, Lamata,
  Rhode, Rinaldi, Razavi, and Smith]{Niederer2011}
Niederer, S.~A., G.~Plank, P.~Chinchapatnam, M.~Ginks, P.~Lamata, K.~S. Rhode,
  C.~A. Rinaldi, R.~Razavi, and N.~P. Smith, 2011.
\newblock Length-dependent tension in the failing heart and the efficacy of
  cardiac resynchronization therapy.
\newblock \emph{Cardiovascular research} 89:336--343.

\bibitem[Orvar et~al.(1993)Orvar, Gregersen, and Christensen]{Orvar1993}
Orvar, K.~B., H.~Gregersen, and J.~Christensen, 1993.
\newblock Biomechanical characteristics of the human esophagus.
\newblock \emph{Digestive diseases and sciences} 38:197--205.

\bibitem[Neuhuber and W{\"o}rl(2016)]{Neuhuber2016}
Neuhuber, W.~L., and J.~W{\"o}rl, 2016.
\newblock Enteric co-innervation of striated muscle in the esophagus: still
  enigmatic?
\newblock \emph{Histochemistry and cell biology} 146:721--735.

\bibitem[Carlson et~al.(2022)Carlson, Prescott, Germond, Brenner, Carns,
  Correia, Tetreault, McMahan, Hinchcliff, Kou,
  et~al.]{Carlson2022heterogeneity}
Carlson, D.~A., J.~E. Prescott, E.~Germond, D.~Brenner, M.~Carns, C.~S.
  Correia, M.-P. Tetreault, Z.~H. McMahan, M.~Hinchcliff, W.~Kou, et~al., 2022.
\newblock Heterogeneity of primary and secondary peristalsis in systemic
  sclerosis: A new model of “scleroderma esophagus”.
\newblock \emph{Neurogastroenterology \& Motility} 34:e14284.

\bibitem[Acharya et~al.(2021{\natexlab{b}})Acharya, Kou, Halder, Carlson,
  Kahrilas, Pandolfino, and Patankar]{Acharya2021Pumping}
Acharya, S., W.~Kou, S.~Halder, D.~A. Carlson, P.~J. Kahrilas, J.~E.
  Pandolfino, and N.~A. Patankar, 2021.
\newblock Pumping Patterns and Work Done During Peristalsis in Finite-Length
  Elastic Tubes.
\newblock \emph{Journal of Biomechanical Engineering} 143.
\newblock \urlprefix\url{http://dx.doi.org/10.1115/1.4050284}.

\bibitem[Lang et~al.(2019)Lang, Medda, and Shaker]{Lang2019}
Lang, I.~M., B.~K. Medda, and R.~Shaker, 2019.
\newblock Characterization and mechanism of the esophago-esophageal contractile
  reflex of the striated muscle esophagus.
\newblock \emph{American Journal of Physiology-Gastrointestinal and Liver
  Physiology} 317:G304--G313.

\bibitem[Sengupta(2000)]{Sengupta2000}
Sengupta, J., 2000.
\newblock An overview of esophageal sensory receptors.
\newblock \emph{The American journal of medicine} 108:87--89.

\bibitem[Omari et~al.(2022)Omari, Zifan, Cock, and Mittal]{Omari2022}
Omari, T.~I., A.~Zifan, C.~Cock, and R.~K. Mittal, 2022.
\newblock Distension contraction plots of pharyngeal/esophageal peristalsis:
  next frontier in the assessment of esophageal motor function.
\newblock \emph{American Journal of Physiology-Gastrointestinal and Liver
  Physiology} 323:G145--G156.

\bibitem[Pehlevan et~al.(2016)Pehlevan, Paoletti, and Mahadevan]{Pehlevan2016}
Pehlevan, C., P.~Paoletti, and L.~Mahadevan, 2016.
\newblock Integrative neuromechanics of crawling in D. melanogaster larvae.
\newblock \emph{Elife} 5:e11031.
\newblock \urlprefix\url{https://doi.org/10.7554/eLife.11031}.

\bibitem[Gjorgjieva et~al.(2013)Gjorgjieva, Berni, Evers, and
  Eglen]{Gjorgjieva2013}
Gjorgjieva, J., J.~Berni, J.~F. Evers, and S.~J. Eglen, 2013.
\newblock Neural circuits for peristaltic wave propagation in crawling
  Drosophila larvae: analysis and modeling.
\newblock \emph{Frontiers in computational neuroscience} 7:24.

\bibitem[Hughes and Thomas(2007)]{Hughes2007}
Hughes, C.~L., and J.~B. Thomas, 2007.
\newblock A sensory feedback circuit coordinates muscle activity in Drosophila.
\newblock \emph{Molecular and Cellular Neuroscience} 35:383--396.

\bibitem[Nash and Panfilov(2004)]{Nash2004}
Nash, M.~P., and A.~V. Panfilov, 2004.
\newblock Electromechanical model of excitable tissue to study reentrant
  cardiac arrhythmias.
\newblock \emph{Progress in biophysics and molecular biology} 85:501--522.

\bibitem[Elisha et~al.(2022{\natexlab{a}})Elisha, Acharya, Halder, Carlson,
  Kou, Kahrilas, Pandolfino, and Patankar]{Elisha2022Regime}
Elisha, G., S.~Acharya, S.~Halder, D.~A. Carlson, W.~Kou, P.~J. Kahrilas, J.~E.
  Pandolfino, and N.~A. Patankar, 2022.
\newblock Peristaltic regimes in esophageal transport.
\newblock \emph{Biomechanics and Modeling in Mechanobiology} 1--19.
\newblock \urlprefix\url{https://doi.org/10.1007/s10237-022-01625-x}.

\bibitem[Elisha et~al.(2022{\natexlab{b}})Elisha, Halder, Carlson, Kou,
  Kahrilas, Pandolfino, and Patankar]{Elisha2022Sphincter}
Elisha, G., S.~Halder, D.~A. Carlson, W.~Kou, P.~J. Kahrilas, J.~E. Pandolfino,
  and N.~A. Patankar, 2022.
\newblock A mechanics--based perspective on the function of human sphincters
  during functional luminal imaging probe manometry.
\newblock (preprint).

\bibitem[Ottesen(2003)]{Ottesen2003}
Ottesen, J., 2003.
\newblock Valveless pumping in a fluid-filled closed elastic tube-system:
  one-dimensional theory with experimental validation.
\newblock \emph{Journal of Mathematical Biology} 46:309--332.
\newblock \urlprefix\url{https://doi.org/10.1007/s00285-002-0179-1}.

\bibitem[Whittaker et~al.(2010)Whittaker, Heil, Jensen, and
  Waters]{Whittaker2010}
Whittaker, R.~J., M.~Heil, O.~E. Jensen, and S.~L. Waters, 2010.
\newblock A Rational Derivation of a Tube Law from Shell Theory.
\newblock \emph{The Quarterly Journal of Mechanics and Applied Mathematics}
  63:465--496.
\newblock \urlprefix\url{https://doi.org/10.1093/qjmam/hbq020}.

\bibitem[Kwiatek et~al.(2011)Kwiatek, Hirano, Kahrilas, Rothe, Luger, and
  Pandolfino]{Kwiatek2011}
Kwiatek, M.~A., I.~Hirano, P.~J. Kahrilas, J.~Rothe, D.~Luger, and J.~E.
  Pandolfino, 2011.
\newblock Mechanical Properties of the Esophagus in Eosinophilic Esophagitis.
\newblock \emph{Gastroenterology} 140:82--90.
\newblock \urlprefix\url{https://doi.org/10.1053/j.gastro.2010.09.037}.

\bibitem[Manopoulos et~al.(2006)Manopoulos, Mathioulakis, and
  Tsangaris]{Manopoulos2006}
Manopoulos, C.~G., D.~S. Mathioulakis, and S.~G. Tsangaris, 2006.
\newblock One-dimensional model of valveless pumping in a closed loop and a
  numerical solution.
\newblock \emph{Physics of Fluids} 18:017106.
\newblock \urlprefix\url{https://doi.org/10.1063/1.2165780}.

\bibitem[Bringley et~al.(2008)Bringley, Childress, Vandenberghe, and
  Zhang]{Bringley2008}
Bringley, T.~T., S.~Childress, N.~Vandenberghe, and J.~Zhang, 2008.
\newblock An experimental investigation and a simple model of a valveless pump.
\newblock \emph{Physics of Fluids} 20:033602.
\newblock \urlprefix\url{https://doi.org/10.1063/1.2890790}.

\bibitem[Pullan et~al.(2004)Pullan, Cheng, Yassi, and Buist]{Pullan2004}
Pullan, A., L.~Cheng, R.~Yassi, and M.~Buist, 2004.
\newblock Modelling gastrointestinal bioelectric activity.
\newblock \emph{Progress in biophysics and molecular biology} 85:523--550.

\bibitem[Chambers et~al.(2008)Chambers, Bornstein, and Thomas]{Chambers2008}
Chambers, J.~D., J.~C. Bornstein, and E.~A. Thomas, 2008.
\newblock Insights into mechanisms of intestinal segmentation in guinea pigs: a
  combined computational modeling and in vitro study.
\newblock \emph{American Journal of Physiology-Gastrointestinal and Liver
  Physiology} 295:G534--G541.

\bibitem[Elisha et~al.(2023{\natexlab{a}})Elisha, Halder, Acharya, Carlson,
  Kou, Kahrilas, Pandolfino, and Patankar]{Elisha2023EGJLoop}
Elisha, G., S.~Halder, S.~Acharya, D.~A. Carlson, W.~Kou, P.~J. Kahrilas, J.~E.
  Pandolfino, and N.~A. Patankar, 2023.
\newblock A mechanics--based perspective on the function of the esophagogastric
  junction during functional luminal imaging probe manometry.
\newblock \emph{Biomechanics and Modeling in Mechanobiology} 1--19.

\bibitem[Gregersen et~al.(2008)Gregersen, Pedersen, and Drewes]{Gregersen2008}
Gregersen, H., J.~Pedersen, and A.~M. Drewes, 2008.
\newblock Deterioration of muscle function in the human esophagus with age.
\newblock \emph{Digestive diseases and sciences} 53:3065--3070.

\bibitem[Kou et~al.(2015)Kou, Pandolfino, Kahrilas, and Patankar]{Kou2015ajpgi}
Kou, W., J.~E. Pandolfino, P.~J. Kahrilas, and N.~A. Patankar, 2015.
\newblock Simulation studies of circular muscle contraction, longitudinal
  muscle shortening, and their coordination in esophageal transport.
\newblock \emph{American Journal of Physiology-Gastrointestinal and Liver
  Physiology} 309:G238--G247.
\newblock \urlprefix\url{https://doi.org/10.1152/ajpgi.00058.2015}.

\bibitem[Halder et~al.(2021)Halder, Acharya, Kou, Kahrilas, Pandolfino, and
  Patankar]{Halder_2021}
Halder, S., S.~Acharya, W.~Kou, P.~J. Kahrilas, J.~E. Pandolfino, and N.~A.
  Patankar, 2021.
\newblock Mechanics informed fluoroscopy of esophageal transport.
\newblock \emph{Biomechanics and Modeling in Mechanobiology} 20.
\newblock \urlprefix\url{https://doi.org/10.1007/s10237-021-01420-0}.

\bibitem[Elisha et~al.(2023{\natexlab{b}})Elisha, Halder, Carlson, Kahrilas,
  Pandolfino, and Patankar]{Elisha2023EsophBodyLoop}
Elisha, G., S.~Halder, D.~A. Carlson, P.~J. Kahrilas, J.~E. Pandolfino, and
  N.~A. Patankar, 2023.
\newblock A mechanics--based perspective on the pressure-cross-sectional area
  loop within the esophageal body.
\newblock \emph{Frontiers in Physiology} 13:2693.

\bibitem[Wang(1999)]{wang1999relaxation}
Wang, D., 1999.
\newblock Relaxation oscillators and networks.
\newblock \emph{Wiley encyclopedia of electrical and electronics engineering}
  18:396--405.

\bibitem[Cherubini et~al.(2008)Cherubini, Filippi, Nardinocchi, and
  Teresi]{cherubini2008electromechanical}
Cherubini, C., S.~Filippi, P.~Nardinocchi, and L.~Teresi, 2008.
\newblock An electromechanical model of cardiac tissue: Constitutive issues and
  electrophysiological effects.
\newblock \emph{Progress in biophysics and molecular biology} 97:562--573.

\bibitem[Grasman(2012)]{grasman2012asymptotic}
Grasman, J., 2012.
\newblock Asymptotic methods for relaxation oscillations and applications,
  volume~63.
\newblock Springer Science \& Business Media.

\bibitem[Daniel et~al.(1994)Daniel, Bardakjian, Huizinga, and
  Diamant]{daniel1994relaxation}
Daniel, E., B.~Bardakjian, J.~Huizinga, and N.~Diamant, 1994.
\newblock Relaxation oscillator and core conductor models are needed for
  understanding of GI electrical activities.
\newblock \emph{American Journal of Physiology-Gastrointestinal and Liver
  Physiology} 266:G339--G349.

\bibitem[Jones and Kopell(2006)]{jones2006local}
Jones, S., and N.~Kopell, 2006.
\newblock Local network parameters can affect inter-network phase lags in
  central pattern generators.
\newblock \emph{Journal of mathematical biology} 52:115--140.

\bibitem[Collins and Stewart(1993)]{collins1993coupled}
Collins, J.~J., and I.~N. Stewart, 1993.
\newblock Coupled nonlinear oscillators and the symmetries of animal gaits.
\newblock \emph{Journal of Nonlinear science} 3:349--392.

\bibitem[Wang(1995)]{wang1995emergent}
Wang, D., 1995.
\newblock Emergent synchrony in locally coupled neural oscillators.
\newblock \emph{IEEE transactions on neural networks} 6:941--948.

\bibitem[Campbell and Wang(1996)]{campbell1996synchronization}
Campbell, S., and D.~Wang, 1996.
\newblock Synchronization and desynchronization in a network of locally coupled
  Wilson-Cowan oscillators.
\newblock \emph{IEEE transactions on neural networks} 7:541--554.

\end{thebibliography}
\end{document}